\global\def\draftcontrol{0}

   \def\versionno{ n=2 quenches }
\catcode`\@=11

\expandafter\ifx\csname draftcontrol\endcsname\relax\global\def\draftcontrol{0}
\fi

{\count255=\time\divide\count255 by 60
\xdef\hourmin{\number\count255}
\multiply\count255 by-60\advance\count255 by\time
\xdef\hourmin{\hourmin:\ifnum\count255<10 0\fi\the\count255}}
\def\draftdate{\number\month/\number\day/\number\year\ \ \ \hourmin }

\newcommand\makepapertitle{\par
  \begingroup
    \renewcommand\thefootnote{\@fnsymbol\c@footnote}%
    \def\@makefnmark{\rlap{\@textsuperscript{\normalfont\@thefnmark}}}%
    \long\def\@makefntext##1{\parindent 1em\noindent
            \hb@xt@1.8em{%
                \hss\@textsuperscript{\normalfont\@thefnmark}}##1}%
     \newpage
     \global\@topnum\z@   % Prevents figures from going at top of page.
     \@makepapertitle
     \thispagestyle{empty}\@thanks
  \endgroup
  \setcounter{footnote}{0}%
  \global\let\thanks\relax
  \global\let\makepapertitle\relax
  \global\let\@makepapertitle\relax
  \global\let\@thanks\@empty
  \global\let\@author\@empty
  \global\let\@date\@empty
  \global\let\@title\@empty
  \global\let\title\relax
  \global\let\author\relax
  \global\let\date\relax
  \global\let\and\relax
  \def\version{\let\version\@version\@gobble}
}
\def\@makepapertitle{%
  \newpage
   \ifnum\draftcontrol=1 {}
   \version\versionno
   \vskip 3em%
   \else
   \hfill\hbox to 3cm {\parbox{4cm}{\@pubnum}\hss}%
   \vskip 3em%
   \fi
   \begin{center}%
   \let \footnote \thanks
     {\LARGE {\@title}}%
     \vskip 1.5em%
     {\normalsize%\large
       \lineskip .5em%
       \begin{tabular}[t]{c}%
         \@author
       \end{tabular}\par}%
     \vskip 1.5em%
     {\@bstract}%
     \end{center}%
     \vskip 1.5em
     \@date%
   \par
}

\gdef\@pubnum{}
\def\pubnum#1{%
  \gdef\@pubnum{#1}}

\gdef\@bstract{}
\def\Abstract#1{%
  \gdef\@bstract{%
   \parbox{\textwidth-0pc}{%
   \centerline{\bf Abstract}\penalty1000%
\kern.2cm%
\noindent%\abstractfont \baselineskip=12pt
\renewcommand\baselinestretch{1.0}%
{#1}}}
}

\def\ps@paper{\let\@mkboth\@gobbletwo%
     \ifnum\draftcontrol=1
    \def\@oddfoot{\hbox to \textwidth{\tiny \versionno \hfil\tiny\draftdate}%
    \hskip -\textwidth \hbox to \textwidth{\hfil\rm\thepage\hfil}}%
     \else\def\@oddfoot{\hbox to \textwidth{\hfil\rm\thepage\hfil}}
     \fi
     \let\@evenfoot\@oddfoot
}
\def\body{\clearpage
          \pagestyle{paper}
    }
\def\@version#1{\ifnum\draftcontrol=1
\typeout{}\typeout{#1}\typeout{}
\vskip3mm\centerline{\hbox{\fbox{\normalsize{\tt DRAFT -- #1 -- }
                   {\draftdate}}}}\vskip3mm
\fi}
\let\version\@version
\long\def\eqlabel#1{\ifnum\draftcontrol=1
                    \tag@false  % there are some problems with multline without this
                    \tag*{(\theequation) \hbox to -0.2cm{\hspace{0cm}\small{#1}\hss}}
                    \refstepcounter{equation}
                    \edef\@currentlabel{\theequation}
                    \ltx@label{#1}          % use old LaTeX \label instead of new definition
                    \else
                    \label{#1}
                    \fi
                    }
\let\st@bibitem\@bibitem
\let\st@lbibitem\@lbibitem
\ifnum\draftcontrol=1
  \def\@bibitem#1{%
    \st@bibitem{#1}\a@@label{#1}\ignorespaces}
  \def\@lbibitem[#1]#2{%
    \st@lbibitem[#1]{#2}\a@@label{#2}\ignorespaces}
  \def\a@@label#1{%
    \gdef\a@lab{\smash{\normalfont\small#1}}
    \ifvmode
      \if@inlabel
        \global\setbox\@labels\hbox{%
          \llap{\a@lab\let\a@lab\relax
                \kern\@totalleftmargin\kern\marginparsep}%
          \box\@labels}%
      \fi
    \fi}
\fi
\documentclass[12pt,letterpaper]{article}
\usepackage{amsmath,amssymb,array,calc,epsfig,rotating,bm,appendix}
\usepackage[sort]{cite}
\usepackage{graphicx}
\usepackage{psfrag,verbatim}

\ifnum\draftcontrol=1
\fi
\renewcommand\baselinestretch{1.25}
\renewcommand\section{\@startsection {section}{1}{\z@}%
                                   {-3.5ex \@plus -1ex \@minus -.2ex}%
                                   {2.3ex \@plus.2ex}%
                                   {\normalfont\large\bfseries}}
\renewcommand\subsection{\@startsection{subsection}{2}{\z@}%
                                   {-3.25ex\@plus -1ex \@minus -.2ex}%
                                   {1.5ex \@plus .2ex}%
                                   {\normalfont\normalsize\bfseries}}
\renewcommand\subsubsection{\@startsection{subsubsection}{3}{\z@}%
                                   {-3.25ex\@plus -1ex \@minus -.2ex}%
                                   {1.5ex \@plus .2ex}%
                                   {\normalfont\normalsize\it}}
\renewcommand\paragraph{\@startsection{paragraph}{4}{\z@}%
                                   {-3.25ex\@plus -1ex \@minus -.2ex}%
                                   {1.5ex \@plus .2ex}%
                                   {\normalfont\normalsize\bf}}

\numberwithin{equation}{section}

\def\revise#1       {\raisebox{-0em}{\rule{3pt}{1em}}%
                     \marginpar{\raisebox{.5em}{\vrule width3pt\
                     \vrule width0pt height 0pt depth0.5em
                     \hbox to 0cm{\hspace{0cm}{%
                     \parbox[t]{4em}{\raggedright\footnotesize{#1}}}\hss}}}}

\newcommand{\ie}{{\it i.e.,}\ }
\newcommand{\eg}{{\it e.g.,}\ }

\newcommand{\beq}{\begin{equation}}
\newcommand{\eeq}{\end{equation}}
\newcommand{\beqa}{\begin{eqnarray}}
\newcommand{\eeqa}{\end{eqnarray}}
\newcommand{\labell}[1]{\eqlabel{#1}}
\newcommand{\eqlabell}[1]{\eqlabel{#1}}
\newcommand{\reef}[1]{\eqref{#1}}

\def\cale         {{\cal E}}

\def\caln         {{\cal N}}
\def\calo         {{\cal O}}
\def\calp         {{\cal P}}

\def\del          {\partial}

\def\sqr#1#2{{\vcenter{\vbox{\hrule height.#2pt
 \hbox{\vrule width.#2pt height#1pt \kern#1pt
 \vrule width.#2pt}\hrule height.#2pt}}}}

\def\a{\alpha}

\def\dd{\delta}
\def\DD{\Delta}

\def\aa1{\phi}
\def\cc1{\psi}

\def\D{\Delta}

\def\l{\lambda}

\def\ra{\Longrightarrow}

\def\t{\tau}

\def\la{\langle}
\def\ra{\rangle}

\def\pp{\alpha_\lambda}
\def\q{\hat{\phi}}

\catcode`\@=12

\begin{document}

%%%
%%%%%% text starts here
%%%%%%%%%

\title{\bf  Quantum quenches
of holographic plasmas} %by an operator of arbitrary dimension} 
\pubnum{UWO-TH-13/2}

%\title{\bf Holographic quantum quenches\\
%at finite temperature}

\date\today

\author{
Alex Buchel,$^{1,2}$ Luis Lehner,$^{1}$ Robert C. Myers,$^1$ and Anton van Niekerk$^{1,3}$\\[0.4cm]
\it $^1$\,Perimeter Institute for Theoretical Physics\\
\it Waterloo, Ontario N2J 2W9, Canada\\[0.2cm]
\it $^2$\,Department of Applied Mathematics,
University of Western Ontario\\
\it London, Ontario N6A 5B7, Canada\\[0.2cm]
\it $^3$\,Department of Physics \& Astronomy and Guelph-Waterloo Physics Institute\\
\it University of Waterloo, Waterloo, Ontario N2L 3G1, Canada\\[0.2cm]
}

\Abstract{We employ holographic techniques to study quantum quenches at finite
temperature, where the quenches involve varying the coupling of the boundary
theory to a relevant operator with an arbitrary conformal dimension $2\leq\D\leq4$.
The evolution of the system is studied by evaluating the expectation value of the
quenched operator and the stress tensor throughout the process. The time
dependence of the new coupling is characterized by a fixed timescale and the
response of the observables depends on the ratio of the this timescale to the
initial temperature. The observables exhibit universal scaling behaviours
when the transitions are either fast or slow, \ie when this ratio is very small
or very large. The scaling exponents are smooth functions of the operator
dimension. We find that in fast quenches, the relaxation time  is set by the
thermal timescale regardless of the operator dimension or the precise quenching
rate.}

\makepapertitle

\body

\version\versionno
\tableofcontents

\section{Introduction}

Recent advances in cold atom experiments have stimulated a vigorous research
program into quantum quenches, processes in which the physical couplings of a
quantum system are abruptly changed \cite{more}. The basic motivation is to
understand the organizing principles governing the far-from-equilibrium
behaviour of such systems. Although such quenches are well understood in the
context of quantum mechanics \cite{LL}, much less is known about such processes
in quantum field theories. Theoretical progress has been made for a variety of
systems, including two-dimensional conformal field theories, (nearly) free
field theories and integrable models -- \eg see \cite{cc1,janet}. However,
broadly applicable theoretical techniques, which provide an efficient
description of these quenches, remain to be found.

Gauge/gravity duality \cite{m1} provides a remarkable new approach to studying
certain strongly coupled field theories. Of course, in this framework,
questions about the field theory are recast into questions about gravity in one
higher dimension. These holographic models seem to be especially well suited
for the study of quantum quenches since, with relatively modest efforts, one is
able to study strongly coupled quantum field theories, real-time processes and
systems at finite temperature, as well as allowing for analysis in general
spacetime dimensions. Hence holographic techniques have recently been applied
to the study of quantum quenches \cite{das4} and the related issue of
`thermalization' \cite{early,thermalize}. However, given the complexities of
the bulk description of rapid changes in the boundary theory, numerical
relativity is increasingly being applied to study these far-from-equilibrium
processes \cite{cy,adsnumerics} --- see also~\cite{NRHEP}.

In this paper, we extend the calculations presented in \cite{blm}, in which the
gauge/gravity duality was used to study `thermal quenches' in a plasma of the
strongly coupled $\caln=2^*$ gauge theory. More specifically, \cite{blm}
studied the response of an initial thermal equilibrium state to variations of
the coupling of the boundary theory to either a dimension two or three
operator. The analysis was restricted to a high temperature regime where the
calculations were carried out to leading order in ${m}/{T}\ll 1$. Here $m$ is
the relevant mass scale introduced by the new coupling. In the present case, we
extend these holographic calculations to consider quenches made by coupling to
a relevant operator with an arbitrary conformal dimension in the range
$2\leq\D\leq4$.  The behaviour of the strongly
coupled boundary theory in the present quenches is very similar to that found
in \cite{blm}. In fact, for many of the results, we are able to identify a
simple function of the conformal dimension which interpolates between the
different cases which are explicitly studied both here and in \cite{blm}. For
example, we find that in fast quenches, the increase in the energy density
scales like $(T_i/\D t)^{2\D-4}$, where $T_i$ is the initial temperature and
$\Delta t$ is the timescale over which the new coupling is turned on --- fast
quenches are then those for which $T_i/\D t\gg1$.

The remainder of the paper is organized as follows:  In section \ref{model}, we
describe the holographic model which is used to study our quenches  and derive
the gravitational equations that are to be solved. Next we examine solutions of
these equations in section \ref{solution}. In particular, by restricting our
attention to the high temperature regime, we show that to leading order we only
need to solve the linearized equation for the bulk scalar. We also consider the
asymptotic boundary expansion for these solutions in Eddington-Finkelstein
coordinates. In section \ref{feffermans}, we translate the latter expansion to
Fefferman-Graham coordinates, which are more suitable to study physical
observables in the boundary theory. In section \ref{renorm}, after finding the
counterterms that renormalize the bulk action, we find expressions for the
expectation value of the stress-energy tensor and the quenched operator in
terms of gravitational variables. We also show that these observables obey the
expected Ward identities. In section \ref{entropy}, we identify the appropriate
translation between the gravitational variables and quantities in the boundary
theory. This dictionary allows us to write expressions for the entropy
production and the change in other thermodynamic quantities induced by the
quench. Section \ref{numerics} provides a brief description of our numerical
procedure. In section \ref{slowquenches}, we provide an independent analysis of
the response in the slow quench limit, which later provides a check of our
numerical results. We present and discuss various aspects of our numerical
solutions in section \ref{results}. We conclude with a summary of the results
and further comments in section \ref{discussion}. Finally, there are various
appendices describing certain technical details. Appendix \ref{coeff} presents
explicit coefficients for the leading terms in the asymptotic of the expansion
in section \ref{solution}. Similarly, appendix \ref{fefferman} presents
coefficients for the asymptotic expansions appearing in section
\ref{feffermans}. Finally, appendix \ref{stress} describes the variations of
the renormalized bulk action constructed in section \ref{renorm}, which yield
the expectation value of the stress tensor and the quenched operator.

\section{Holographic model} \label{model}

We will apply holographic techniques to study quantum quenches in a strongly
coupled four-dimensional QFT. The quantum quench is implemented by adding a
relevant operator with time-dependent coupling to the Lagrangian of the QFT, as
follows \cite{blm}:
\begin{equation}
\mathcal{L}_0 \rightarrow \mathcal{L}_0 + \l(t)\,\mathcal{O}_{\Delta}.
\end{equation}
In our calculations, the theory described by $\mathcal{L}_0$ is in fact a
conformal field theory. The operator $\mathcal{O}_{\Delta}$ is relevant,
meaning that it has conformal dimension $\Delta < 4$. We will only consider
$\Delta > 2$ here, as natural in our holographic framework --- see below. In
our analysis, we start with the theory in a thermal state with $\lambda=0$ and
quench the system by switching on the coupling to some non-zero value. To
further simplify our analysis, we will focus on quenches in the high
temperature regime, where the temperature $T$ provides the dominant scale in
the problem. That is, we will only study quenches where $\lambda\ll
T^{4-\Delta}$ at all stages. Note, however, that we will allow the rate of
change of coupling $\lambda$ to be arbitrarily large. In particular, we allow
$\partial_t\l(t)\gtrsim T^{5-\Delta}$.

As the unperturbed QFT is a four-dimensional conformal field theory, the
gravitational dual of the vacuum state is five-dimensional anti-de Sitter
(AdS$_5$) spacetime. Since we are interested instead in a thermal state of the
boundary CFT, the appropriate dual spacetime is an asymptotically AdS$_5$
planar black hole \cite{wit99} --- we consider the boundary QFT in $R^{1,3}$,
hence the `planar' geometry for the black hole horizon. Switching on the
coupling $\l$ is dual to switching on a massive scalar field in the
gravitational theory. More precisely, we are modifying the asymptotic boundary
conditions for the bulk scalar field in a way that matches the profile $\l(t)$.
Using holographic methods, we can easily determine the response of the QFT by
examining the response of the scalar, which yields $\langle\mathcal{O}_{\Delta}
\rangle$, as well as the response of the spacetime metric, which yields the
energy density, pressure and entropy density of the boundary field theory.
Since we are considering the high temperature regime, our calculations will be
perturbative in the amplitude of the bulk scalar. That is, the scalar will only
produce `small' perturbations on the AdS$_5$ black hole background.

The dimension $\Delta$ of the operator $\mathcal{O}_{\Delta}$ is related to the
AdS length scale $L$ and the mass $m$ of the bulk scalar field by \cite{m2}
\begin{equation}
\Delta = 2 + \sqrt{4+L^{2}m^{2}}.
\labell{delta}
\end{equation}
Notice that a relevant operator is dual to a scalar field with $m^{2}\leq0$. Of
course, such a tachyonic mass is still consistent in five-dimensional AdS space
as long as it respects the Breitenlohner-Freedman bound \cite{BF}, \ie
$m^{2}\geq -4/L^{2}$. In eq.~\reef{delta}, this imposes the constraint
$\Delta\geq 2$. The unitarity bound for a scalar operator in the
four-dimensional CFT allows for $\Delta\ge 1$, however, to study operators in
the range $2>\Delta\ge1$, we must use the `alternative quantization' of the
dual bulk scalar set forward in \cite{igor9}. However, we will not consider
this possibility in the following and restrict our attention to $\Delta> 2$.

The dual gravitational theory is Einstein gravity coupled to a cosmological
constant and a massive scalar field, \ie
\begin{equation}
S_{bulk} = \frac{1}{16\pi G^{(5)}_{N}}\int d^{5}x \sqrt{-g}\left(R + 12 -
\frac{1}{2}\left(\partial \Phi\right)^{2} - \frac{1}{2}m^{2}\Phi^{2}\right)\, .
\labell{eq:action}
\end{equation}
Since Newton's constant appears in an overall factor in front of the action,
the scalar field $\Phi$ is dimensionless. Further, we have also implicitly set
the AdS curvature scale to one, \ie $L=1$, as can be inferred from the
cosmological constant term. With this convention, it follows that $m^{2}$ will
also be a dimensionless number, which is implicitly given in units of $1/L^2$.
As explained in \cite{blm}, the scalar field might have further interactions,
\eg a $\Phi^4$ potential, but any such higher order terms will not play a role
in the following analysis describing the high temperature regime.

As an aside, let us comment that it is natural to think of the unperturbed
boundary theory as the  $\mathcal{N}=4$ super-Yang-Mills (SYM) theory, in
the limit of large $N_c$ and large 't Hooft coupling. In this case, our
conventions are such that the five-dimensional Newton's constant is given by
\begin{equation}
G^{(5)}_{N}\equiv \frac{\pi}{2\, N_c^2}\,.
\eqlabell{g5}
\end{equation}
However, we are slightly liberal in our analysis here in that we allow the
conformal dimensions of $\mathcal{O}_{\Delta}$ to take arbitrary values, rather
than restricting ourselves to the spectrum of $\mathcal{N}=4$ SYM. In this more
general context, we can relate Newton's constant to the central charge of the
boundary CFT with
\begin{equation}
C_T\equiv \frac{5}{\pi\,G^{(5)}_{N}}\,.
\eqlabell{g5b}
\end{equation}
where $C_T$ is the central charge defining the leading singularity in the
two-point correlator of the stress tensor --- \eg see \cite{ct}.

Varying the action (\ref{eq:action}) with respect to the metric $g$ and the
scalar field $\Phi$, one obtains respectively Einstein's equations and the
curved-space Klein-Gordon equation
\begin{eqnarray}
0 &=& E_{\mu \nu} \equiv R_{\mu\nu} - \frac{1}{2}\,\partial_{\mu}
\Phi\,\partial_{\nu}\Phi - g_{\mu\nu}\left(\frac{1}{2}R + 6  -
\frac{1}{4}\left(\partial \Phi\right)^{2} - \frac{1}{4}m^{2}\Phi^{2}\right)\,,
\labell{eq:einstein}\\
0 &=& \frac{1}{\sqrt{-g}}\,\partial_{\mu}\!\left(\sqrt{-g}g^{\mu\nu}\partial_{\nu}
\Phi\right)-m^{2}\Phi\,. \labell{eq:kg}
\end{eqnarray}
We express our metric ansatz using infalling Eddington-Finkelstein (EF)
coordinates
\begin{equation}
ds^{2} = -A(v,y)\,dv^{2} + \Sigma^{2}(v,y)\,d\vec{x}^{2} + 2\, dv dy\,,
\labell{eq:efmetric}
\end{equation}
as was used in \cite{cy,gfluid,blm} in the context of holographic thermal
systems. For the scalar in this background, we take $\Phi=\Phi(v,y)$ (\ie it is
independent of the spatial directions $x^{i}$). This choice allows us to
describe homogeneous quenches where the coupling $\lambda$ is spatially
constant but varies in time. The above is a convenient gauge \reef{eq:efmetric}
for numerically evolving the scalar field within a characteristic formulation. The
resulting radial vector $\frac{\partial}{\partial y}$ is null and all points on
a line with constant $v$ (and $x^i$) are causally connected. The resulting system
of (partial differential) equations provide a nested system of
(with both radial and time integrations)  that can be
evolved the spacetime radially from the boundary at $y=\infty$
inwards and in forward in time.  We will return to this discussion when we describe the numerics in
section \ref{numerics}.

With this metric ansatz \reef{eq:efmetric}, the Klein-Gordon equation
\reef{eq:kg} and Einstein's equations \reef{eq:einstein} become \cite{blm}
 \beqa
0&=&2\Sigma \del_y\!(\dot{\Phi})+ 3\,(\del_y\!\Sigma)
\dot{\Phi}+3\dot{\Sigma}\del_y\Phi -m^2\Sigma
\Phi\,,
 \eqlabell{eoms3}\\
0&=&\Sigma\, \del_y\!(\dot{\Sigma})+2\dot{\Sigma}\,\del_y\!\Sigma
-2\Sigma^2+\frac{1}{12}m^2\Phi^2\Sigma^2\,,
 \eqlabell{eoms1}\\
0&=&4+\del_y^2A-\frac{12}{\Sigma^2}\dot{\Sigma}\,\del_y\!\Sigma
+\dot{\Phi}\,\del_y\Phi-\frac 16 m^2\Phi^2\,,
 \eqlabell{eoms2}\\
0&=&\ddot{\Sigma}-\frac 12\,\dot{\Sigma}\, \del_y\!A+\frac 16 \Sigma\,
(\dot{\Phi})^2\,,
 \eqlabell{eoms4}\\
0&=&\del_y^2\Sigma+\frac 16\Sigma\, (\del_y\Phi)^2\,,
 \eqlabell{eoms5}
 \eeqa
where we have defined for any function $h(v,y)$,
\begin{equation}
\dot{h}\equiv \del_v h+\frac 12 A\, \del_y h\,.
\eqlabell{der}
\end{equation}
More precisely, the above equations are obtained as:
\begin{itemize}
\item Eq.~(\ref{eoms3}) is equivalent to the Klein-Gordon equation
    (\ref{eq:kg}) multiplied by $\Sigma$.
\item Eq.~(\ref{eoms1}) corresponds to the combination
\begin{equation}
\frac{1}{3}\Sigma^{2}E_{vy}+\frac{1}{6}A\Sigma^2E_{yy}=0\,.
\end{equation}
\item Eq.~(\ref{eoms2}) corresponds to the combination
\begin{equation}
\frac{1}{3\Sigma^2}\left(6 E_{ii}- 8\Sigma^2E_{vy}-4 A\Sigma^2E_{yy}\right)=0\,.
\end{equation}
Note that $E_{ii}$ denotes one of the diagonal components of $E_{\mu\nu}$
with $\mu=\nu=i$, \ie there is no implicit sum over $i$ in this expression.
\item Eq.~(\ref{eoms4}) corresponds to the combination
\begin{equation}
-\frac{1}{3}\Sigma E_{vv}-\frac{1}{3}A\Sigma E_{vy}-\frac{1}{12}A^2 \Sigma E_{yy}=0\,.
\end{equation}
\item Eq.~(\ref{eoms5}) corresponds to $\Sigma\,E_{yy}=0$.
\end{itemize}
Note that eqs.~\reef{eoms4} and (\ref{eoms5}) are constraint equations, implied
by the previous three equations \cite{blm}.

\section{Solutions to the equations} \label{solution}

\subsection{Static solutions} \label{staticsol}

As noted above, because we study quenches of the boundary QFT from an initial
thermal state, we consider the dual AdS spacetime initially containing a black
hole. With $\Phi=0$, the spacetime will have the static solution
\begin{eqnarray}
A(v,y) &=& y^{2} - \frac{\mu^{4}}{y^{2}}, \nonumber \\
\Sigma(v,y) &=& y, \labell{eq:static}
\end{eqnarray}
where the black hole horizon is located at $y=\mu$ and the asymptotic boundary
of the spacetime is located at $y=\infty$. This black hole solution gives the
gravity description of the original (conformal) boundary theory in thermal
equilibrium. The QFT temperature is given by the temperature of the black hole,
namely $T={\mu}/{\pi}$.\footnote{Our conventions below will introduce a small
correction to this result -- see section \ref{entropy}.}

Now following \cite{blm}, our analysis will be limited to considering a high
temperature regime, where $\l(t) \ll T^{4-\Delta}$. As noted above, this means
that our calculations in the dual gravitational description are perturbative in
the amplitude of the bulk scalar. In other words, we assume that the AdS
spacetime contains a `large' black hole and the scalar only makes `small'
perturbations on this background geometry. If we parameterize the amplitude of
scalar field by the small parameter $\ell$, if follows from the Einstein
equations \reef{eq:einstein} that the scalar only backreacts on the metric at
order $\ell^{2}$. At the lowest order in $\ell$, the scalar and the metric can
therefore be written as \cite{blm}
\begin{eqnarray}
\Phi(v,y) &=& \ell\,\Phi_{\textrm{p}}(v,y) + o\left(\ell^{3}\right)\,, \nonumber \\
A(v,y) &=& y^{2} - \frac{\mu^{4}}{y^{2}} + \mu^{2}\ell^{2}A_{\textrm{p}}(v,y)
+ o\left(\ell^{4}\right)\,,
\labell{hfire}\\
\Sigma(v,y) &=& y + \mu\, \ell^{2} \Sigma_{\textrm{p}}(v,y)+ o\left(\ell^{4}\right)\,,
\nonumber
\end{eqnarray}
where factors of $\mu$ were introduced above to make both metric functions,
$A_{\textrm{p}}(v,y)$ and $\Sigma_{\textrm{p}}(v,y)$, dimensionless.

As a matter of convenience, we now change to the dimensionless coordinates
$\rho\equiv\mu/y$, $\t\equiv\mu v$, as well as $\vec{x}'\equiv\mu\vec{x}$.  For
this choice of radial coordinate, the boundary lies at $\rho=0$ and the black
hole horizon lies at $\rho=1$. The scalar field and the metric coefficients are
then written as
\begin{eqnarray}
\Phi(\t,\rho) &=& \ell\,\Phi_{\textrm{p}}(\t,\rho) + o\left(\ell^{3}\right)\,,
\nonumber\\
A(\t,\rho) &=& \mu^{2}\left(\rho^{-2} - \rho^{2} +
\ell^{2}A_{\textrm{p}}(\t,\rho)+ o\left(\ell^{4}\right)\right)\,,
\labell{hfire2} \\
\Sigma(\t,\rho) &=& \mu\left(\rho^{-1} + \ell^{2} \Sigma_{\textrm{p}}(\t,\rho)+ o\left(\ell^{4}\right)
\right)\, . \nonumber
\end{eqnarray}
In these coordinates, the metric then becomes
\begin{equation}
ds^{2} = \mu^{-2}\left(-A(\t,\rho)\,d\t^{2} + \Sigma^{2}(\t,\rho)\,d\vec{x}'^{2}\right)
- 2 \frac{d\t d\rho}{\rho^{2}}\,.
\labell{eq:efle}
\end{equation}
Note that the factor of $\mu^{-2}$ cancels with the $\mu^2$ contained in the
metric coefficients $A$ and $\Sigma^2$.  The metric, and therefore the
equations of motion will be independent of the black hole mass parameter $\mu$
in these coordinates.

If we consider the Klein-Gordon equation (\ref{eoms3}) to order $\ell$, the
field $\Phi_{\textrm{p}}$ decouples from the metric functions $A_{\textrm{p}}$
and $\Sigma_{\textrm{p}}$ and we are left with the linearized equation
\cite{blm}
\begin{equation}
-\frac{m^{2}  \Phi_{\textrm{p}}}{\rho }+3   \partial_{\t}{\Phi}_{\textrm{p}}-\left(3  +
\rho ^4 \right)\partial_{\rho}\Phi_{\textrm{p}}-2   \rho  \partial_{\t}\partial_{\rho}\Phi_{\textrm{p}}
+ \left( \rho -  \rho ^5 \right)\partial^{2}_{\rho}\Phi_{\textrm{p}}=0. \labell{eq:phi}
\end{equation}
The metric perturbations can then be determined from eqs.~(\ref{eoms1}) and
(\ref{eoms2}) at order $\ell^{2}$ \cite{blm}:
\begin{eqnarray}
0&=& \left[-2 \left(3-\rho^{4}\right)+\rho^{2}  \left(1-\rho^{4}\right) \partial^{2}_{\rho}
+ \rho  \left(4 \partial_{\tau}-4 \partial_{\rho}-2\rho \partial_{\t}\partial_{\rho}\right)\right]
\Sigma_{\textrm{p}} \nonumber \\
 &&\quad + \rho\left[2 -\rho  \partial_{\rho}\right]A_{\textrm{p}} + \frac{m^{2}}{6 \rho}
 \Phi_{\textrm{p}}^{2}\,, \labell{eq:eoml21}\\
&&\nonumber\\
0&=& 24 \left[\partial_{\t}-\frac{1}{\rho}\left(1-\rho ^4\right)\left(1+\rho \partial_{\rho}\right)
\right]\Sigma_{\textrm{p}}+2\left[6-2 \rho  \partial_{\rho}-  \rho ^2 \partial^{2}_{\rho}\right]A_{\textrm{p}}
\nonumber \\
&&\quad+  \left[2  \partial_{\t}\Phi_{\textrm{p}}-\left(1-\rho ^4\right)  \partial_{\rho}\Phi_{\textrm{p}}
\right]\partial_{\rho}\Phi_{\textrm{p}}
+\frac{m^{2}}{ 3 \rho^{2}}  \Phi_{\textrm{p}}^{2}\,. \labell{eq:eoml22}
\end{eqnarray}
Again, note that the mass parameter $\mu$ does not appear in these equations
(\ref{eq:phi})--(\ref{eq:eoml22}).

In the case of a static or equilibrium configuration, eq.~(\ref{eq:phi}) can be
solved for the leading order scalar field
\begin{eqnarray}
\Phi_{\textrm{p}}\left(\rho\right) &=& c_{1}\, \rho ^{4-\Delta }\ {}_{2}\textrm{F}_{1}\left(\frac{4-\Delta }{4},
\frac{4-\Delta }{4},\frac{4-\Delta }{2},\rho ^4\right)\nonumber \\
&&\quad- {c_{1}}\,\frac{ \Gamma\left(\frac{4-\Delta }{2}\right) \Gamma\left(\frac{\Delta }{4}\right)^2}{\Gamma\left(
\frac{4-\Delta }{4}\right)^2 \Gamma\left(\frac{\Delta }{2}\right)}\, \rho ^{\Delta } \ {}_{2}\textrm{F}_{1}\left(
\frac{\Delta }{4},\frac{\Delta }{4},\frac{\Delta }{2},\rho ^4\right)\,, \labell{eq:stsol}
\end{eqnarray}
where $_{2}\textrm{F}_{1}$ denotes a hypergeometric function. The constant
$c_{1}$ is arbitrary but the coefficient of the second term above is chosen to
ensure regularity of the scalar at the horizon. Separately, both
$_{2}\textrm{F}_{1}\left(\frac{4-\Delta }{4},\frac{4-\Delta }{4},\frac{4-\Delta
}{2},\rho ^4\right)$ and $_{2}\textrm{F}_{1}\left(\frac{\Delta
}{4},\frac{\Delta }{4},\frac{\Delta }{2},\rho ^4\right)$ have a logarithmic
divergence near $\rho=1$ but with the relative factor above, these logarithmic
terms cancel in eq.~\reef{eq:stsol}. This static bulk solution will describe
the system (to leading order in $\ell$) after it has equilibrated after the
quench with a finite coupling $\lambda$. Hence it will be useful to extract the
relative magnitude of the normalizable and the non-normalizable modes of the
bulk scalar in this new equilibrium configuration --- see the next section.

\subsection{Time-dependent solutions}

In this subsection, we write down the asymptotic expansion for the leading
order scalar $\Phi_{\textrm{p}}\left(\tau,\rho\right)$ and metric functions,
$A_{\textrm{p}}(\tau,\rho)$ and $\Sigma_{\textrm{p}}(\tau,\rho)$, in a
time-dependent solution. Note that when $\Delta\in\mathbb{Z}$ or
$\Delta\in\mathbb{Z}_{n+\frac{1}{2}}$ (\eg $\Delta=2$ or 3 as in \cite{blm}),
logarithmic terms appear in these asymptotic expansions. However, generically
these expansions do not contain any logarithmic terms and this is the case that
we consider in the following.

The time-dependent solution $\Phi_{\textrm{p}}(\tau,\rho)$ has an asymptotic
expansion close to $\rho=0$ of the form:
\begin{eqnarray}
&&\Phi_{\textrm{p}}\left(\t,\rho\right)\ = \nonumber \\
&& \rho ^{4-\Delta } \left(\phi_{(0)}(\t)+\rho  \dot{\phi}_{(0)}+\frac{(2 \Delta -7)
\rho ^2 }{4 (\Delta -3)}\ddot{\phi}_{(0)}+\frac{(2 \Delta -9) \rho ^3 }{12 (\Delta -3)}
\dddot\phi_{(0)}+\textrm{o}\left(\rho^{4}\right)\right) \labell{eq:phip}\\
&&+\rho ^{\Delta } \left(\phi_{(2\Delta-4)}(\t)+\rho  \dot{\phi}_{(2\Delta-4)}
+\frac{(2 \Delta -1) \rho ^2 }{4 (\Delta -1)}\ddot{\phi}_{(2\Delta-4)}+\frac{(2 \Delta +1)
\rho ^3 }{12 (\Delta -1 )}\dddot\phi_{(2\Delta-4)}+\textrm{o}\left(\rho^{4}\right)\right)\,,
\nonumber
\end{eqnarray}
where the coefficients $\phi_{(0)}$ and $\phi_{(2\Delta-4)}$ are now functions
of $\tau$. Here $\dot{h}\equiv\partial_{\t}h$, for any $\t$-dependent function
$h$.    In the following, we will choose
some function for the coefficient of the non-normalizable mode,
$\phi_{(0)}(\tau)$, and then the normalizable coefficient
$\phi_{(2\Delta-4)}(t)$ is determined by numerically integrating
eq.~\reef{eq:phi}.  However, from the static solution \reef{eq:stsol}, we have
an analytic solution
\begin{equation}
{\rm equilibrium:}\qquad
\phi_{(2\Delta-4)} = -\frac{\Gamma\left(\frac{4-\Delta }{2}\right) \Gamma\left(
\frac{\Delta }{4}\right)^2}{\Gamma\left(\frac{4-\Delta }{4}\right)^2
\Gamma\left(\frac{\Delta }{2}\right)}
\,\phi_{(0)}
\labell{eq:phinfty}
\end{equation}
for the late-time configuration describing the boundary theory after it has
equilibrated with finite $\lambda$.

The solutions for the metric perturbations at order $\ell^{2}$ take the form
\begin{eqnarray}
A_{\textrm{p}}(\t,\rho) &=& \sum_{n=4}\left[a_{2,n}(\t)\rho^{n-2}+
\alpha_{2,n}(\t)\rho^{2-2\Delta+n}+\beta_{2,n}(\t)\rho^{2\Delta-6+n}\right]\,,
  \labell{eq:sola}\\
\Sigma_{\textrm{p}}(\t,\rho) &=& \sum_{n=5}\left[s_{2,n}(\t)\rho^{n-2}+
\sigma_{2,n}(\t)\rho^{2-2\Delta+n}+\theta_{2,n}(\t)\rho^{2\Delta-6+n}\right]\, ,
\labell{eq:sols}
\end{eqnarray}
where (most of) the coefficients can be determined by solving
eqs.~\reef{eq:eoml21} and \reef{eq:eoml22} order by order in powers of $\rho$.
However, the coefficient $a_{2,4}$ enters these equations as a free parameter.
Now taking the limit $\rho\to0$, we simplify eq.~(\ref{eoms4}) using results
for the expansion coefficients from the other equations of motion to produce
the following constraint:
\begin{align}
&&\dot{a}_{2,4} &= \frac{1}{9}\left(\Delta\left(2\Delta-5\right)\phi_{(2\Delta-4)}
\dot{\phi}_{(0)}-\left(4-\Delta\right)\left(2\Delta-3\right)
\phi_{(0)}\dot{\phi}_{(2\Delta-4)}\right), \labell{eq:adot}\\
\text{and hence}&&\nonumber\\
&&a_{2,4}(\t) &= \mathcal{C} - \frac{1}{9}\left(4-\Delta\right)\left(2\Delta-3\right)
\,\phi_{(0)}(\t)\,\phi_{(2\Delta-4)}(\t) \nonumber \\
&& &\qquad\quad + \frac{2}{3}\left(\Delta-2\right)\int^{\t}_{-\infty}d\t'\
\phi_{(2\Delta-4)}(\t')\,\dot{\phi}_{(0)}(\t')\,, \labell{eq:a24t}
\end{align}
where $\mathcal{C}$ is an integration constant. Following \cite{blm}, we will
choose $\mathcal{C}$ at a later stage so that the entropy production in the
quench is proportional to $a_{2,4}(\t=\infty)$. Note that since initially we
have $\phi_{(0)}(\t=-\infty)=0=\phi_{(2\Delta-4)}(\t=-\infty)$, it follows that
$a_{2,4}(-\infty) = \mathcal{C}$. Further if we set $\phi_{(0)}(\t=\infty)=1$,
then $a_{2,4}$ asymptotes to
\begin{eqnarray}
a_{2,4}(\infty) &=& a_{2,4}(-\infty) - \frac{1}{9}\left(4-\Delta\right)
\left(2\Delta-3\right)\phi_{(2\Delta-4)}(\infty) \nonumber \\
 && \qquad+ \frac{2}{3}\left(\Delta-2\right)\int^{\infty}_{-\infty}d\t'
 \ \phi_{(2\Delta-4)}(\t')\,\dot{\phi}_{(0)}(\t')\,. \labell{eq:a24}
\end{eqnarray}
All the remaining coefficients appearing in eqs.~\reef{eq:sola} and
\reef{eq:sols} can be determined in terms of $\phi_{(0)}$, $\phi_{(2\Delta-4)}$
and $a_{2,4}$. Explicit expressions of some of the leading coefficients are
given in appendix \ref{coeff}.

\section{Fefferman-Graham coordinates} \labell{feffermans}

We would like to evaluate the entropy density, the expectation value of the
stress-energy tensor and of the operator $\mathcal{O}_{\Delta}$ in the boundary
theory during a quench. Following the standard approach \cite{count,skenderis,sk1}, we need to
vary the on-shell gravitational action \reef{eq:action} with respect to the
asymptotic boundary value of the appropriate fields --- see section
\ref{stress}. While EF coordinates are useful for evaluating the equations of
motion, they are not as useful for determining the boundary one-point
functions. The reason for the latter is that the ``radial'' direction
$\partial_\rho$ is not orthogonal to the spacetime boundary located at
$\rho=0$, which is clear from the fact that the metric has off-diagonal $\t$
and $\rho$ components.  It will therefore be useful to transform to
Fefferman-Graham (FG) coordinates \cite{fefferman}, in which the radial
coordinate is orthogonal to the boundary of the spacetime. The FG coordinates
have a spacelike radial coordinate $r$ in contrast to the EF coordinates, with
the null radial coordinate $\rho$.  The FG coordinates are more appropriate for
holographic renormalization, since we can choose a planar cut-off surface by
simply fixing $r$ to some small parameter $\epsilon$.

In FG coordinates, the (asymptotically) AdS spacetime has the line-element
\begin{equation} \label{eq:fgle}
ds^{2} = \frac{G_{ab}(x,r)\,dx^{a}\,dx^{b}}{r^{2}} + \frac{dr^2}{r^2},
\end{equation}
$a$ and $b$ running from $0$ to $3$. By equating this FG line-element
(\ref{eq:fgle}) to the previous EF line-element (\ref{eq:efle}) and writing the
Eddington-Finkelstein coordinates $\tau$ and $\rho$ as functions of the
Fefferman-Graham coordinates $t$ and $r$, we obtain a set of three equations
from which we can solve for $\tau(t,r)$ and $\rho(t,r)$, as well as the metric
component $G_{00}$. The set of equations is
\begin{eqnarray}
0 &=&\mu^{-2}A\rho^{2} \dot{\t}\t' + \left(\dot{\rho} \t' + \rho'\dot{\t}\right)\,,
\labell{eq1001}\\
-1 &=& r^{2}\left(\mu^{-2}A\left(\t'\right)^{2} + \frac{2}{\rho^{2}}\rho'\t'\right)\,,
\labell{eq1002}\\
G_{00} &=&  r^{2}\left(-\mu^{-2}A\dot{\t}^{2}-\frac{2}{\rho^{2}}\dot{\t}\dot{\rho}\right)\,,
\labell{eq1003}
\end{eqnarray}
where primes denote $\partial_{r}$ and dots denote $\partial_{t}$. We solve
eqs.~\reef{eq1001} and \reef{eq1002} by writing $\t$ and $\rho$ as power series
in $r$, with $t$-dependent coefficients:
\begin{eqnarray}
\frac{\t(t,r)}{\mu} &=& t + \sum_{n=1}v_{(n)}(t)r^{n} +\nonumber\\
&& \ell^{2}\left(\sum_{n=5}\vartheta_{(n)}(t)r^{n} + r^{9-2\Delta}\sum_{n=0}\nu_{(n)}(t)r^{n}
 + r^{2\Delta}\sum_{n=1}\omega_{(n)}(t)r^{n}\right)\,, \labell{eq:v}\\
\rho(t,r) &=& \mu r + \sum_{n=1}\rho_{(n)}(t)r^{n} + \nonumber\\
&&\ell^{2}\left(\sum_{n=5}\chi_{(n)}(t)r^{n} + r^{9-2\Delta}\sum_{n=0}\xi_{(n)}(t)r^{n} +
r^{2\Delta}\sum_{n=1}\zeta_{(n)}(t)r^{n}\right)\,. \labell{eq:rho}
\end{eqnarray}
 Upon solving for the above, we can also
determine the metric $G_{ab}$ and scalar field $\Phi_{\textrm{p}}$ in terms of
similar asymptotic expansions in $r$
\begin{eqnarray}
G_{ab}(t,r) &=& g^{(0)}_{ab} + g^{(4)}_{ab}\,r^{4} \nonumber\\
&&+ \ell^{2}\left(\sum_{n=4}c_{(n)ab}(t)r^{n}   + r^{8-2\Delta}
\sum_{n=0}d_{(n)ab}(t)r^{n} + r^{2\Delta}\sum_{n=0}e_{(n)ab}(t)r^{n}
\right)\,, \labell{eq:fgmetric}\\
\Phi_{\textrm{p}}(t,r) &=& \left(r^{4-\Delta}\sum_{n=0}f_{(n)}(t)\,r^{n}
 + r^{\Delta}\sum_{n=0}g_{(n)}(t)\,r^{n} \right)\,. \labell{eq:fgscalar}
\end{eqnarray}
Explicit expressions of the leading coefficients are given in appendix
\ref{fefferman}. For an asymptotic solution of the nonlinear equations of
motion in FG coordinates, see \cite{Hung:2011ta}.

\section{Holographic renormalization} \labell{renorm}

Given the metric and scalar field written in FG coordinates, we must evaluate
the on-shell gravitational action (\ref{eq:action}). However, a naive
evaluation yields a number of divergences associated with integrating out to
the asymptotic boundary at $r=0$. Hence following the standard approach
\cite{count,skenderis,sk1}, we first regulate the calculation by introducing a
cut-off surface $r=\epsilon$ and then the divergences are eliminated by adding
boundary counterterms. Actually these counterterms are added in addition to the
usual Gibbons-Hawking-Brown-York term
\begin{equation}
S_{GHBY} = -\frac{1}{8\pi G^{(5)}_{N}}\int d^{4}x\sqrt{-\gamma}K \Big{|}_{r=\epsilon},
\end{equation}
where $\gamma_{ab}(\epsilon)$ is the induced metric on the cut-off surface and
$K$ is the trace of the extrinsic curvature of this surface. Recall that in our
study, we choose the boundary geometry to be flat, \ie
\begin{equation}
g^{(0)}_{ab} = \lim_{r\to0}G_{ab}(t,r) = \eta_{ab}\,,
\labell{flatg}
\end{equation}
and so the counterterm action turns out to be
\begin{align}
S_{count} = \frac{1}{16\pi G^{(5)}_{N}}&\int d^{4}x\sqrt{-\gamma}
\Bigg{(}-6 - \frac{4-\Delta }{2}\,\Phi^2  \labell{scountx}\\
&\qquad+ \frac{1}{4(\Delta-3)}\,\left(\partial\Phi\right)^{2}
+\frac{1}{24(\Delta-3)}\,R\left(\gamma\right)\,\Phi^2
\Bigg{)}\Bigg{|}_{r=\epsilon}\,,
\nonumber
\end{align}
where $R\left(\gamma\right)$ corresponds to the Ricci scalar constructed with
$\gamma_{ab}$. The $\left(\partial\Phi\right)^{2}$ and
$R\left(\gamma\right)\,\Phi^2$ terms only cancel divergences which occur when
$\Delta > 3$ and so they should be discarded when $\D\leq3$. Although the term
with $R\left(\gamma\right)\,\Phi^2$ vanishes to leading order when evaluated on
a planar cut-off surface, it is required to cancel a divergence that arises in
varying the metric to determine the stress tensor \cite{cascade}. In
particular, it cancels a divergent contribution to the pressure $\mathcal{P}$
for $\D>3$ at order $\ell^2$. Also note that for the special cases $\Delta=2$,
$3$ and $4$, there are also further logarithmic and finite counterterms, but we
do not concern ourselves with these here. The interested reader can find a
complete discussion of these cases in \cite{blm,papa}.

The holographic action $S_{reg}=S_{bulk} + S_{GHBY} + S_{count}$ can now be used
to calculate the one-point correlators of the stress tensor and operator
$\mathcal{O}_{\Delta}$. In order to calculate these expectation values, we need
to vary $S_{reg}$ with respect to the boundary metric and the scalar field,
respectively. The details of these calculations are given in appendix
\ref{stress} and the final results are:
\begin{eqnarray}
8 \pi G^{(5)}_{N}\, \mathcal{E} &=& \frac{3}{2}\mu^{4} - \ell^{2}\mu^{4}\left(
\frac{3}{2}a_{2,4} + \frac{1}{6} \left(2\D-3 \right)\left(4-\D \right)
\phi_{(0)}\phi_{(2\Delta-4)} \right)\,,
\labell{eq:endens} \\
8 \pi G^{(5)}_{N}\, \mathcal{P} &=& \frac{1}{2}\mu^{4} - \ell^{2}\mu^{4}\left(
\frac{1}{2}a_{2,4} - \frac{1}{18} \left(4\D-9 \right)\left(4-\D \right)
\phi_{(0)}\phi_{(2\Delta-4)}  \right)\,,
\labell{eq:press}\\
16 \pi G^{(5)}_{N}\,\langle \mathcal{O}_{\Delta} \rangle &=& 2\mu^{\Delta} \ell\,\pp
 \left(\Delta-2 \right) \phi_{(2\Delta-4)}\,.
 \labell{eq:vev}
\end{eqnarray}
Here $\mathcal{E}$ and $\mathcal{P}$ denote the energy density and pressure in
the boundary theory, \ie $\langle T^{00} \rangle=\mathcal{E}$ and $\langle
T^{ij} \rangle = \delta^{ij}\, \mathcal{P}$. Further, $\pp$ is a
proportionality constant relating the leading coefficient in the expansion
\reef{eq:fgscalar} of the bulk scalar with the coupling in the boundary theory,
\ie $\ell f_{(0)}=\pp\,\lambda$. We fix the precise value of this constant in
section \ref{finalt} --- see eq.~\reef{ppx}.

These one-point correlators must respect certain Ward identities \cite{sk1}. In
particular, one has the diffeomorphism Ward identity
\begin{equation}
\del^i \la\, T_{ij}\ra=\la\calo_\D\ra\ \del_j \lambda\,,
\eqlabell{warddiffeo}
\end{equation}
Of course, when the coupling $\lambda$ is constant, this expression reduces to
the conservation of energy and momentum in the boundary theory. In the present
case with a time-dependent coupling, the $j=t$ component of
eq.~\reef{warddiffeo} yields
 \beq
\partial_t\,\cale = -\la\calo_\D\ra\ \del_t \lambda\,.
 \labell{worky}
 \eeq
Here the expression on the right-hand side describes the work done by varying
the coupling in the boundary theory.\footnote{Note that a minus sign appears
here in accord with our conventions, which differ slightly from those in
\cite{blm}.} Let us verify that eqs.~\reef{eq:endens} and \reef{eq:vev} satisfy
this constraint: First, comparing the expansions of the bulk scalar in
eqs.~(\ref{eq:phip}) and \reef{eq:fgscalar} and recalling the relation
$\ell f_{(0)}=\pp\,\lambda$ from appendix \ref{stress}, we find to leading order
\begin{equation}
\phi_{(0)} = \mu^{\Delta - 4}\,\pp\, \frac{\lambda}{\ell}.
\labell{eq:p0}
\end{equation}
Then
differentiating eq.(\ref{eq:endens}), we find
\begin{eqnarray}
8 \pi G^{(5)}_{N}\, \partial_{t}\mathcal{E}
&=&  \ell^{2}\mu^{4}\left( -\frac{3}{2}\dot{a}_{2,4} - \frac{1}{6}
\left(2\D-3 \right)\left(4-\D \right) \left(\dot{\phi}_{(0)}\phi_{(2\Delta-4)}
+\phi_{(0)}\dot{\phi}_{(2\Delta-4)}\right) \right) \nonumber \\
&=& -\ell^{2}\mu^{4}\left(\Delta-2\right)\,\phi_{(2\Delta-4)}\,\dot{\phi}_{(0)}\,,
\labell{cheqx}
\end{eqnarray}
where we simplified the expression by substituting for $\dot{a}_{2,4}$ from
eq.~(\ref{eq:adot}).  Now using eqs.~(\ref{eq:vev}) and (\ref{eq:p0}), we see
that this expression precisely matches the expected Ward identity \reef{worky}.
Let us comment that this match should be no surprise since the constraint
\eqref{eoms4} (which was used to derive eq.~\reef{eq:adot}) reduces to
precisely this Ward identity \reef{worky} on the asymptotic boundary $r=0$
\cite{blm}.

We also have the conformal Ward identify
\begin{equation}
T^{a}{}_{a} = \left(4-\Delta\right)\,\langle \mathcal{O}_{\Delta} \rangle\,\lambda
\,,\labell{eq:tmm}
\end{equation}
which follows from taking the trace of the stress-energy tensor with
eqs.~(\ref{eq:endens}) and \reef{eq:press} and substituting eqs.~(\ref{eq:vev})
and (\ref{eq:p0}). Here we do not find any anomalous terms (at quadratic order
in $\ell$), since we are assuming that the operator $\mathcal{O}_{\Delta}$ has
a fractional conformal dimension. This result can be contrasted with the
discussion in \cite{blm} which considered $\D=2$ and 3.

\section{Temperature and entropy density} \labell{entropy}

In this section we will calculate the temperature of the boundary theory before
and after the quench, as well as the entropy produced during the quench. As
described above, we are assuming that the quench takes the scalar field from a
vanishing initial value with $\phi_{(0)}=0$ and $\phi_{(2\D-4)}=0$) to a final
equilibrium solution where $\phi_{(0)}=1$ and
$\phi_{(2\D-4)}=\phi_{(2\D-4)}(\infty)$. In section \ref{reverse}, we will
consider `reverse' quenches which instead take the system from $\phi_{(0)}=1$
to 0. In our perturbative calculations for high temperature quenches, we find
that if the profile for the `reverse' quench is given by
$\tilde{\phi}_{(0)}(\t)=1 - \phi_{(0)}(\t)$, where $\phi_{(0)}(\t)$ describes
some `forward' quench, then we find that $\tilde{\phi}_{(2\D-4)}(\t)=
\phi_{(2\D-4)}(\infty) -\phi_{(2\D-4)}(\t)$, where $\phi_{(2\D-4)}(\t)$ is the
response for the corresponding `forward' quench. Similarly, we will find that
the entropy production is the same in the forward and reverse quenches.
Further, in the case of an adiabatic quench, no entropy is created and the
process is reversible.

As discussed in section \ref{staticsol}, the initial configuration before the
quench is the well-known planar AdS black hole described by
eq.~\reef{eq:static}. The calculation of the corresponding temperature is a
straightforward exercise with the result $T={\mu}/{\pi}$. However, recall that
in eq.~\reef{eq:a24t} we established a convention where $a_{2,4}(-\infty) =
\mathcal{C}$. That is, our metric perturbation is nonvanishing even at
$\tau=-\infty$. The effect of this convention is to shift the black hole mass
parameter, \ie $\mu\to\mu\xi$ where $\xi^{4} = 1 - \ell^{2}a_{2,4}(-\infty)$.
Hence, to quadratic order in the expansion in $\ell$, the initial temperature
becomes
\begin{equation}
T_{i} =  \frac{\mu\, \xi}{\pi} =
\frac{\mu}{\pi}\left(1 - \frac{\ell^{2}}{4}\,a_{2,4}(-\infty)\right)\,.
\labell{eq:ti}
\end{equation}

\subsection{Final temperature} \labell{finalt}

Next we wish to determine the final equilibrium temperature of the system after
the quench has taken place. This calculation is more subtle as with our
perturbative calculations, since we will not have the full metric describing the
final black hole geometry. Instead then, we turn to the thermodynamics of the
boundary theory to determine the final temperature. That is, we will compare
the energy density and pressure in QFT variables (already in terms of the final
temperature $T_f$ and the coupling $\lambda$) to the energy density and
pressure calculated holographically in terms of gravitational variables.  In
doing so, we are able to derive meaningful relations between the field theory
coupling and temperature and the bulk parameters $\mu$ and $\ell$.  Of course,
by assuming a form for $\mathcal{E}$ and $\mathcal{P}$, our final temperature
and entropy production will necessarily depend on the conventions used to
define our coupling. This cannot be helped, because we do not know the
Lagrangian for the boundary theory when the quench is by an operator of
arbitrary dimension $\Delta$. This can be contrasted with the discussion in
\cite{blm} for the cases of $\Delta=2,3$, where the exact equilibrium
expressions for $\mathcal{E}$ and $\mathcal{P}$ are known from
\cite{Buchel:2007vy}. Nonetheless, we will find physically meaningful
interpretations for our results.

To begin, we make the following ansatz for the energy density and pressure in
the final equilibrium of the boundary theory,
\begin{eqnarray}
\mathcal{E}_{f} &=& \mathcal{A}\, T^{4}_{f} \left (1 - \alpha_{f}
\left(\frac{\l_{f}}{T^{4-\Delta}_{f}}\right)^{2}\right)\,, \labell{eq:fted}\\
\mathcal{P}_{f} &=& \frac{\mathcal{A}}{3}\, T^{4}_{f} \left (1 -
\left(\frac{\l_{f}}{T^{4-\Delta}_{f}}\right)^{2}\right)\,, \labell{eq:ftpr}
\end{eqnarray}
where $\lambda_f=\lambda(\tau=\infty)$ denotes the final value of the coupling.
To leading order our ansatz reduces to the expressions expected for a conformal
theory and is in accord with our analysis, the perturbation of these conformal
terms is quadratic in the coupling. Further, we have expressed the
perturbations in terms of the dimensionless ratio $\l_f/T_f^{4-\D}$. Setting
the pre-factor for this term in the pressure \reef{eq:ftpr} really defines our
normalization for the coupling. We can compare these expressions with those
given in \cite{blm,Buchel:2007vy}. For example, we find for $\D=3$,
\begin{equation}
\l^{2}_{f} = \frac{2\Gamma\left(\frac{3}{4}\right)^{4}}{\pi^{4}}\,m^{2}_{f}\,,
%= 2\phi_{(2\D-4)}m^{2}_{f}.
\end{equation}
where $m_f$ was the fermion mass in the boundary theory.  Using this
expression, we can confirm the results derived below for the equilibrium values

of the observables agree with those given in \cite{blm,Buchel:2007vy}.

Now we need to determine the constant of proportionality $\alpha_{f}$ in
eq.~(\ref{eq:ftpr}). To proceed, we only assume that the boundary theory obeys
standard thermodynamics, following \cite{Buchel:2003ah}. First, we write the
free energy density as
\begin{equation}
F = \mathcal{E} - T\, S\,, \labell{free}
\end{equation}
where $S$ is the entropy density.  In the absence of any chemical potentials,
$F = -\mathcal{P}$.  Therefore combining these
expressions with eqs.~\reef{eq:fted} and \reef{eq:ftpr}, the final entropy
density is given by
\begin{equation}
S_{f} = \frac{\mathcal{A}}{3}\, T^{3}_{f} \left (4 -
\left(3\alpha_{f} + 1\right)\left(\frac{\l_{f}}{T^{4-\Delta}_{f}}\right)^{2}\right)\,.
\labell{entrof}
\end{equation}
We use the first law of thermodynamics (with fixed volume) to write
\begin{equation}
\frac{d\mathcal{E}_{f}}{d T_{f}} = T_{f} \frac{d S}{d T_{f}}.
\label{eq:fixedpressure}
\end{equation}
The left-hand side of eq.~(\ref{eq:fixedpressure}) is
\begin{equation}
\frac{d\mathcal{E}_{f}}{d T_{f}} = \mathcal{A}\, T^{3}_{f} \left (
4 - \left(2\Delta-4\right)\alpha_{f}\left(\frac{\l_{f}}{T^{4-\Delta}_{f}}\right)^{2}
\right) \nonumber
\end{equation}
whereas the right-hand side is
\begin{equation}
T_{f}\frac{d{S}_{f}}{d T_{f}} = \mathcal{A}\, T^{3}_{f} \left (4 -
\frac{1}{3}\left(3\alpha_{f} + 1\right)\left(2\Delta-5\right)\left(
\frac{\l_{f}}{T^{4-\Delta}_{f}}\right)^{2}\right)\,. \nonumber
\end{equation}
By comparing these two expressions, we solve for $\alpha_{f}$ as
\begin{equation}
\alpha_{f} = \frac{2\Delta-5}{3}\,.
\labell{finalpha}
\end{equation}
Note that it may seem that the quench has no effect on the energy density for
$\D=\frac{5}{2}$ (when $\alpha_f=0$), but even in this case, the initial and
final temperatures will differ by a term of order $\lambda^{2}_{f}$. Hence,
there will still be a change in $\mathcal{E}$ in this case, contained in the
$T^{4}_{f}$ term in eq.~(\ref{eq:fted}).

Next, we compare these results for the boundary theory with the corresponding
expression in the gravitational dual. In particular, we would like to find
$\ell$ in terms of the temperature $T_f$ and the coupling $\lambda_f$. However,
first we fix the normalization factor $\mathcal{A}$ appearing in
eqs.~\reef{eq:fted} and \reef{eq:ftpr}. This factor would be the unchanged in
the initial equilibrium of the conformal boundary theory, \ie at $t=-\infty$,
we would have $\mathcal{E}_i=\mathcal{A}\,T_i^4$. Comparing the latter
expression with eq.~(\ref{eq:endens}) then yields
\begin{equation}
\mathcal{A}\,T^{4}_{i} = \frac{3}{16 \pi G^{(5)}_{N}}\,
\mu^{4}\left(1-\ell^{2}a_{2,4}(-\infty)\right)\,.
\labell{nonamex}
\end{equation}
Given the expression for the initial temperature in eq.~(\ref{eq:ti}), we see
that
\begin{equation}
\mathcal{A} = \frac{3\pi^{4}}{16 \pi G^{(5)}_{N}}\,. \labell{splat}
\end{equation}

Next, we take the trace of the stress tensor in both the field theory and the
gravitational dual:
\begin{eqnarray}
\left(T_{\textrm{QFT}}\right)^{a}{}_{a} &=& -\frac{2}{3}\,\mathcal{A}\, T^{4}_{f}\left(4-\D
\right)\left(\frac{\l_{f}}{T^{4-\Delta}_{f}}\right)^{2}\,,
\labell{traceQFT}\\
\left(T_{\textrm{GR}}\right)^{a}{}_{a}  &=& \frac{\mu^{4}\ell^{2}}{8 \pi G^{(5)}_{N}}
\left(4-\Delta\right)
\left(\Delta-2\right)\phi_{(0)}\phi_{(2\Delta-4)} \nonumber\\
&\xrightarrow[t \rightarrow \infty]{}& \frac{\mu^{4}\ell^{2}}{8 \pi G^{(5)}_{N}}
\left(4-\Delta\right)
\left(\Delta-2\right)\phi_{(2\Delta-4)}(\infty)\,,
\labell{traceGR}
\end{eqnarray}
where $\phi_{(2\Delta-4)}(\infty)$ is given by eq.~(\ref{eq:phinfty}) with
$\phi_{(0)}=1$, \ie
\begin{equation}
\phi_{(2\Delta-4)}(\infty) = -\frac{\Gamma\left(\frac{4-\Delta }{2}\right) \Gamma\left(
\frac{\Delta }{4}\right)^2}{\Gamma\left(\frac{4-\Delta }{4}\right)^2
\Gamma\left(\frac{\Delta }{2}\right)}\,.
\labell{eq:black}
\end{equation}
Equating the two expressions above and using eq.~\reef{splat},  we find
\begin{eqnarray}
\ell^{2}
%&=&  \frac{\pi^{4}}{\left(\Delta-2\right)\mu^{4}\phi_{(2\Delta-4)}(\infty)}\
%\frac{\l^{2}_{f}}{T^{4-2\Delta}_{f}} \nonumber \\
 &=& \frac{1}{\left(\Delta-2\right)|\phi_{(2\Delta-4)}(\infty)|}\
 \left(\frac{\l_{f}}{T^{4-\Delta}_{f}}\right)^{2} + o\left(\lambda^{4}_{f}\right)\,,
 \labell{eq:l2}
\end{eqnarray}
to leading order in $\l_{f}/T^{4-\D}_{f}$. Note that here we have also used
eq.~\reef{eq:ti} to substitute $\mu^{4}=\pi^{4}T^{4}_{f} + o(\l_f^2)$ since the
initial and final temperatures will only differ by $o(\l_f^2)$ in our
perturbative calculations. Further, the above expression takes account of the
fact that $\phi_{(2\Delta-4)}(\infty)$ is always negative in the range of
interest, \ie $2<\D<4$ --- see eq.~\reef{eq:black} above. Recalling that we set
$\phi_{(0)}(\infty)=1$, we note that implicitly the right-hand side of
eq.~(\ref{eq:l2}) is actually $\ell^2 \phi_{(0)}^2$ and so this equation fixes
the normalization between the leading coefficient in the asymptotic expansion
of the bulk scalar and the boundary coupling, \ie
\begin{equation}
\ell \phi_{(0)} =  \frac{ 1}{\sqrt{\left(\Delta-2\right)\lvert\phi_{(2\Delta-4)}(\infty)\rvert}}\,
\frac{\l}{T^{4-\Delta}}+ o\left(\lambda^{3}\right).
\end{equation}
Alternatively in appendix \ref{stress}, we introduced the
proportionality constant $\pp$ in $\ell f_{(0)}=\pp\,\lambda$. So comparing the
expansions of the bulk scalar in eqs.~(\ref{eq:phip}) and \reef{eq:fgscalar} using eq. \reef{eq:source} ,
we now have
\begin{equation}
\pp=\frac{
\pi^{4-\D}}{\sqrt{\left(\Delta-2\right)\lvert\phi_{(2\Delta-4)}(\infty)\rvert}}+ o(\l^2)\,,
\labell{ppx}
\end{equation}
where as above, we used $\mu^{4}=\pi^{4}T^{4} + o(\l^2)$.

\subsection{Entropy production during the quench} \label{entprod}

Here we extend the previous analysis to determine the entropy production during
the quench. First using the expression for the free energy density \reef{free},
as well as $F = -\mathcal{P}$, we find
\begin{equation}
\frac{S_{f}}{S_{i}} = \frac{T_{i}}{T_{f}}\ \frac{\mathcal{E}_{f}+
\mathcal{P}_{f}}{\mathcal{E}_{i}+\mathcal{P}_{i}}\,.
\labell{ratio9}
\end{equation}
Initially the boundary theory is conformal and the vanishing trace of the
stress tensor requires $\mathcal{E}_{i} = 3\mathcal{P}_{i}$. Now the latter can
be used to re-express eq.~\reef{ratio9} as
\begin{equation}
\frac{S_{f}}{S_{i}} = \frac{T_{i}}{T_{f}}\
\left(\frac{3}{4}\,\frac{\mathcal{E}_{f}}{\mathcal{E}_{i}}
+ \frac{1}{4}\,\frac{\mathcal{P}_{f}}{\mathcal{P}_{i}}\right)\,.
\labell{eq:sratio}
\end{equation}

First, we determine the ratio of the temperatures by equating the final energy
densities given in terms of the gravitational variables \reef{eq:endens} and of
the boundary theory \reef{eq:fted}. The initial temperature is introduced here
by substituting for $\mu$ using eq.~(\ref{eq:ti}), which then yields
\begin{eqnarray}
\frac{T_{i}}{T_{f}} &=&  1 + \frac{\ell^2}{4}\left(a_{2,4}(\infty) - a_{2,4}(-\infty) + \frac{2}{9}
\left(2\D^{2}-8\D+9\right)
\phi_{(2\D-4)}(\infty)\right)
\,. \labell{eq:tiotf}
\end{eqnarray}
Now using the expressions for the energy density and pressure in
eqs.~(\ref{eq:endens}) and \reef{eq:press} at the initial and final times, we
find:
\begin{eqnarray}
\frac{\mathcal{E}_{f}}{\mathcal{E}_{i}} &=& 1 - \ell^{2}\left(
a_{2,4}(\infty) - a_{2,4}(-\infty) + \frac{1}{9} \left(2\D-3 \right)\left(4-\D \right)
 \phi_{(2\Delta-4)}(\infty)\right)\,, \labell{eq:efoei}\\
\frac{\mathcal{P}_{f}}{\mathcal{P}_{i}} &=& 1 - \ell^{2}\left(
a_{2,4}(\infty) - a_{2,4}(-\infty) - \frac{1}{9} \left(4\D-9 \right)\left(4-\D \right)
\phi_{(2\Delta-4)}(\infty)\right)\,. \labell{eq:pfopi}
\end{eqnarray}
Combining these results in eq.~\reef{eq:sratio} then yields
\begin{equation}
\frac{S_{f}}{S_{i}} = 1 - \frac{3\ell^{2}}4\left(
a_{2,4}(\infty) - a_{2,4}(-\infty) - \frac{2}{9}\left(\Delta-3\right)
\left(\Delta-1\right)\phi_{(2\Delta-4)}(\infty)\right)\,.
\labell{eq:sratio2}
\end{equation}
Now recall from eq.~\reef{eq:a24t} that $a_{2,4}(-\infty)=\mathcal{C}$, where
the latter is an arbitrary integration constant. Hence following \cite{blm}, we
choose this constant to simplify the above ratio of entropies, \ie
\begin{equation}
a_{2,4}(-\infty) = -\frac{2}{9}\left(\Delta-3\right)
\left(\Delta-1\right)\phi_{(2\Delta-4)}(\infty)\,.
\labell{eq:ic}
\end{equation}

Hence, after substituting for $\ell^{2}$ and $a_{2,4}(-\infty)$ from
eqs.~(\ref{eq:l2}) and (\ref{eq:ic}), respectively, the ratio of the final and
initial entropies \reef{eq:sratio2} becomes
\begin{equation}
\frac{S_{f}}{S_{i}} = 1 + \frac{3\, a_{2,4}(\infty)}{4\left(\Delta-2\right)
\phi_{(2\Delta-4)}(\infty)}\ \left(\frac{\l_{f}}{T^{4-\Delta}_{f}}\right)^{2}\,.
\labell{eq:sfosi}
\end{equation}
Further substituting for $\ell^{2}$ and $a_{2,4}(-\infty)$ in
eqs.~\reef{eq:tiotf}--\reef{eq:pfopi}, we find the change in temperature,
energy density and pressure are given by
\begin{eqnarray}
\frac{\Delta T}{T_{i}} &=&   \left[\frac{\D-2}6+\frac14\,
\frac{ a_{2,4}(\infty)}{\left(\D-2\right)
\phi_{(2\D-4)}(\infty)}\right]
\left(\frac{\l_{f}}{T^{4-\D}_{f}}\right)^{2}\,,
\labell{eq:tiotf2}\\
\frac{\Delta\mathcal{E}}{\mathcal{E}_{i}} &=& \left[\,\frac{1}3+
\frac{ a_{2,4}(\infty)}{\left(\D-2\right)
\phi_{(2\D-4)}(\infty)}\right] \left(\frac{\l_{f}}{T^{4-\Delta}_{f}}\right)^{2}\,,
 \labell{eq:efeifw}\\
\frac{\Delta\mathcal{P}}{\mathcal{P}_{i}} &=& \left[\frac{2\D-7}3+
\frac{ a_{2,4}(\infty)}{\left(\D-2\right)
\phi_{(2\D-4)}(\infty)}\right] \left(\frac{\l_{f}}{T^{4-\Delta}_{f}}\right)^{2}\,,
\labell{eq:pfpifw}
\end{eqnarray}
where our notation is \eg $\Delta\mathcal{P} =\mathcal{P}_f- \mathcal{P}_i$.

The second law of thermodynamics demands that the ratio $S_{f}/S_{i}$ must always be greater
than one. Hence requiring $a_{2,4}(\infty)\le0$ becomes a test of our numerical
solutions and we successfully confirm that this inequality is satisfied in the
obtained numerical solutions. Since $\phi_{(2\Delta-4)}(\infty)$ is always negative in
our analysis (and we restrict our attention to $\D>2$), eqs.~\reef{eq:tiotf2}
and \reef{eq:efeifw} indicate that the changes in the temperature and the
energy density are always positive. However, from eq.~\reef{eq:pfpifw}, the
change in pressure is only guaranteed to be positive for $\Delta\ge7/2$.
Otherwise, the pressure can either increase or decrease depending on the
precise value of $\D$ and the magnitude of $a_{2,4}(\infty)$. A more detailed
discussion is given in section \ref{behaviour}, where we consider the effect of
the numerically determined values of $a_{2,4}(\infty)$ on the shifts of these
quantities.

Another check of the present analysis comes from considering the adiabatic
limit. As we discuss in section \ref{slowquenches} in this case, the system
remains in a quasi-static equilibrium with
$\phi_{(2\Delta-4)}(t)=\phi_{(2\Delta-4)}(\infty)\,\phi_{(0)}(t)$. Substituting
this expression into eq.~\reef{eq:a24}, as well as using the integration
constant chosen in eq.~\reef{eq:ic}, it is straightforward to show that
$a_{2,4}(\infty)$ vanishes. Hence as expected for an adiabatic transition, no
entropy is produced, as discussed in \cite{blm}.

As a final consistency check, we consider the speed of sound in the thermal
plasma, which is given by
\begin{eqnarray}
c^{2}_{s} = \frac{d\mathcal{P}}{d\mathcal{E}} &=&
\left(\frac{d\mathcal{P}}{d T_{f}}\right)\Big/\left(\frac{d\mathcal{E}}{d T_{f}}\right)
 \nonumber \\
&=& \frac{1}{3} - \frac{1}{9}\left(4-\D\right)\left(\D-2\right)
\left(\frac{\l_{f}}{T^{4-\Delta}_{f}}\right)^{2}\,.
\labell{fasts}
\end{eqnarray}
Note the second term is negative for all $\D$ in the range $2<\D<4$. Hence we
find $c^{2}_{s}<{1}/{3}$, as required by \cite{Hohler:2009tv}. While
$c^2_s=1/3$  for $\D=2$ and $4$, our analysis only applies for the conformal
dimension strictly limited within the range $2<\D<4$.

\subsection{Reverse quenches} \label{reverse}

Up until now we have assumed that the quenches begin with the boundary theory
being conformal, \ie $\lambda=0$ and then end with some finite $\lambda$. In
the gravitational description then, they involve some profile $\phi_{(0)}(\t)$
which begins with $\phi_{(0)}=0$ at $\t=-\infty$ and ends with $\phi_{(0)}=1$
at $\t=\infty$. In this section, we consider `reverse' quenches in which the
coupling is initially finite and is brought down to zero. In particular, we can
readily repeat the analysis for reverse quenches where the non-normalizable
coefficient of the bulks scalar is chosen to be
\begin{equation}
\tilde{\phi}_{(0)}(\t)=1-\phi_{(0)}(\t)\,.
\labell{backward}
\end{equation}
Because our analysis is limited to the perturbative high temperature regime,
the equation of motion \reef{eq:phi} for the bulk scalar is linear and hence we
can add any two solutions to produce a third solution. In particular then,
adding the scalar field solutions for the forward and reverse quench must yield
the equilibrium solution with $\phi_{(0)}(\t)=1$. Alternatively, the reverse
quench produced from eq.~\reef{backward} is simply the equilibrium solution
\reef{eq:stsol} (with $c_1=1$) minus the time-dependent solution describing the
forward quench. In the equilibrium case, the normalizable coefficient in the
bulk scalar is $\phi_{(2\D-4)}(\infty)$ and so the corresponding coefficient in
the reverse quench must be
\begin{equation}
\tilde{\phi}_{(2\D-4)}(\t)=\phi_{(2\D-4)}(\infty)-\phi_{(2\D-4)}(\t)
\,, \labell{backward2}
\end{equation}
where $\phi_{(2\D-4)}(\t)$ denotes the response produced in the original
(forward) quench.

In the reverse quench, the metric coefficient $\tilde{a}_{2,4}$ still satisfies
eq.~(\ref{eq:a24t}) and so we have the solution
\begin{eqnarray}
\tilde{a}_{2,4}(\t) &=& \tilde{\mathcal{C}} -
\frac{1}{9}\left(4-\Delta\right)\left(2\Delta-3\right)\tilde{\phi}_{(0)}(\t)\,
\tilde{\phi}_{(2\Delta-4)}(\t) \nonumber \\
 &&\qquad + \frac{2}{3}\left(\Delta-2\right)\int^{\t}_{-\infty}d\t'
 \tilde{\phi}_{(2\Delta-4)}(\t')\,\dot{\tilde{\phi}}_{(0)}(\t')\,.
 \labell{solrever}
\end{eqnarray}
where $\tilde{\mathcal{C}}$ is a new integration constant. Note that in this
case, the second term vanishes for $\t\to\infty$ but as $\t\to-\infty$, it is
proportional to $\tilde{\phi}_{(2\Delta-4)}(-\infty)
=\phi_{(2\Delta-4)}(\infty)$. Repeating the analysis of the previous section
for our reverse quench and demanding that the entropy production is now
proportional to $\tilde{a}_{2,4}$, we find that the integration constant must
be chosen as
\begin{equation}
\tilde{\mathcal{C}} = \frac{1}{3}(\Delta -2)\tilde{\phi}_{(2\Delta-4)}(-\infty)\,.
\labell{cccc}
\end{equation}
With this choice then, we have
\begin{equation}
\frac{S_{f}}{S_{i}} = 1 + \frac{3\tilde{a}_{2,4}(\infty)}{4\left(\Delta-2\right)
\tilde{\phi}_{(2\Delta-4)}(-\infty)}\
\left(\frac{\l_{f}}{T^{4-\Delta}_{f}}\right)^{2}\,.
\labell{cccc2}
\end{equation}
Further, when we compare the expression for $\tilde{a}_{2,4}(\infty)$ with
eq.~\reef{eq:a24} for $a_{2,4}(\infty)$, it is straightforward to show that
these two constants are equal, \ie
\begin{equation} \label{eq:fwbw}
\tilde{a}_{2,4}(\infty) = a_{2,4}(\infty)\,,
\end{equation}
just as was found in \cite{blm}. Hence comparing to eqs.~\reef{eq:sfosi} and
\reef{cccc2} and noting that $\tilde{\phi}_{(2\Delta-4)}(-\infty)
=\phi_{(2\Delta-4)}(\infty)$, we see that the entropy production is identical
in the forward and reverse quenches.

We can also find the changes in the temperature, the energy density and
pressure as before:
\begin{eqnarray}
\frac{\Delta T}{T_{i}} &=&  \left[-
\frac{\D-2}6+\frac14\, \frac{ \tilde{a}_{2,4}(\infty)
}{\left(\D-2\right) \tilde{\phi}_{(2\D-4)}(\infty)}\right]
\left(\frac{\l_{f}}{T^{4-\D}_{f}}\right)^{2}\,,
 \labell{eq:tiotfbw}\\
\frac{\Delta \mathcal{E}}{\mathcal{E}_{i}} &=& \left[-
\frac{1}3+ \frac{ \tilde{a}_{2,4}(\infty)
}{\left(\D-2\right) \tilde{\phi}_{(2\D-4)}(\infty)}\right]
\left(\frac{\l_{f}}{T^{4-\Delta}_{f}}\right)^{2}\,,
 \labell{eq:efeibw}\\
\frac{\Delta\mathcal{P}}{\mathcal{P}_{i}} &=& \left[-
\frac{2\D-7}3+ \frac{ \tilde{a}_{2,4}(\infty)
}{\left(\D-2\right) \tilde{\phi}_{(2\D-4)}(\infty)}\right]
\left(\frac{\l_{f}}{T^{4-\Delta}_{f}}\right)^{2}\,,
\labell{eq:pfpibw}
\end{eqnarray}
where again \eg $\Delta\mathcal{P} =\mathcal{P}_f- \mathcal{P}_i$. Comparing
these expressions for the reverse quench with the corresponding results in
eqs.~\reef{eq:tiotf2}--(\ref{eq:pfpifw}) for the forward case, we see that
$a_{2,4}(\infty)$ and $\tilde{a}_{2,4}(\infty)$ have  the same coefficients
while the constant terms are equal but with opposite sign.  Hence for the
reverse quenches, whether any of these physical quantities increases or
decreases depends on both the magnitude of $\tilde{a}_{2,4}(\infty)$ and the
value of $\D$ --- see section \ref{behaviour} for a further discussion. We also
note that in the case of an adiabatic quench (see section \ref{slowquenches}),
we find $a_{2,4}(\infty)=0$ and hence $\tilde{a}_{2,4}(\infty) =0$. That is,
for adiabatic transitions, no entropy is produced. However, we would also find
that the changes in the temperature, energy density and pressure are exactly
opposite in the forward and reverse cases.

Since the forward and reverse quenches are simply related in our perturbative
high temperature analysis, we will continue to focus on the forward quenches in
the rest of the paper.

\section{Numerical procedure} \label{numerics}

We now briefly describe the numerical simulations which we used to understand
the quenches. Essentially we implemented the same approach as in \cite{blm},
using numerical techniques developed in \cite{Isaac, diss}. For details on
our numerical implementation the interested reader can see the
appendix in \cite{blm}. The primary purpose of our simulations was to find
the response $\phi_{(2\Delta-4)}(\t)$ for a given source $\phi_{(0)}(\t)$, as
described earlier, it is convenient to work with infalling characteristics
and in particular we adopt
Eddington-Finkelstein coordinates as in eq.~\reef{eq:efmetric} or
\reef{eq:efle}. These coordinates are regular at the horizon allowing us
to excise the black hole from the computational domain (by integrating some distance
inwards and stopping the integration as this region is causally disconnected from the
outside).

We choose the source term in the asymptotic expansion (\ref{eq:phip}) of the
bulk scalar to be
\begin{equation}
\phi_{(0)}(\t) = \frac{1}{2} + \frac{1}{2}\tanh(\frac{\t}{\alpha})\,.
\labell{eq:tanh}
\end{equation}
This describes a family of quenches where, as desired, the source starts at
zero in the asymptotic past and ends at one in the asymptotic future. Here
$\alpha$ controls the rate at which the quench takes place. In terms of the
dimensionful time, we have $\t=\mu v\simeq \mu t$ and hence the timescale on
which the transition from zero to finite coupling is made is
\begin{equation}
\Delta t = \frac{\alpha}{\mu}\simeq \frac{\alpha}{\pi T_i}\,.
\labell{time}
\end{equation}
Hence for $\a\gg 1$, the quenches are `slow' which means that the transition
occurs on a timescale that is much longer than the thermal relaxation
timescale. Alternatively for $\alpha\ll1$, the quenches are `fast', meaning the
transition timescale is much shorter than the thermal timescale. The limit
$\alpha\rightarrow0$ would correspond to an `instantaneous' quench, \ie
$\phi_{(0)}$ becomes a step-function. The limit $\alpha\rightarrow\infty$
corresponds to an adiabatic transition --- see section \ref{slowquenches}. We
note that with eq.~\reef{eq:tanh}, the source profile has a continuous
derivative for all time. More general quenches which are not infinitely
differentiable with respect to time will be studied in a later paper
\cite{blmn}.

Again the goal of our simulations is to find the response
$\phi_{(2\Delta-4)}(\t)$ for a given source $\phi_{(0)}(\t)$. Practically, it
is more convenient to solve the linearized scalar equation \reef{eq:phi} in
terms of $\q(\t,\rho)$, which is defined by
\begin{eqnarray}
\Phi_{\textrm{p}}\left(t,\rho\right)
&=& \rho ^{4-\Delta } \left(\phi_{(0)}+\rho  \dot{\phi}_{(0)}+\frac{(2 \Delta-7 ) \rho ^2
 }{4 (\Delta -3)}\ddot{\phi}_{(0)}+\cdots\right) + \rho^{\kappa}\,\q(\t,\rho)\,.
 \labell{eq:numphi}
\end{eqnarray}
Here the term in brackets contains the leading terms in the asymptotic expansion
(\ref{eq:phip}) of $\Phi_{p}$. In particular, we include any terms which are
leading compared to $\rho^{\Delta}$. Formally then the leading solution takes
the form $\q = \rho^{\D-\kappa}\phi_{(2\D-4)}+\cdots$ and so the leading
behaviour in $\q$ contains the desired response function. However, as a
practical matter, we obtained noticeably better accuracy for $\phi_{(2\D-4)}$
by fitting $\q$ (at each timestep) with an expansion of the form 
\begin{equation}
\q(\t,\rho)= \epsilon_{0}\rho^{4-\D-\kappa} + \epsilon_{1}\rho^{5-\D-\kappa} + \cdots
+\rho^{\D-\kappa}\phi_{(2\D-4)}+ \delta_{n}\rho^{4-\D-\kappa+n} + o\left(\rho^{\D-\kappa+1}\right)\,,
\labell{truefit}
\end{equation}
where $\rho^{4-\D-\kappa+n}$ is the next leading order after $\rho^{\D-\kappa}$.
The coefficients $\epsilon_{i}$ are, of course, all very small since the terms
at these orders are already included in eq.~\reef{eq:numphi}.  The factor
$\delta_{n}$ is typically not small, but including this term nonetheless
gives better results when fitting $\phi_{(2\D-4)}$.
In the cases that we choose $\kappa>4-\D$, we only include terms
in eq.~\reef{truefit} with nonnegative powers of $\rho$.

We introduce $\kappa$
for convenience in the fit and choose it so that $\q$ still vanishes at
the asymptotic boundary, \ie $\rho=0$.  We further choose $\kappa$ so that 
the relative power of $\rho$ between the two contributions in eq.~\reef{eq:numphi} 
is an integer, \ie $\D+\kappa$ is an integer.  This ensures that upon substituting into the equation of motion 
  \reef{eq:phi}, after simplification, only integer powers of $\rho$ appear in the coefficients of $\q$ and its derivatives,
as well as in the source terms.  Our choices of $\kappa$, for each of the conformal dimensions considered in our simulations, are shown in table \ref{kappa}.
\begin{table}[ht]
\caption{Choices made for $\kappa$ while simulating $\q(\t,\rho)$ for various $\D$. \label{kappa}}
\centering
\begin{tabular}{c c c c c c c}
&&&\\
\hline
 $\D$ &\vline\vline& 7/3 & 8/3 & 10/3 & 11/3 & 4 \\ [0.5ex]
\hline
 $\kappa$ &\vline\vline& 5/3 & 4/3 & 5/3 & 1/3 & 2 \\ [0.5ex]
\hline
\end{tabular}
\end{table}

Once the numerical solution has been obtained at each timestep, we fit $\q$ with a
series in $\rho$ with rational exponents, as described above, to determine the
coefficient $\phi_{(2\Delta-4)}$.  Repeating this process for each timestep
then generates the full profile for $\phi_{(2\Delta-4)}(\t)$.

%Given the definition of $\q$ in eq.~\reef{eq:numphi}, we plug this expression
%into the linearized equation (\ref{eq:phi}). We evolve $\Phi_{\textrm{p}}$
%forward in time while imposing the boundary conditions that it vanishes on some
%(very) early time slice and that it vanishes at the asymptotic boundary
%$\rho=0$.  With these specifications, we can find the radial profile of the
%response function $\q$ on a given time slice by making use of the radial
%profiles of $\q$ at the two previous timesteps in the discretized spacetime.
%Further for $\q$ at a given radius, we also use the (known) values at the two
%radial points closer to the boundary on the current timestep. This is done by
%replacing the derivatives in eq.~(\ref{eq:phi}) by the discretized versions
%described in the appendix in \cite{blm}. To implement this scheme, we set
%$q(\t,0) = q(\t,0+d\rho)=0$. With $q(\t+ 2d\t,\rho)$ and $q(\t+ 2d\t,\rho +
%d\rho)$ known, as well as  $q$ at each position $\rho$ for the timesteps $\t$
%and $\t + d\t$, we algebraically solve for $q(\t+2 d\t,\rho + 2 d\rho)$ and
%then repeat the process to determine $q(\t+2 d\t,\rho + 3 d\rho)$, until we
%have the entire radial profile of $q$ at the timestep $\t + 2d\t$.  We
%determine the profile of $q$ up to some distance inside the black hole horizon,
%such that errors that occur there cannot propagate back to the boundary and
%lead to errors in $\phi_{(2\D-4)}$.
%

\section{Slow quenches and the adiabatic limit} \label{slowquenches}

In this section, we consider the case where the transition between the initial
and final theories is made arbitrarily slow. In fact, we find an analytic
solution of $a_{2,4}$ for such slow quenches below. The results derived from
this approach provide an independent check for our numerical solutions --- see the discussion
in section \ref{discussion}.

Let us first consider the adiabatic limit: Given our choice for the integration
constant $a_{2,4}(-\infty)$ in eq.~(\ref{eq:ic}), the expression \reef{eq:a24}
for $a_{2,4}(\infty)$ after equilibration becomes
\begin{equation}
a_{2,4}(\infty) = -\frac{1}{3}\left(\Delta-2\right)\phi_{(2\Delta-4)}(\infty)
+ \frac{2}{3}\left(\Delta-2\right)\int^{\infty}_{-\infty}d\t' \phi_{(2\Delta-4)}(\t')\,
\dot{\phi}_{(0)}(\t')\,. \labell{eq:a24new}
\end{equation}
As noted in \cite{blm}, in the adiabatic limit, the system remains in a
quasi-static equilibrium throughout the transition. Hence the ratio of the
normalizable and non-normalizable coefficients in the bulk scalar are precisely
as given in eq.~\reef{eq:phinfty} at every stage of the transition, \ie
\begin{equation}
{\rm adiabatic:}\qquad
\phi_{(2\Delta-4)}(\t)= \phi_{(2\Delta-4)}(\infty)\,\phi_{(0)}(\t)\,.
\labell{adiabatic}
\end{equation}
In this case, the integral in eq.~(\ref{eq:a24new}) becomes
\begin{eqnarray}
\int^{\infty}_{-\infty}d\t' \phi_{(2\Delta-4)}(\t')\,\dot{\phi}_{(0)}(\t')
 &=& \frac{1}{2}\phi_{(2\Delta-4)}(\infty)\ \int^{\infty}_{-\infty}d\t'
\ \partial_{\t'}\!\left({\phi}_{(0)}(\t')\right)^{2} \nonumber \\
&=& \frac{1}{2}\phi_{(2\Delta-4)}(\infty)\,.
\labell{adiabatic2}
\end{eqnarray}
Given this result, we easily see that $a_{2,4}(\infty)$ (and hence the entropy
production) vanishes in the adiabatic limit.

For arbitrarily slow quenches, the response should approach the adiabatic
profile \reef{adiabatic} as a limit. Given the profile in eq.~\reef{eq:tanh},
this slow limit is achieved by taking $\a$ large. Then following \cite{blm}, we
can write the bulk scalar $\Phi_{\textrm{p}}$ in an expansion in inverse powers
of $\a$, \ie
\begin{equation}
\Phi_{\textrm{p}} = \phi_{(0)}\left(\tau\right)\,\Phi_{\textrm{e}}(\rho)
+ \sum^{\infty}_{n=1}\a^{-n}\,\phi^{(n)}_s(\tau)\,R^{(n)}_s(\rho)
\,, \labell{eq:phislow}
\end{equation}
where $\phi_{(0)}\left(\tau\right)$ is the profile in eq.~\reef{eq:tanh} and
$\Phi_{\textrm{e}}(\rho)$ is the equilibrium solution given in
eq.~(\ref{eq:stsol}) (with $c_1=1$). Hence the first term above is the desired
solution describing an adiabatic transition, \ie this term yields the response
in eq.~\reef{adiabatic}. We have assumed that each of the subsequent terms in
the series are separable and this form is easily confirmed, \eg see below.

Let us solve for the first correction in eq.~\reef{eq:phislow},
$\phi^{(1)}_s(\tau)\,R^{(1)}_s(\rho)$, since this term produces the leading
contribution to $a_{2,4}$ in the slow quench limit. Substituting our ansatz
\reef{eq:phislow} into the decoupled Klein-Gordon equation (\ref{eq:phi}), we
find at order $\a^{-1}$
\begin{eqnarray}
&& \phi^{(1)}_s(\tau)\,\left[-\rho^{2}\left(1-\rho^{4}\right)\partial^{2}_{\rho}+
\rho\left(\rho^{4}+3\right)\partial_{\rho}+m^{2}\right]R^{(1)}_s(\rho)\nonumber\\
&&\qquad\qquad\qquad =\ \alpha\,\dot{\phi}_{(0)}\ \rho\left(3-2\rho\,\partial_{\rho}
\right)\Phi_{\textrm{e}}(\rho)\,.
\labell{leading}
\end{eqnarray}
Note that $\alpha\,\dot{\phi}_{(0)}$ is order $\a^0$ in our expansion because
the derivative of $\phi_{(0)}$ brings out a factor of $1/\a$. Now solving
eq.~\reef{leading}, requires that the time-dependent function takes the form
$\phi^{(1)}_s(\tau)=\alpha\,\dot{\phi}_{(0)}$ while the radial profile
satisfies the inhomogeneous equation:
\begin{equation}
\left[-\rho^{2}\left(1-\rho^{4}\right)\partial^{2}_{\rho}+\rho\left(\rho^{4}+3\right)
\partial_{\rho}+\D(\D-4)\right]\!R^{(1)}_s(\rho)
=\rho\left(3-2\rho\,\partial_{\rho}
\right)\Phi_{\textrm{e}}(\rho)\,.
\labell{eq:inhom}
\end{equation}

The solution to the above equation can be written in series form as
\begin{equation}
R^{(1)}_s(\rho) = \rho^{4-\D}\sum^{\infty}_{n=0} a_{(n)}\rho^{n} +
\rho^{\D}\sum^{\infty}_{n=0} b_{(n)}\rho^{n}\,.
\labell{eq:w}
\end{equation}
Here we have two independent integration constants, $a_{(0)}$ and $b_{(0)}$.
The first coefficient $a_{(0)}$ is set to zero, since the order $\rho^{4-\D}$
contribution to $\Phi_{\textrm{p}}$ defines the source and we do not want any
new sources beyond $\phi_{(0)}$ to appear at higher orders in the $1/\alpha$
expansion in eq.~\reef{eq:phislow}. To determine $b_{(0)}$, we demand that
$R^{(1)}_s(\rho)$ is regular on the event horizon at $\rho=1$. To analyze the
profile close to the horizon, we make the change of coordinates $z=1-\rho$ and
solve for $R^{(1)}_s(z)$. Near $z=0$, we may write $R^{(1)}_s(z)$ as the series
\begin{equation}
R^{(1)}_s(z) = \sum^{\infty}_{n=0} c_{(n)} z^{n}\,,
\end{equation}
which includes only the solution that is well-behaved at $z=0$. In this series,
we now have the undetermined coefficient $c_{(0)}$.

To proceed further, we resorted to solving eq.~\reef{eq:inhom} numerically. In
particular, we produced independent solutions integrating in from $z=1$ (or
$\rho=0$) and integrating out from $z=0$ (or $\rho=1$). Then by matching the
two solutions for $R^{(1)}_s$ at an intermediate point between the asymptotic
boundary and horizon, we solved for both $b_{(0)}$ and $c_{(0)}$
simultaneously. This shooting method
was used to solve for the undetermined coefficients in the cases
$\D=\frac{7}{3}$, $\frac{8}{3}$, $\frac{10}{3}$ and $\frac{11}{3}$. We should
emphasize that although this approach is again numerical, it is an independent
approach very different in spirit from the numerical approach described
previously.\footnote{It is straightforward to write a formal Green's function solution for eq.~\reef{eq:inhom}.  Given this form of the solution, the coefficient $b_{(0)}$ can be
determined by numerically evaluating a specific integral.  The results produced this way
agree well with those given in eq.~\reef{eq:a24analytic}.}

Comparing the present solution with the expansion in eq.~\reef{eq:phip}, we see
that the corrected response takes the form
\begin{equation}
\phi_{(2\Delta-4)}(\t)= \phi_{(2\Delta-4)}(\infty)\,\phi_{(0)}(\t)-b_{(0)}\,
\dot{\phi}_{(0)}(\t)+o(1/\a^2)\,.
\labell{correctd}
\end{equation}
Again the second term above is of order $1/\a$ because the derivative acting on
$\phi_{(0)}(\t)$ brings out this factor. Now upon substituting this expression
into eq.~\reef{eq:a24new}, the adiabatic response yields zero and so we are
left with
\begin{eqnarray}
a_{2,4}(\infty)& =&-
\frac{2}{3}\left(\Delta-2\right)b_{(0)}\int^{\infty}_{-\infty}d\t'
 \dot{\phi}_{(0)}^{\,2}(\t')+o(1/\a^2)
 \nonumber\\
& =&-
\frac{2}{9\a}\left(\Delta-2\right)b_{(0)}+o(1/\a^2)\,,
\labell{eq:a24newb}
\end{eqnarray}
where the final result is produced by inserting eq.~\reef{eq:tanh} for
$\phi_{(0)}$. Hence given the numerical solution for $b_{(0)}$ for each value
of $\D$ listed above, we can determine the leading contribution to $a_{2,4}$
for slow quenches. Our results are as follows:
\begin{eqnarray}
\D=\frac{7}{3}:&&a_{2,4}\left(\infty\right) = -0.01958\,\frac1\alpha\,,\nonumber\\
\D=\frac{8}{3}:&&a_{2,4}\left(\infty\right) = -0.05205\,\frac1\alpha\,,\nonumber\\
\D=\frac{10}{3}:&&a_{2,4}\left(\infty\right) = -0.09838\,\frac1\alpha\,,\nonumber\\
\D=\frac{11}{3}:&&a_{2,4}\left(\infty\right) = -0.1083\ \,\frac1\alpha\,.
\labell{eq:a24analytic}
\end{eqnarray}
These results compare well to our numerical results, as discussed
below in section \ref{discussion}.

\section{Results} \label{results}

We now turn to the results of our numerical simulations.  We determined the
response functions for various values of $\a$ and with several different masses
of the bulk scalar. Explicitly, the scalar masses for which we studied the
quenches are: $m^{2}=-\frac{35}{9}$ (with $\Delta=\frac{7}{3}$),
$m^{2}=-\frac{32}{9}$ (with $\Delta=\frac{8}{3}$), $m^{2}=-\frac{20}{9}$ (with
$\Delta=\frac{10}{3}$) and $m^{2}=-\frac{11}{9}$.  The masses were chosen so
that the conformal dimension lies in the desired range $2<\D<4$ and so that
$4-2\D$ is not an integer. The latter ensures that the asymptotic expansion
(\ref{eq:phip}) for the scalar does not contain any logarithmic terms. We
further comment on $\Delta=4$ case, which together with results of \cite{blm}
would form a complete picture of perturbative quenches in strongly couple gauge
theories induced by the coupling of a relevant operator.

In  subsection \ref{fast}, we extract the response function
$\phi_{(2\Delta-4)}$ for fast quenches (\ie $\a<1$) from our numerical data for
each $\Delta$.  From that we see that $a_{2,4}$ and the maximum displacement of
$\phi_{(2\Delta-4)}$ follows a scaling behaviour for fast quenches. In
subsection \ref{universal}, we see that the scaling of these quantities follow
a universal behaviour, determined by $\D$ and $\a$ only.  In subsection
\ref{numslow}, we show the numerical results for slow quenches (\ie $\a> 1$).
We show that for very large $\a$, $\phi_{(2\Delta-4)}$ approaches the adiabatic
response \reef{adiabatic}. We also find that $a_{2,4}$ scales as $\alpha^{-1}$,
as expected from the analysis in section \ref{slowquenches}. Finally, in
subsection \ref{excitation}, we study the excitation times of the response, by
considering when it first deviates from the adiabatic response by more than
$5\%$ (an arbitrary threshold).  We find that for fast quenches, the scaled
excitation time $\t_{ex}/\a$ scales logarithmically with the quenching
parameter $\a$.  We also see a universal behaviour in the slopes
$-\left(\frac{\partial(\t_{ex}/\a)}{\partial \log\a}\right)$ in the fast quench
limit for fractional $\DD$.  In the same subsection we study the relaxation
times for the quenches, defined by the last time that $\phi_{(2\Delta-4)}$
falls below a $5\%$ deviation from its final equilibrium value.

\subsection{Response for fast quenches}\labell{fast}

Here we list the results for intermediate to fast quenches (\ie $\a<1$).  We
first find the response $\phi_{(2\Delta-4)}$ and then determine the coefficient
$a_{2,4}(\infty)$ using equation (\ref{eq:a24}), as well as the integration
constant $a_{2,4}(-\infty)$ in eq.~(\ref{eq:ic}).  By doing multiple
simulations of the time-dependent behaviour of $\phi_{(2\Delta-4)}$ with increasingly
finer discretizations, we can extrapolate for the continuum value of
$a_{2,4}(\infty)$.  Knowing $a_{2,4}(\infty)$, we can determine the response of
various physical quantities in the boundary theory to the quench, using
eqs.~\reef{eq:sfosi}--\reef{eq:pfpifw}. Note that these results would also
apply to the reverse quenches, as discussed in section \ref{reverse}. We also
study the behaviour of $\phi_{(2\Delta-4)}$ and $a_{2,4}(\infty)$ as functions
of $\a$. In the next subsection, we compare this $\a$ dependence for different
values of $\Delta$ and identify the universal behaviour for operators of
different dimension.

\begin{figure}[t]
\begin{center}
\psfrag{at}[Br][tl]{{${\scriptstyle\a^{-1}\t}$}}
\psfrag{aphi}[c]{{$\scriptstyle{\a^{2\Delta-4}\phi_{(2\Delta-4)}}$}}
  \includegraphics[width=3in]{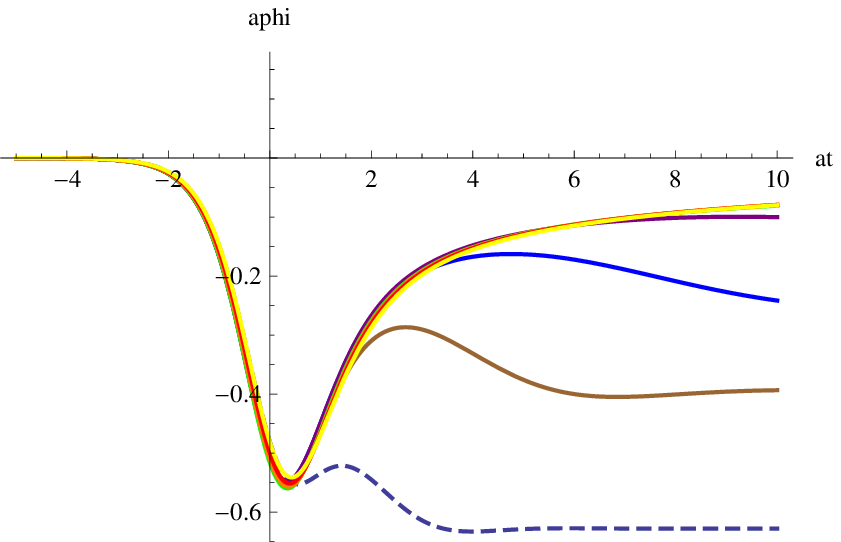}
    \includegraphics[width=3in]{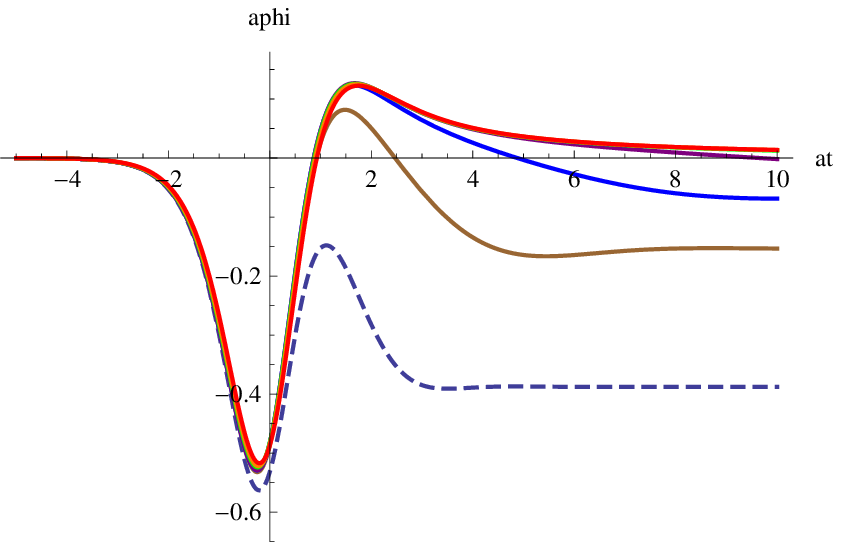}
      \includegraphics[width=3in]{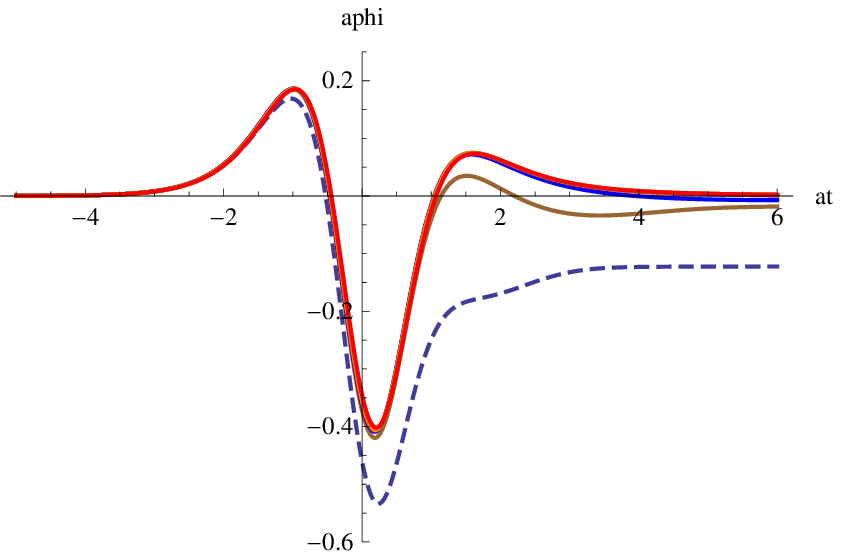}
        \includegraphics[width=3in]{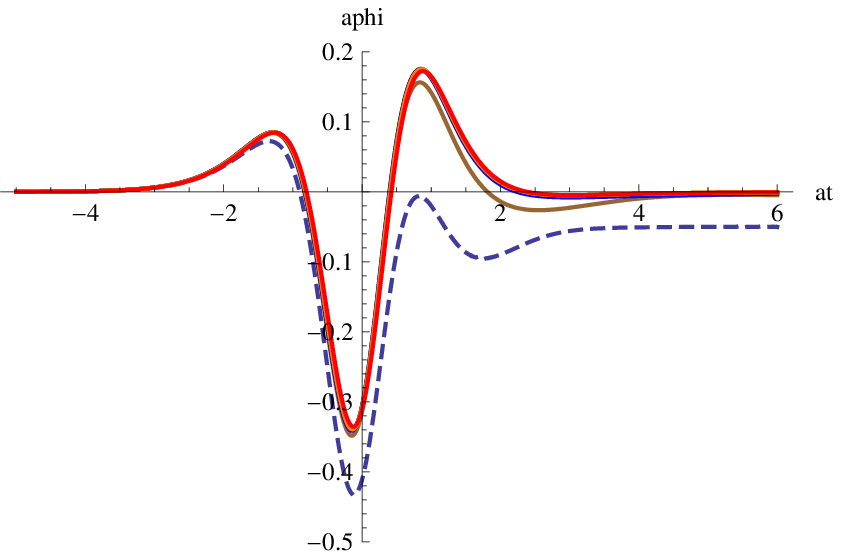}
\end{center}
  \caption{(Colour online)
Plots of the response coefficient $\phi_{(2\Delta-4)}$ for quenches of
different speeds, in the fast quench regime. Time is rescaled by a factor of
$1/\a$ and the value of $\phi_{(2\DD-4)}$ is rescaled by $\a^{2\D-4}$.
Clockwise from the top left, the plots are for $\DD=7/3$, $8/3$, $11/3$
and $10/3$. In each case, the response is presented for $\a=1$ (dashed),
$1/2$(brown), $1/4$ (blue), $1/8$ (purple), $1/16$ (green), $1/32$ (orange) and
$1/64$ (red) (as well as $\a=1/128$ (yellow) for $\DD=7/3$). }
\label{p12fast}
\end{figure}

In figure \ref{p12fast}, we plotted $\phi_{(2\DD-4)}$ against time for various values of $\a$. The time is rescaled by a
factor of $1/\a$ so that the different plots equilibrate in approximately the
same distance on the horizontal axis. In each of the plots, we have also
rescaled the vertical axis by $\a^{2\D-4}$. Hence the maximum displacement of
$\phi_{(2\DD-4)}$ is actually growing with decreasing $\a$. With the rescaled
axes, the peaks corresponding to smaller $\a$ lie close together, which seems
to indicate that there is a universal scaling depending on the operator
dimension. Further with these scalings, the response seems to converge to a
particular limit with decreasing $\a$.  For fast enough quenches, we expect
these plots to coincide exactly.

\begin{figure}[t]
\begin{center}
\psfrag{phi}{{$\scriptstyle{\phi_{(0)}}$}}
\psfrag{aphi}{{$\scriptstyle{\a^{2\Delta-4}\phi_{(2\Delta-4)}}$}}
  \includegraphics[width=2.4in]{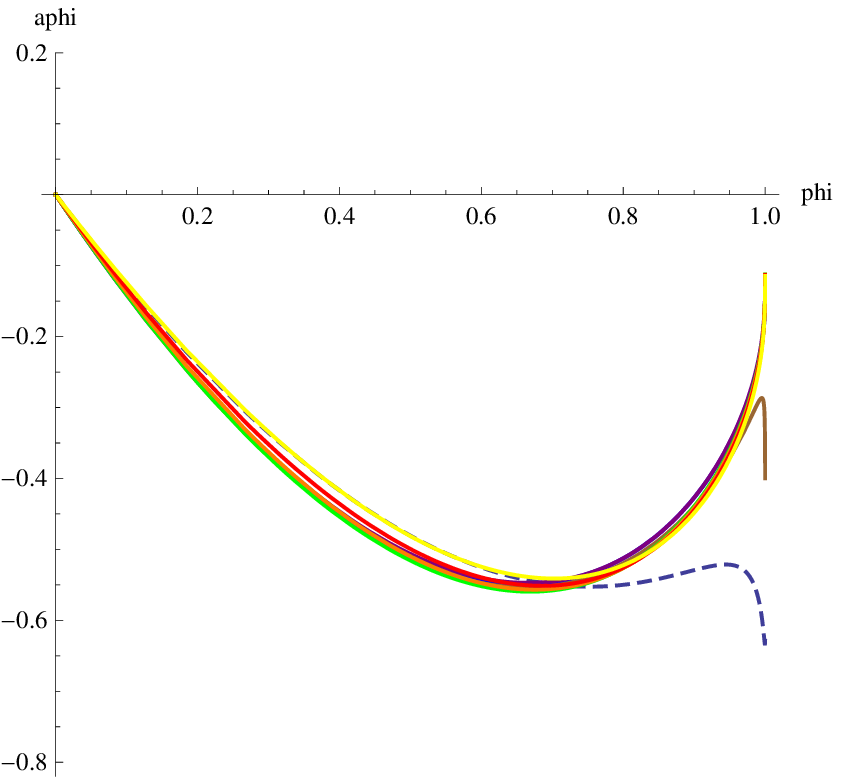}
    \includegraphics[width=2.4in]{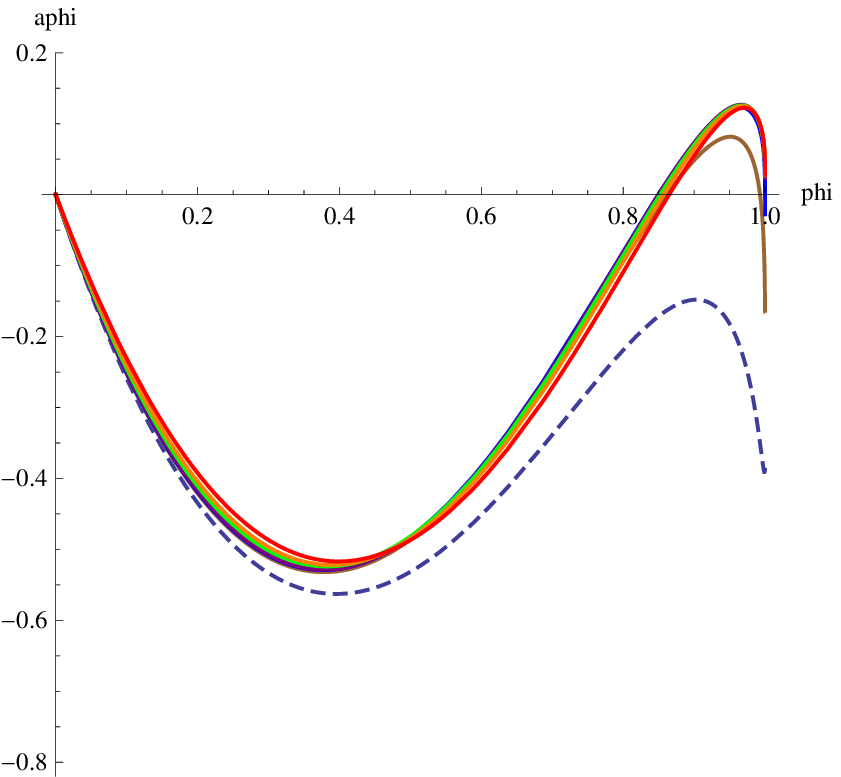}
      \includegraphics[width=2.4in]{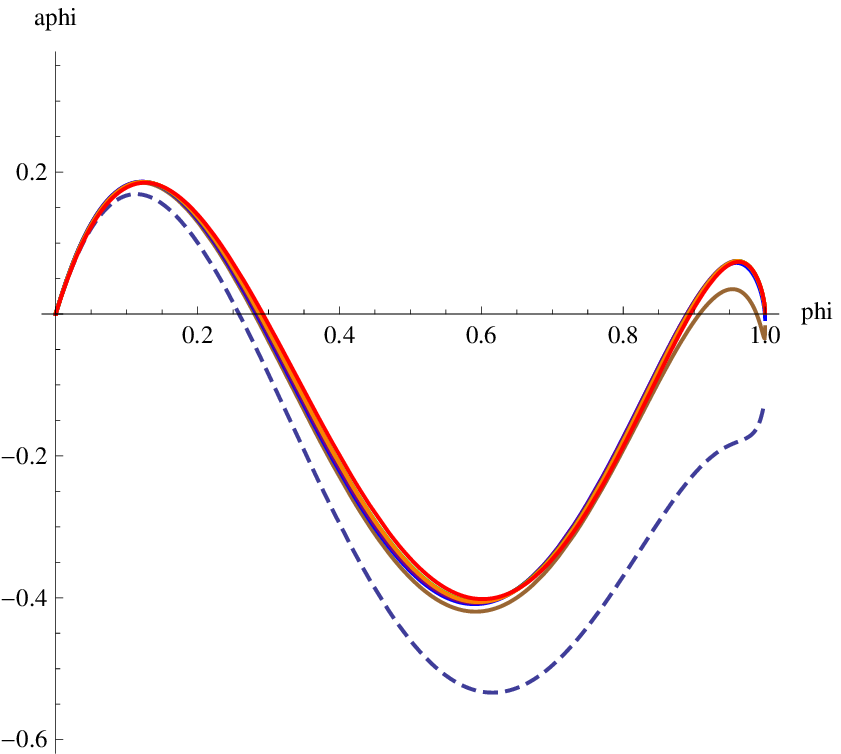}
        \includegraphics[width=2.4in]{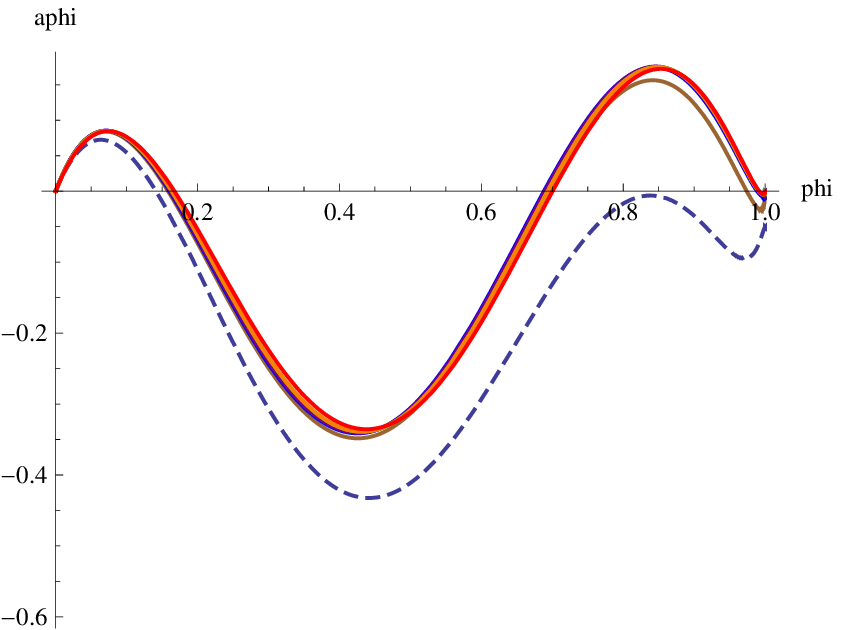}
\end{center}
  \caption{(Colour online)
Plots of $\phi_{(2\Delta-4)}$ for quenches of different speeds, in the fast
quench regime. $\phi_{(2\DD-4)}$ is rescaled by $\a^{2\D-4}$ and is plotted
against the actual value of the source $\phi_{(0)}$. Clockwise from the top
left, the plots are for $\DD=7/3$, $8/3$, $11/3$ and $10/3$. In
each case, the response is presented for various values of $\a$, which are
indicated using the same colour scheme as in figure \ref{p12fast}.}
\label{p12p0fast}
\end{figure}

In figure \ref{p12p0fast}, we show a different visualization of the same fast
quenches.  Here, $\a^{2\D-4}\phi_{(2\D-4)}$ is plotted as a function of the
source $\phi_{(0)}$.  It is perhaps easier to see the convergence of the faster
quenches to a limiting curve. Note that because of the growth of
$\phi_{(2\D-4)}$ for fast quenches, the expression (\ref{eq:a24}) for
$a_{2,4}(\infty)$ is dominated by the integral. Further the latter can be
re-expressed as $\int^{1}_{0}d\phi_{(0)}\phi_{(2\D-4)}$ and so for these fast
quenches, $a_{2,4}(\infty)$ is essentially given by the area under the curves
in figure \ref{p12p0fast}.

\begin{figure}[t]
\begin{center}
\psfrag{lna}[B][tl]{{$\scriptstyle{\log\a}$}}
\psfrag{lna24}[c]{{$\scriptstyle{\log(-a_{2,4}(\infty))}$}}
  \includegraphics[width=3in]{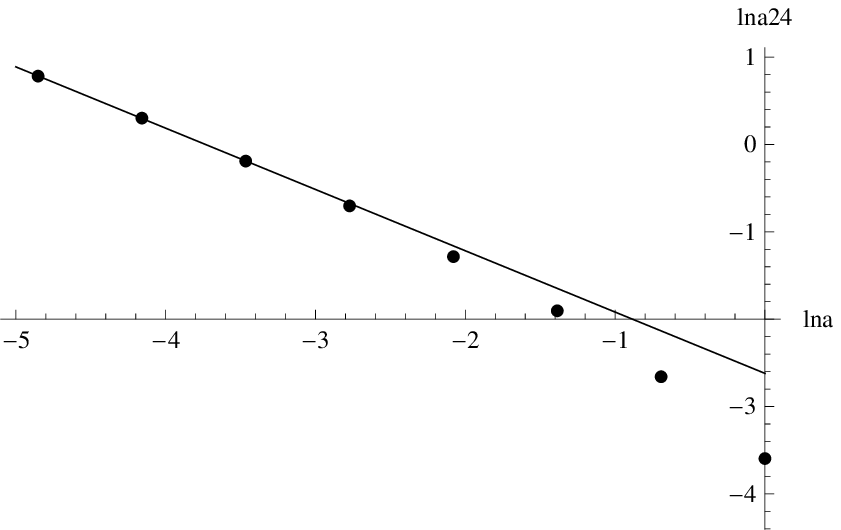}
  \includegraphics[width=3in]{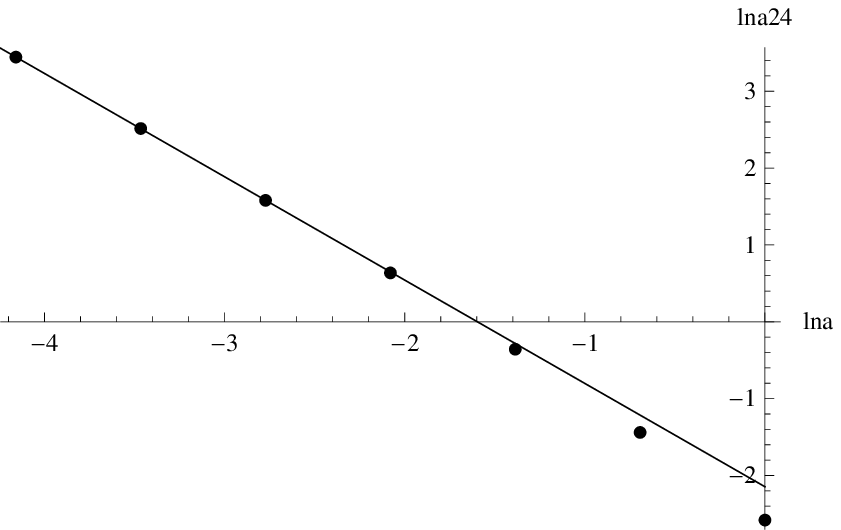}
  \includegraphics[width=3in]{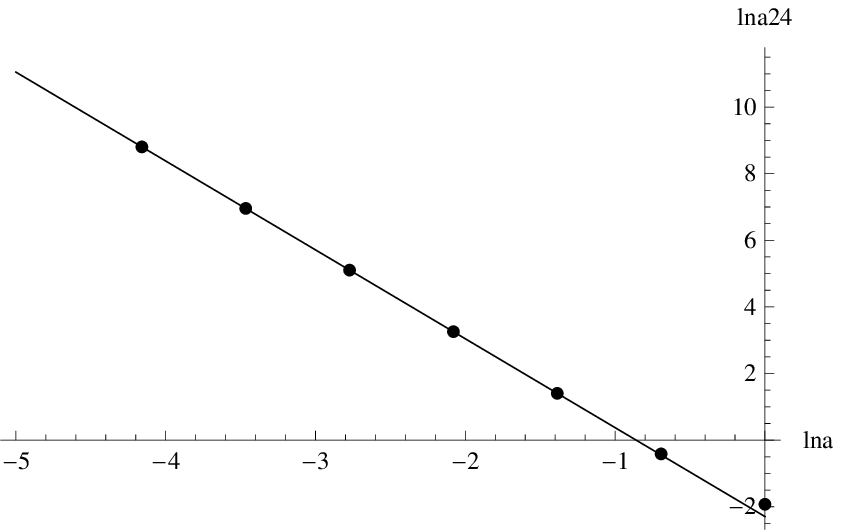}
  \includegraphics[width=3in]{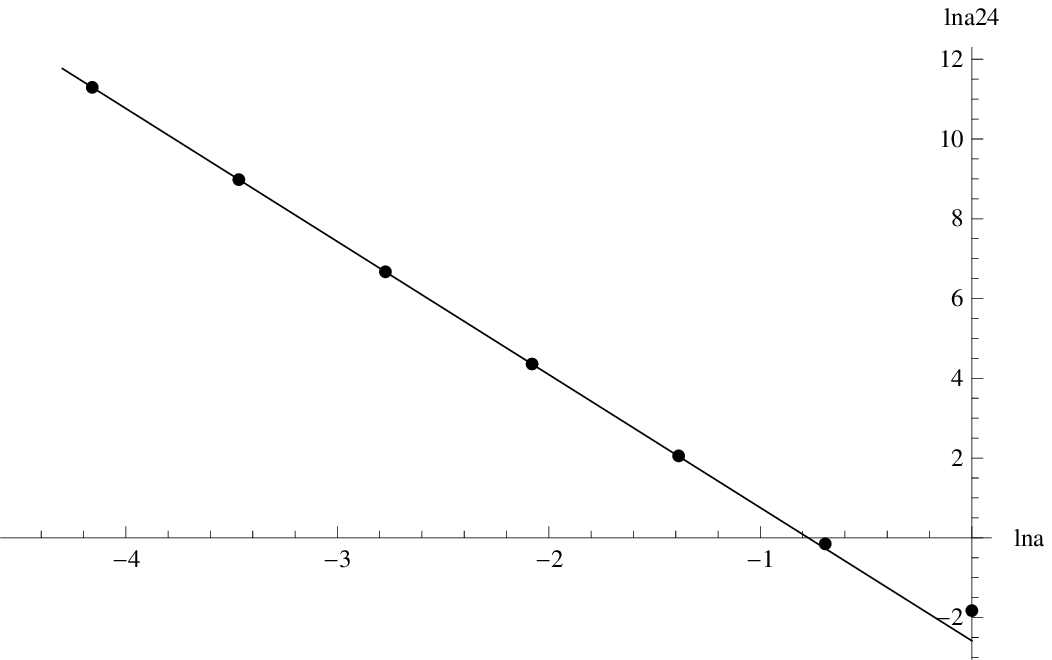}
\end{center}
  \caption{
$\log$-$\log$ plots for $-a_{2,4}(\infty)$ versus $\a$ for various $\DD$, in
the fast quench regime. The straight lines shown are least-squares fits through
the three leftmost data-points in each case. The fact that the plots tend to
straight lines for negative values of $\log\a$ means that $a_{2,4}(\infty)$
scales as a power law for small $\a$. Clockwise from the top left, the plots
are for $\DD=7/3$, $8/3$, $11/3$ and $10/3$.}  \label{loga24fast}
\end{figure}

In figure \ref{loga24fast},\footnote{Note that we plot the logarithm of
$-a_{2,4}(\infty)$ since $a_{2,4}(\infty)$ is always negative.} we see that
$\log(-a_{2,4}(\infty))$ plotted against $\log\a$ tends to a straight line for
small values of $\a$. This indicates that $a_{2,4}(\infty)$ scales as some
power law of the quenching parameter $\a$ for very fast quenches. A fit of this
linear behaviour suggests the slope matches $4-2\Delta$ in each case. The lines
shown in the plot are the linear fits through the points corresponding to the
three fastest quenches, thus showing the asymptotic behaviour of
$a_{2,4}(\infty)$ for fast quenches. Although it is not shown, for small values
of $\a$, the logarithm of $\max\{|\phi_{(2\D-4)}|\}$ plotted against $\log\a$
also tends to a straight line with the same slope as in the plot of
$\log(-a_{2,4}(\infty))$.

As discussed in \cite{blm,blmn}, the scaling of the response $\phi_{(2\D-4)}$
for fast quenches is more subtle when $\D=2,$ 3 and 4. Specifically, it is
rather the `subtracted' response, defined by
\begin{equation}
\hat{\phi}_{(2\D-4)}\equiv \phi_{2\D-4}+
\begin{cases}
\ \ \ \ln\a\ \phi_{(0)}\,,\qquad\quad\  {\rm for}\ \D=2\,,\cr
\ \ \frac12 \ln\a\ \ddot{\phi}_{(0)}\,,\qquad \quad
{\rm for}\ \D=3\,,\cr
-\frac{1}{16} \ln\a\ \ddddot\phi_{(0)}\,,\qquad
{\rm for}\ \D=4\,,
\end{cases}
\labell{subresp}
\end{equation}
which scales faithfully in the limit of fast quenches, \ie
\begin{equation}
\lim_{\a\to 0}\ \a^{2\D-4}\hat{\phi}_{(2\D-4)}\ =\ {\rm constant}\,.
\labell{subscale}
\end{equation}
Notice, however, that the additional $\ln\a$ terms in $\phi_{2\D-4}$ above do
not contribute to $a_{2,4}(\infty)$ for $\D=3$ or 4. For these cases, we have
\begin{eqnarray}
\D=3\,:&&
\int_{-\infty}^\infty\ d\t\ \ddot{\phi}_{(0)}(\t)\,\dot{\phi}_{(0)}(\t)
=\left[\frac 12 \dot{\phi}_{(0)}(\t)^{\,2}\right]_{-\infty}^\infty=0
\labell{vanish}\\
\D=4\,:&&
\int_{-\infty}^\infty\ d\t\ \ddddot{\phi}_{(0)}(\t)\,\dot{\phi}_{(0)}(\t)
=\left[ \dddot{\phi}_{(0)}(\t)\,\dot{\phi}_{(0)}(\t)-\frac 12
\ddot{\phi}_{(0)}(\t)^{\,2}
\right]_{-\infty}^\infty=0
\nonumber
\end{eqnarray}
which, as we have indicated, both vanish for a generic source $\phi_{(0)}$ as
long as the profile becomes constant as $\t\to\pm\infty$. Hence for $\D=3$ and
4, we still  have $a_{2,4}(\infty)\sim 1/\a^{2\D-4}$ for fast quenches as
$\a\to 0$. On the other hand, as shown in \cite{blm}, the logarithmic term in
eq.~\reef{subresp} gives the dominant contribution in eq.~\reef{eq:a24} for
fast quenches with $\D=2$ and we have instead
\begin{equation}
a_{2,4}(\infty)=\frac 16\ \ln\a +o(1)\,,\qquad{\rm as}\  \a\to 0\,.
\labell{a24dim2}
\end{equation}

\subsection{Universal behaviour for fast quenches}\label{universal}

\begin{figure}[t]
\begin{center}
\psfrag{dd}{{$\Delta$}}
\psfrag{sl}{{$-\frac{d\log(-a_{2,4}(\infty))}{d\log\a}$}}
  \includegraphics[width=4in]{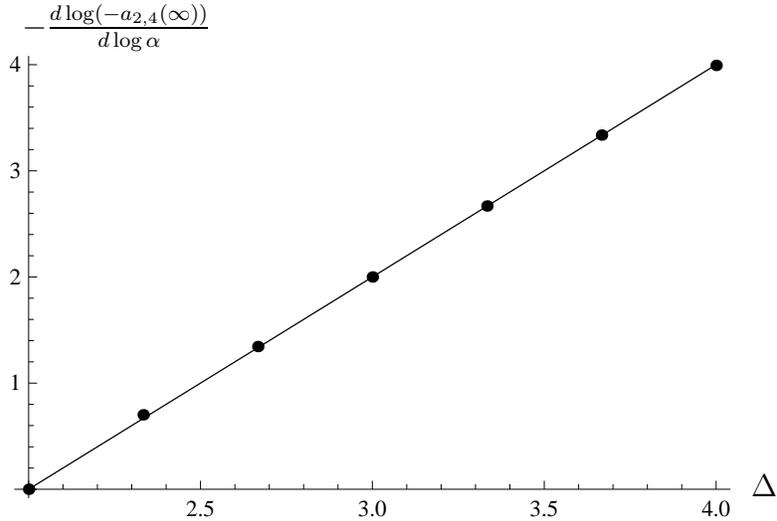}
\end{center}
  \caption{(Colour online)
Plot of the asymptotic scaling of $-a_{2,4}(\infty)$ as a power of $\a$, in the
fast quench limit. The line shown is the predicted theoretical trend $\frac{-d\log(-a_{2,4}(\infty))}{d\log\a}=2\DD-4$. Points
shown are for $\DD=2$, $7/3$, $8/3$, $3$, $10/3$, $11/3$ and $4$. The
datapoints for $\D=2,3$ are taken from \cite{blm}. The datapoint for $\D=4$ is
taken from \cite{blmn}.} \label{a24meta}
\end{figure}

For fast quenches, our numerics above suggested that $-a_{2,4}(\infty)$ grows
as $\a^{-(2\DD-4)}$ as $\a$ becomes arbitrarily small. This behaviour is
confirmed in figure \ref{a24meta}. This plot shows
$\frac{-d\log(-a_{2,4}(\infty))}{d\log\a}$ versus $\DD$ for $\DD=2$, $\DD=7/3$,
$\DD=8/3$, $\DD=3$, $\DD=10/3$, $\DD=11/3$ and $\DD=4$. The data-points are the
slopes of the straight line fits for the plots shown in figure
\ref{loga24fast}. The data-points for $\DD=2$ and $\DD=3$ are taken from
\cite{blm} while the data-point for $\D=4$ is taken from \cite{blmn}. We set
the log derivative to zero for $\DD=2$ but, as noted above in
eq.~\reef{a24dim2}, $a_{2,4}(\infty)$ actually scales as a logarithm of $\a$
\cite{blm}.

As well as the individual data-points, we have plotted the line
\begin{equation}
-\frac{d\log[-a_{2,4}(\infty)]}{d\log\a}=2\DD-4\,. \labell{complicated2}
\end{equation}
In fact, a least-squares fit through the data-points yields
\begin{equation}
-\frac{d\log[-a_{2,4}(\infty)]}{d\log\a}=1.99\DD-3.96\,,
\labell{complicated}
\end{equation}
which matches the expected trend within our numerical accuracy. Hence figure
\ref{a24meta} confirms the universal scaling
\begin{equation}
|a_{2,4}(\infty)|\propto\frac{1}{\a^{2\DD-4}}
\labell{eq:propto}
\end{equation}
for fast quenches by an operator with $2<\DD\le4$.

As noted above, the maximum displacement of $\phi_{(2\Delta-4)}$ seems to
exhibit the same scaling behaviour as above when $\D$ is fractional. However,
as also commented above for $\D=2$, both $a_{2,4}(\infty)$ and
$\phi_{(2\Delta-4)}$ grow as $-\log\a$ for small $\a$ \cite{blm}. Similarly,
for small $\a$ with $\D=3$ and 4, $a_{2,4}(\infty)$ has the above scaling but
the maximum displacement of $\phi_{(2\Delta-4)}$ exhibits an additional
$\log\a$ growth on top of this simple scaling --- see further discussion around
eq.~\reef{subresp} and in \cite{blmn}.

\subsection{Response for slow quenches} \label{numslow}

\begin{figure}[t]
\begin{center}
\psfrag{at}[Br][tl]{{$\scriptstyle{\a^{-1}\t}$}}
\psfrag{phi}[c]{{$\scriptstyle{\phi_{(2\Delta-4)}}$}}
  \includegraphics[width=3in]{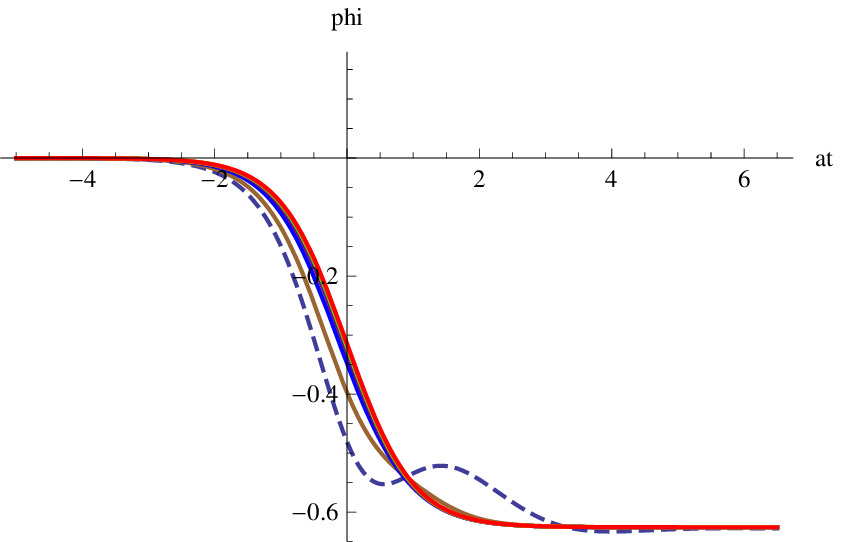}
    \includegraphics[width=3in]{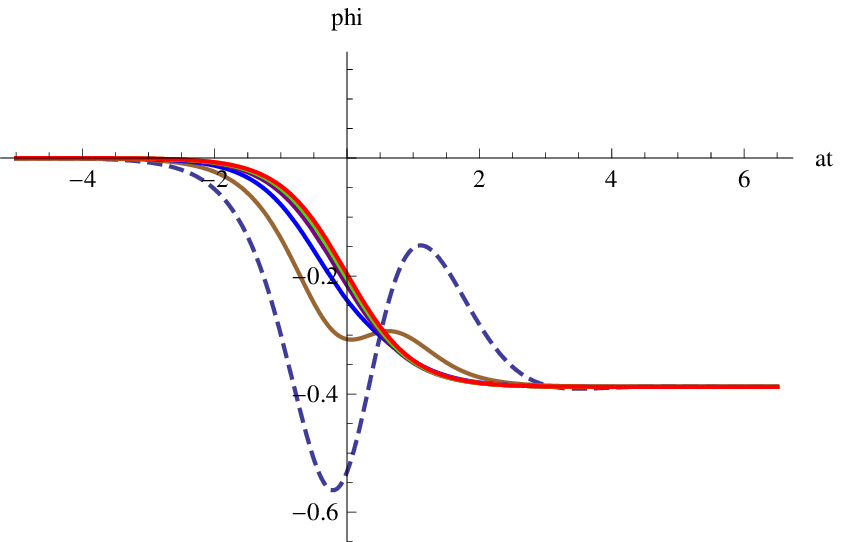}
      \includegraphics[width=3in]{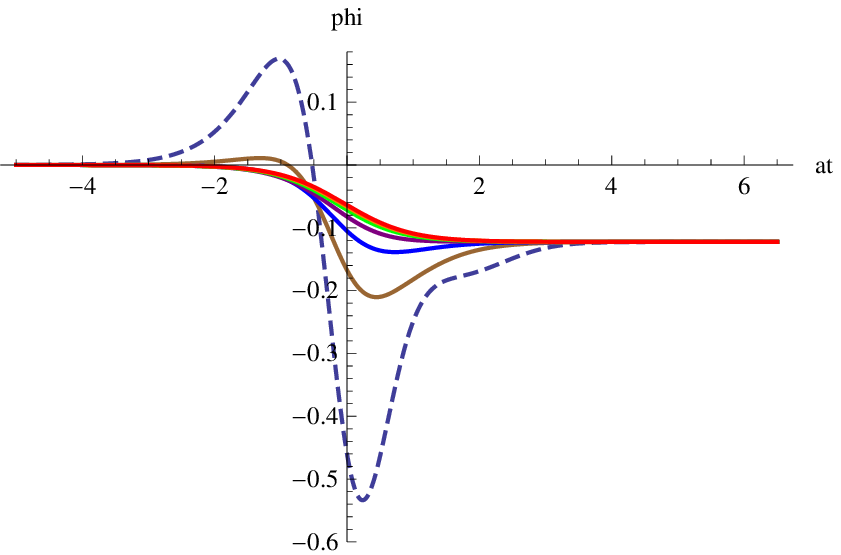}
        \includegraphics[width=3in]{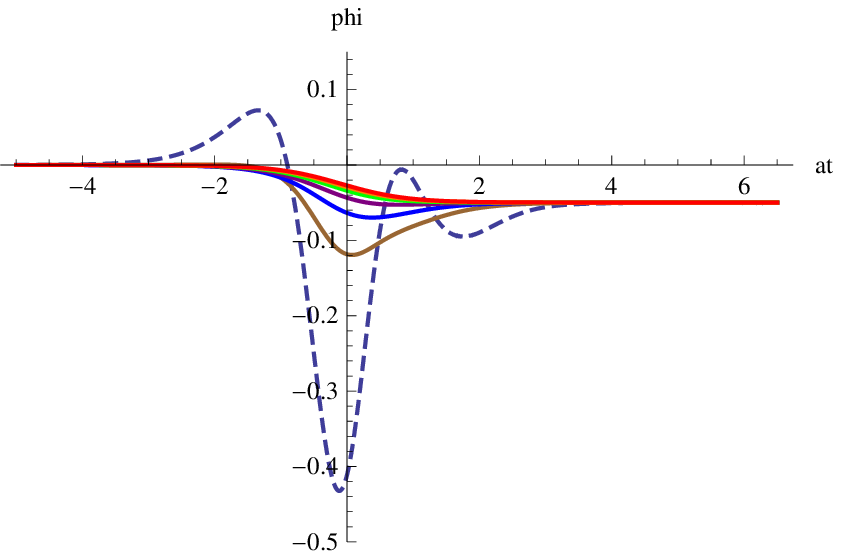}
\end{center}
  \caption{(Colour online)
Plots of $\phi_{(2\Delta-4)}$ for quenches of different speeds in the slow
quench regime. Time is rescaled by a factor of $\a^{-1}$ and the value of
$\phi_{(2\DD-4)}$ is rescaled by $\a^{2\D-4}$.  Plots for larger $\a$ follow
an inverted $\tanh$-profile more closely, which is a negative constant times
the source. Clockwise from the top left, the plots are for $\DD=7/3$,
$8/3$, $11/3$ and $10/3$. In each case, the response is presented
for $\a=1$ (dashed), $2$ (brown), $4$ (blue), $8$ (purple), $16$ (green), $32$
(orange) and $64$  (red).}  \label{p12slow}
\end{figure}

In figure \ref{p12slow}, we plotted $\phi_{(2\DD-4)}$ as a function of $\t$
for the various values of $\D$ and $\a$.  The time is scaled by $\a^{-1}$ so
that the different plots would equilibrate in approximately the same distance
on the horizontal axis. As $\a$ grows large, the curves approach an inverted
$\tanh$ graph, which is the expected adiabatic limit, \ie $\phi_{
(2\DD-4)}(\infty)\, \phi_{(0)}(\t)$ --- see eq.~\reef{adiabatic}. Note that in
this case there is no need to rescale the vertical axis since $\phi_{(2\DD-4)}$ is
of the same order in all cases.

\begin{figure}[t]
\begin{center}
\psfrag{lna}[B][tl]{{$\scriptstyle{\log\a}$}}
\psfrag{lna24}[c]{{$\scriptstyle{\log(-a_{2,4}(\infty))}$}}
  \includegraphics[width=3in]{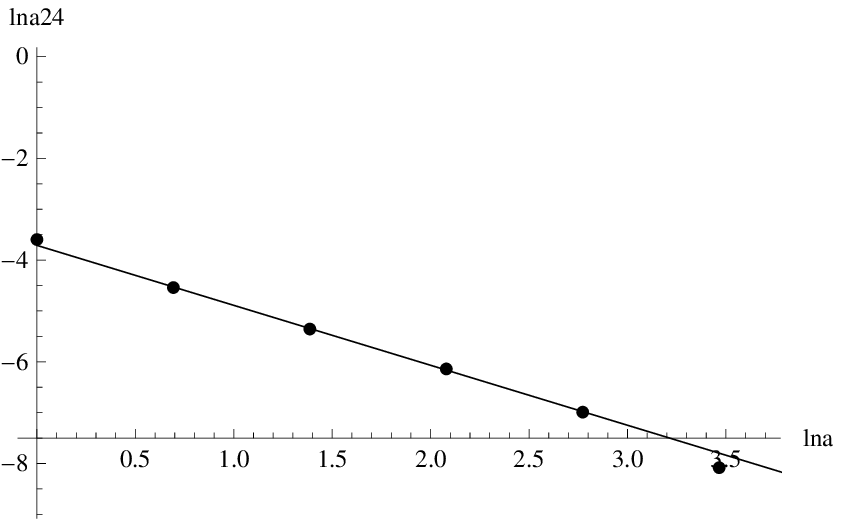}
  \includegraphics[width=3in]{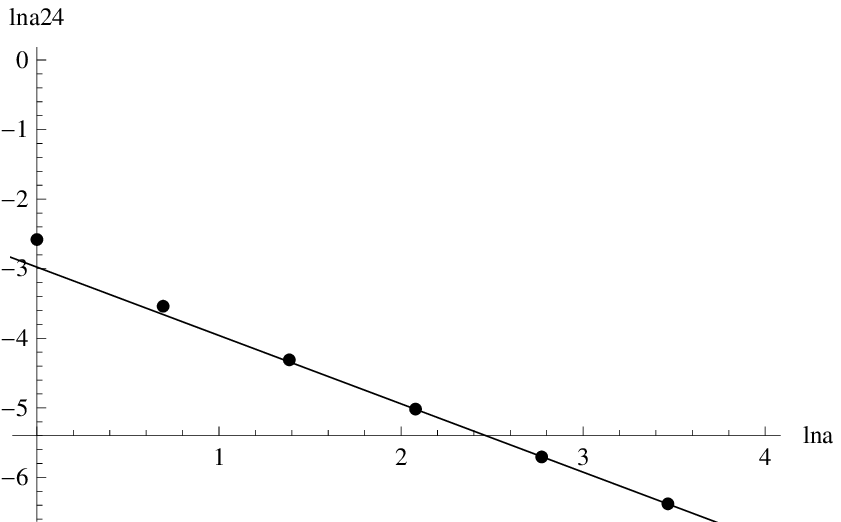}
  \includegraphics[width=3in]{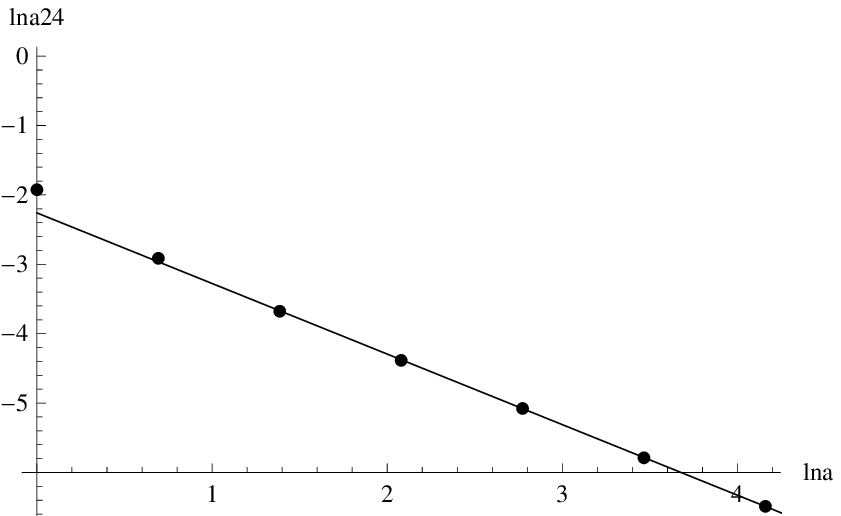}
  \includegraphics[width=3in]{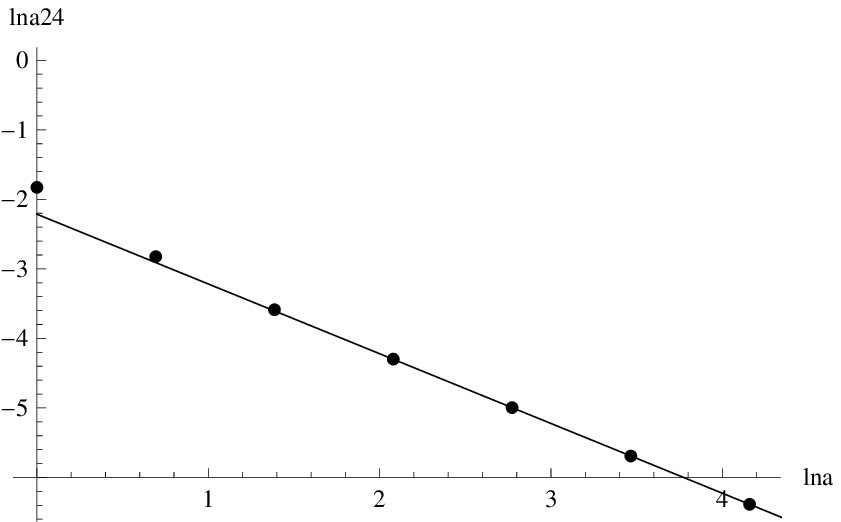}
\end{center}
  \caption{
$\log$-$\log$ plots for $|a_{2,4}|$ versus $\a$ for various $\DD$ in the slow
quench regime.  The fact that the plots tend to straight lines for positive
values of $\log\a$ means that $a_{2,4}$ scales as a power law for large $\a$.
Clockwise from the top left, the plots are for $\DD=7/3$, $8/3$, $11/3$ and
$10/3$.}  \label{loga24slow}
\end{figure}

As discussed in section \ref{slowquenches}, $a_{2,4}(\infty)$, which controls
the entropy production, goes to zero in the adiabatic limit.  Further, the
analysis there showed that the leading contribution gave
$a_{2,4}(\infty)\propto 1/\a$ for slow quenches --- see eq.~\reef{eq:a24newb}.
This behaviour is revealed in our numerical results in figure \ref{loga24slow}.
There $\log(-a_{2,4}(\infty))$ is shown as a function of $\log\a$ and we see that for
large $\a$, the results can be fit with a straight line with a slope of
approximately $-1$ in all the plots shown.  Similar to the fast quench case,
the straight lines are fit through the last three data-points
in each plot.\footnote{Note that in the
case $\D=7/3$ we fit the line through three intermediate points,
since our result for $a_{2,4}(\infty)$ contained a significant numerical error
for the largest value of $\alpha$ shown.} Further the intercepts of the
straight lines in figure \ref{loga24slow} should correspond to
(minus the logarithm of) the coefficients given in eq.~\reef{eq:a24analytic}. We defer the detailed
comparison of the results derived in section \ref{slowquenches} and with the
numerical simulations here until section \ref{discussion}.

Again, although not shown, the same behaviour was also found for slow quenches
in the case $\D=4$. This behaviour was also found to hold for $\D=2$ and $3$ in
\cite{blm}.

\begin{figure}[t]
\begin{center}
\psfrag{phi}{{${\phi_{(0)}}$}}
\psfrag{hphi}{{${\hat{\phi}_{(2\Delta-4)}}$}}
        \includegraphics[width=4in]{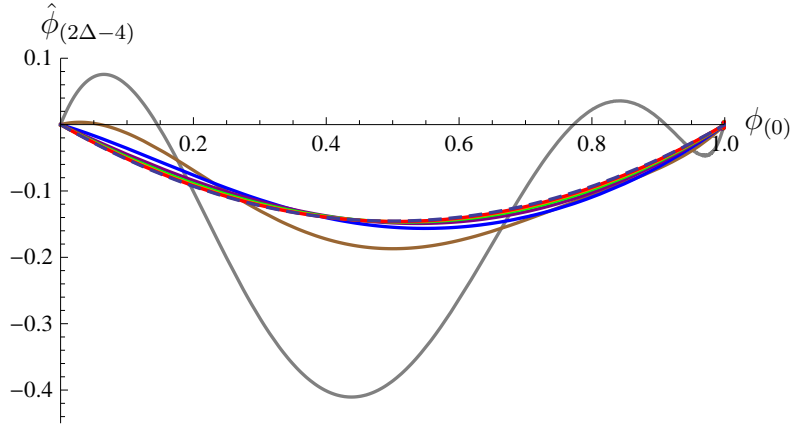}
\end{center}
  \caption{(Colour online)
Plots of the deviation from the adiabatic response as a function of
$\phi_{(0)}$ for slow quenches with different speeds and $\D=11/3$ --- see
eq.~\reef{pow9} for the definition of $\hat{\phi}_{(2\Delta-4)}$. The curves
correspond to $\a=1$ (grey), $2$ (brown), $4$ (blue), $8$ (purple), $16$
(green), $32$ (orange) and $64$ (red). The dashed curve corresponds to
$b_{(0)}\dot{\phi}_{(0)}$.} \label{p12hatp0slow}
\end{figure}

For slow quenches, let us define the deviation of the response from the
adiabatic limit \reef{adiabatic} as
\begin{equation}
\hat{\phi}_{(2\Delta-4)}(\t) = \alpha \left(\,\phi_{(2\Delta-4)}(\t) -
\phi_{(2\Delta-4)}\left(\infty\right)\,\phi_{(0)}(\t)\,\right).
\labell{pow9}
\end{equation}
As discussed in section \ref{slowquenches}, this function should be
approximately given by $b_{(0)}\dot{\phi}_{(0)}$, where $b_{(0)}$ was the
coefficient of the normalizable mode in the radial profile of the $1/\alpha$
contribution. Figure  \ref{p12hatp0slow} shows the deviation $\hat{\phi}_{
(2\Delta-4)}$ as a function of $\phi_{(0)}$ for $\D=11/3$ and different values
of $\alpha$. The dashed curve shows $b_{(0)}\dot{\phi}_{(0)}$, where $b_{(0)}$
was determined by the shooting method in section \ref{slowquenches}. As we can
see in the figure, as $\a$ grows large, the deviation determined by our
numerical simulations is converging on the expected curve. Those curves
corresponding to larger $\alpha$ fit the dashed curve best. The curve for $\a=64$
lies practically on top of the limiting dashed curve.

\subsection{Excitation and relaxation times}\label{excitation}

\begin{figure}[t]
\begin{center}
\psfrag{tex}{{$\t_{ex}$}}
\psfrag{teq}{{$\t_{eq}$}}
\psfrag{toa}{{$\scriptstyle{\t}$}}
\psfrag{delta}{{$\scriptstyle{\delta}$}}
\includegraphics[width=4in]{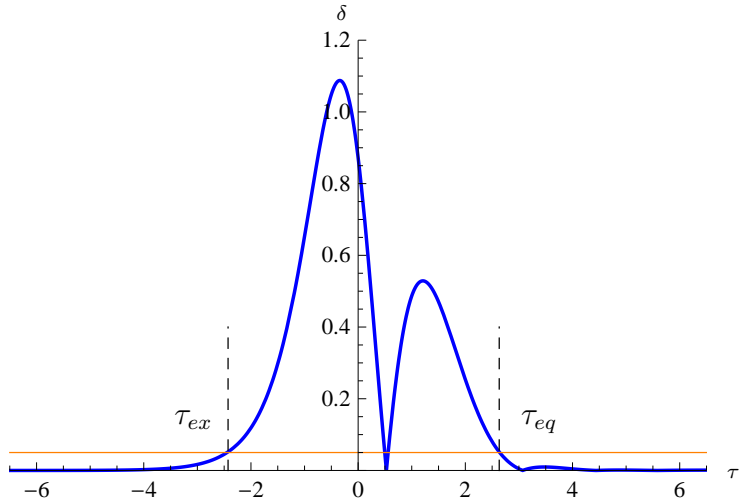}
  \end{center}
  \caption{(Colour online)
Plot of $\dd$ as a function of ${\t}$ for $\a=1$ and $\DD=8/3$. The excitation
time $\t_{ex}$ and relaxation time $\t_{eq}$ are shown as the first and
final times, respectively, at which $\dd$ crosses the threshold of
$\varepsilon=.05$ (shown as the orange line).} \label{relax}
\end{figure}

Next, we consider the excitation and relaxation times following the approach
presented in \cite{blm}. First, we define
\begin{equation}
\dd \equiv \left|\frac{\phi_{(2\DD-4)}-\phi_{(2\DD-4)}(\infty)\phi_{(0)}
}{\phi_{(2\DD-4)}(\infty)}\right|\,,
\labell{definedd}
\end{equation}
\ie the absolute value of the relative deviation of the response
$\phi_{(2\DD-4)}$ from the adiabatic limit $\phi_{(2\DD-4)}(\infty)\,
\phi_{(0)}$. Then we define the excitation time $\t_{ex}$ as the first time at
which $\dd$ reaches $\varepsilon=.05$, where the latter was chosen as an
arbitrary threshold. Similarly, the relaxation or equilibration time $\t_{eq}$
is the latest time at which $\dd$ drops below the $\varepsilon=.05$ threshold.
As an example, $\dd$ is shown in figure \ref{relax} for $\a=1$ and $\DD=8/3$.
The vertical grid lines indicate the excitation time $\t_{ex}$ and relaxation time
$\t_{eq}$. Note that our definitions of $\t_{ex}$ and $\t_{eq}$ is only
expected to be meaningful for relatively small $\a$. As $\a$ grows large, the
response approaches the adiabatic profile \reef{adiabatic} and so for
sufficiently large $\a$, $\dd$ will never exceed the chosen threshold.

\begin{figure}[t]
\begin{center}
\psfrag{toa}[c]{{$\scriptstyle{\frac{|\t_{ex}|}{\a}}$}}
\psfrag{lna}[B][tl]{{$\scriptstyle{\log\a}$}}
  \includegraphics[width=3in]{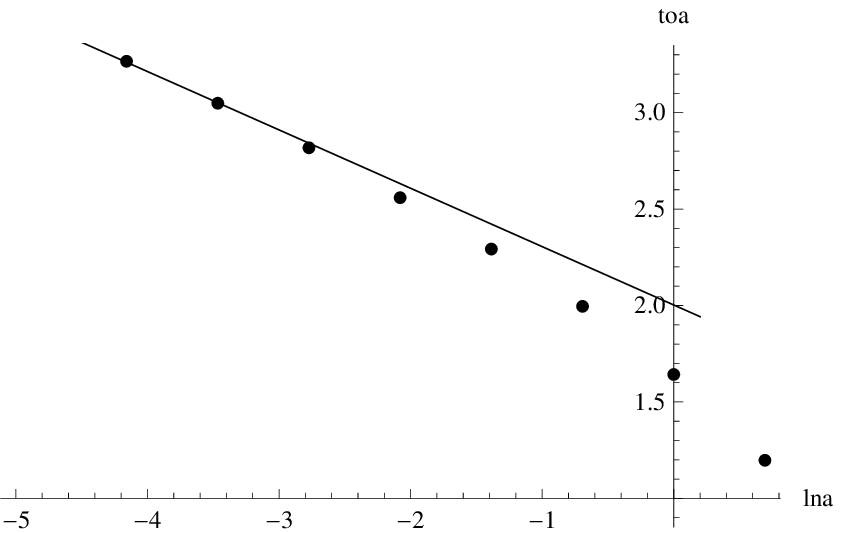}
  \includegraphics[width=3in]{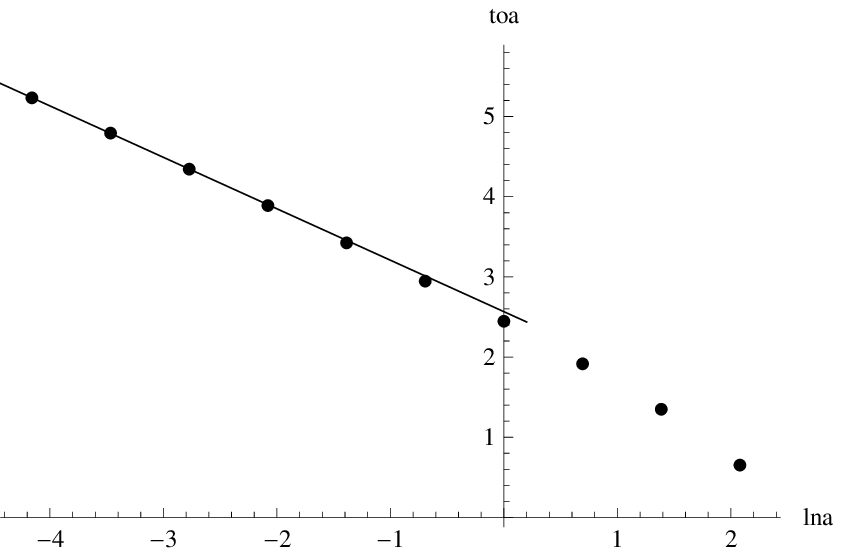}
  \includegraphics[width=3in]{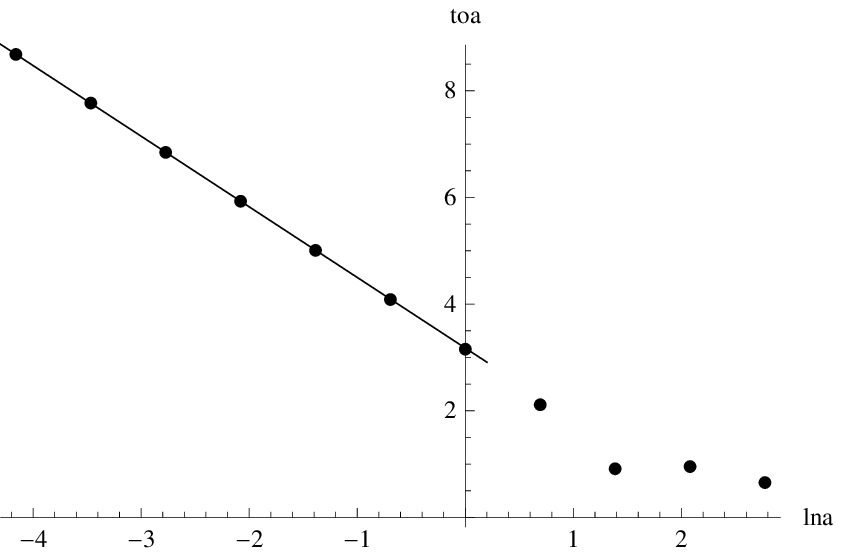}
  \includegraphics[width=3in]{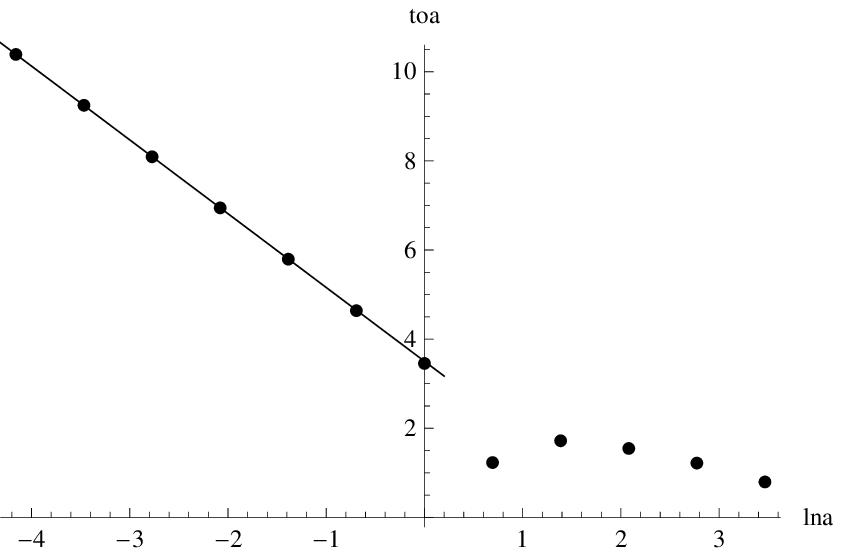}
\end{center}
  \caption{
Plots of $\frac{|\t_{ex}|}{\a}$ as a function of $\log\a$.
The straight lines indicate the asymptotic behaviour for fast quenches.  Clockwise from the
top left, the plots are for $\DD=7/3$, $8/3$, $11/3$ and $10/3$.}
\label{excite}
\end{figure}

Figure \ref{excite} shows the rescaled excitation time $|\t_{ex}|/\a$ as a
function of $\log\a$ for different values of $\D$. For fast quenches (\ie
$\log\a\leq0$), we see that $|\t_{ex}|/\a$ approaches a straight line,
indicating that the excitation time scales as $\beta_\D\,\a\log\a$.
Once again the straight lines are fit through the three points with the
points corresponding to the fastest quenches, for each $\D$.
The constants $\beta_\Delta$ can be determined as the slope of the fitted line in
these plots. The plots also show that, as expected, the behaviour becomes
irregular for $\log\a>0$, in particular for $\D=10/3$ and $11/3$.

\begin{figure}[t]
\begin{center}
\psfrag{Delta}[c]{{${\DD}$}}
\psfrag{y}[B][tl]{{${-\frac{\partial|\t_{ex}/\a|}{\partial\log\a}}$}}
 \includegraphics[width=5in]{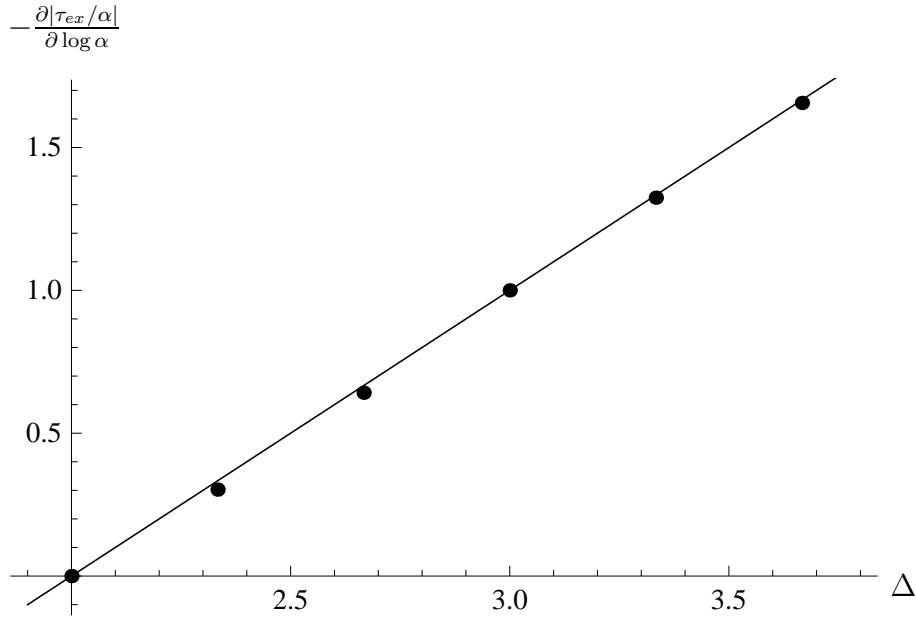}
\end{center}
  \caption{
A plot of (minus) the slope of the fitted straight lines in figure
\ref{excite}.  The datapoints lie approximately on the predicted line
$-\frac{\partial|\t_{ex}/\a|}{\partial\log\a}=\DD-2$ shown. Also shown are the data
for $\DD=2$ and 3 from \cite{blm}.} \label{excitemeta}
\end{figure}

Figure \ref{excitemeta} shows the slopes of the straight-line fits in the
previous plots as a function of $\D$. This plot also includes data-points for
$\D=2$ and $3$ using the results in \cite{blm}.\footnote{For these two points,
we have used the expressions for $|\t_{ex}|/\alpha$ after the
$\log\left(-\log\a\right)$ terms have been subtracted.} The fact that
$\frac{\partial|\t_{ex}|/\a}{\partial\log\a}=0$ for $\D=2$ means that in this
case $\t_{ex}$ has no $\log\a$-dependence for this type of quench. We find that
the datapoints lie approximately on the line
\begin{equation}
-\frac{\partial|\t_{ex}|/\a}{\partial\log\a} = \DD-2\,,
\labell{outside}
\end{equation}
which is numerically almost identical to the fitted line
through all the data-points shown in the figure, namely
\begin{equation}
-\frac{\partial|\t_{ex}|/\a}{\partial\log\a} = 1.003\DD-2.02\,.
\labell{outside2}
\end{equation}
Hence for fast quenches, \ie $\a\ll1$, the excitation time scales as
\begin{equation}
\t_{ex} \simeq (\DD-2)\,\a\,\log\a\,.
\labell{eq:extime}
\end{equation}

Using the same approach, we also studied the relaxation time $\t_{eq}$. In this
case for all of the various $\D$, we found
\begin{equation}
\t_{eq} \approx \a^{0}\,,
\labell{one}
\end{equation}
for fast quenches. That is, $\t_{eq}$ is constant for small $\a$, rather than
scaling with $\a$ in some way, for all (fractional) $\DD$. This behaviour is consistent
with the results for $\DD=2$ and $3$, found in \cite{blm}. In terms of the dimensionful time
coordinate $t$, the relaxation time is $t_{eq}\sim1/\mu$. In terms of the
boundary theory then, the relaxation time is set by the thermal timescale for
fast quenches.

\subsection{Behaviour of the energy and pressure} \label{behaviour}

Recall that our analysis was restricted to considering conformal dimensions in
the range $2<\D<4$. With this restriction, the change in the temperature
$\Delta T$ and energy density $\Delta \mathcal{E}$ will always be positive in
our holographic quenches, as is evident from eqs.~(\ref{eq:tiotf2}) and
(\ref{eq:efeifw}). On the other hand, eqs.~(\ref{eq:tiotfbw}) and
(\ref{eq:efeibw}) show that $\Delta T$ and $\Delta \mathcal{E}$ may have either
sign for the reverse quenches considered in section \ref{reverse}. In
particular, in the adiabatic limit, $\tilde{a}_{2,4}(\infty)$ vanishes and so
we have both $\Delta T<0$ and $\Delta \mathcal{E}<0$. Then as $\a$ becomes
smaller, $\tilde{a}_{2,4}(\infty)$ grows and eventually $\Delta T$ and $\Delta
\mathcal{E}$ become positive. Specifically, eqs.~(\ref{eq:tiotfbw}) and
(\ref{eq:efeibw}) indicate:
\begin{eqnarray}
\Delta T>0\ \ {\rm for}\qquad&&
|\tilde{a}_{2,4}(\infty)| > \frac{2}{3}\left(\D-2\right)^{2}\,
|\tilde{\phi}_{(2\D-4)}(-\infty)|\,,
\labell{wham1}\\
\Delta
\mathcal{E}>0\ \ {\rm for}\qquad&&
|\tilde{a}_{2,4}(\infty)| > \frac{1}{3}\left(\D-2\right)|\tilde{\phi}_{(2\D-4)}
(-\infty)|\,. \labell{wham2}
\end{eqnarray}
Recall that $\tilde{\phi}_{(2\D-4)}(-\infty)={\phi}_{(2\D-4)}(\infty)$
corresponds to the equilibrium response given in eq.~\reef{eq:black}. With our
numerical simulations, we determined the value of $\a$ at which these
thresholds are reached for various values of $\D$ and the results are shown in
table \ref{table1}. We have also included the analogous results for $\D=2$ and
$3$ from \cite{blm}.  Note the qualitative trend is that the threshold value of
$\alpha$ grows monotonically as $\Delta$ increases. Note that eqs.~\reef{wham1}
and \reef{wham2} imply that the threshold for positive $\D\mathcal{E}$ is
greater than that for positive $\D T$ (\ie larger $\a$) for $\D>2.5$, while the
thresholds are reversed for $\D<2.5$. Clearly, this behaviour is reflected in
the results shown in table \ref{table1}.

\begin{table}[ht]
\caption{Approximate upper bounds on $\a$ for which $\D T>0$ and
$\D\mathcal{E}>0$ for reverse quenches. The upper bounds on $\a$ for $\D\mathcal{P}>0$ is for forward
quenches when $\D<3.5$, with different values of $\D$. For $\D=2$ the values of $\D\cale$ and $\D\calp$ are renormalization scheme dependent, see \cite{blm}. \label{table1}}
\centering
\begin{tabular}{c c c c}
&&&\\
\hline\hline
$\D$ & $\D T$ & $\D\mathcal{E}$ & $\D\mathcal{P}$ \\ [0.5ex]
\hline
2   & 0.58 &      &      \\
7/3 & 0.68 & 0.50 & 0.23 \\
8/3 & 0.77 & 0.92 & 0.68 \\
3   & 0.86 & 1.32 & 1.32     \\
10/3 & 1.00 & 2.00 & 5.5 \\
11/3 & 1.41 & 4.00 & -- \\ [1ex]
\hline
\end{tabular}
\label{table:nonlin}
\end{table}

Turning now to the change in the pressure $\D\mathcal{P}$, we have
eqs.~(\ref{eq:pfpifw}) and \reef{eq:pfpibw} for forward and reverse quenches,
respectively. In this case, $\D\mathcal{P}>0$ in all forward quenches as long
as $\D>7/2$ and in all reverse quenches for $\D<7/2$. Otherwise, the sign of
$\D\mathcal{P}$ will depend on the rate of the quench. Here we focus on the
forward quenches with $\D<7/2$. In this case, $\D\mathcal{P}<0$ for slow
quenches and as $\a$ decreases, the change in the pressure reverses its sign
when
\begin{equation}
\Delta
\mathcal{P}>0\ \ {\rm for}\qquad
|{a}_{2,4}(\infty)| > \frac{1}{3}\left(7-2\D\right)
\left(\D-2\right)|{\phi}_{(2\D-4)}
(\infty)|\,. \labell{wham3}
\end{equation}
The value of $\a$ at which these thresholds is reached for various values of
$\D$ is shown in table \ref{table1}. Again, we have also included an analogous
result for $\D=3$  \cite{blm}.  The qualitative trend is that the
threshold value of $\alpha$ grows monotonically as $\Delta$ increases. Further,
we may compare the above threshold to those in eqs.~\reef{wham1} and
\reef{wham2}.\footnote{Note that here we are comparing a threshold for the forward
quenches to thresholds in the reverse quenches.} In particular, the threshold
for positive $\D\mathcal{P}$ is greater than that for positive $\D\mathcal{E}$
for $\D>3$, while the thresholds are reversed for $\D<3$. We also find the
threshold for positive $\D\mathcal{P}$ is less than that for positive
$\D T$ when $\D<2.75$. Clearly, this behaviour is
reflected in the results shown in table \ref{table1}.

\section{Discussion} \label{discussion}

In this paper, we continued the program initiated in \cite{blm} of studying
quantum quenches in strongly coupled quantum field theories using holography.
The process studied here was the quench of a four-dimensional conformal field
theory made with a rapid transition in the coupling  of a relevant operator
$\mathcal{O}_\D$ from zero to some finite value $\lambda_f$. Ref.~\cite{blm}
had considered the special cases where the conformal dimension of the operator
was $\D=2$ and 3. Our holographic analysis allowed for operators with general
conformal dimensions in the range $2<\D<4$. Through the gauge/gravity
correspondence \cite{m1}, this quench was translated to a classical problem in
five-dimensional Einstein gravity coupled to a negative cosmological constant
and a (free) massive scalar field. In particular, the quench was implemented by
introducing a time-dependent boundary condition on the scalar field in the
asymptotically AdS$_5$ spacetime. Our discussion also only considered quenches
of a thermal plasma with an initial temperature $T_i$ in the boundary theory
and was limited to a high temperature regime\footnote{In fact, this inequality
is satisfied by the coupling and temperature throughout the quench.} where
$\lambda_f\ll T_f^{4-\D}$. Hence our calculations were perturbative in the
ratio $\l_f/T_f^{4-\D}$, which in the gravitational description meant that they
were perturbative in the (dimensionless) amplitude of the bulk scalar. However,
there was no restriction on the timescale $\D t$ governing the transition rate
of the coupling. In particular, with the transition profiles in
eq.~\reef{eq:tanh}, our results were described in terms of the dimensionless
parameter: $\a=\pi T_i\,\D t$, \ie the ratio of the transition timescale to the
relaxation timescale of the thermal plasma. In our analysis, we paid special
attention to the limit of adiabatic transitions with $\a\to\infty$ and of very
fast quenches with $\a\to0$. A detailed discussion of the results was given in
section \ref{results}.

In all of these quenches, our computations implicitly gave the response in the
one-point correlators of the relevant operator $\la \calo_{\Delta}\ra$ and the
stress tensor $\la T_{ij}\ra$, as described by
eqs.~\reef{eq:endens}--\reef{eq:vev}, to leading order in $\l_f/T_f^{4-\D}$. In
section \ref{results}, our discussion of the results was presented in terms of
two gravitational parameters, $\phi_{{2\D-4}}(\t)$ and $a_{2,4}(\infty)$. With
eq.~\reef{eq:vev}, we see the first is directly related to
$\la\mathcal{O}_\D\ra$ during the quench. We can rewrite this expression as
\begin{equation}
\langle \mathcal{O}_{\Delta} \rangle = \frac{\pi^4\,C_T}{40}\left(\Delta-2 \right)
\frac{\l_f}{T_f^{4-2\D}}\ \frac{\phi_{{2\D-4}}(t/\pi T_f)}{|\phi_{{2\D-4}}(\infty)|}
\,,
\labell{newvev}
\end{equation}
using eq.~\reef{g5b} and various results from section \ref{entropy}. Further
$a_{2,4}(\infty)$ controls the entropy production \reef{eq:sfosi}, as well as
the changes in the temperature, energy and pressure as given in
eqs.~\reef{eq:tiotf2}--\reef{eq:pfpifw}. One confirmation of our numerical
simulations was that we found $a_{2,4}(\infty)\le0$ for all of our quenches
with any values of $\a$ and $\D$. With eq.~\reef{eq:sfosi}, the latter ensures
that the entropy production was always positive, in accord with the second law
of thermodynamics.

Another confirmation comes from our analysis of slow quenches in section
\ref{slowquenches}. In this case with $\a\gg1$, it was shown that the
linearized equation \reef{eq:phi} for the bulk scalar can be solved using a
power series in $1/\alpha$. While $a_{2,4}(\infty)$ vanishes for the leading
adiabatic solution, we showed that a $1/\a$ contribution appears at the next
order in eq.~\reef{eq:a24newb}. Using a shooting method, we explicitly solved
for this contribution and the results were given in eq.~\reef{eq:a24analytic}.
This same leading order contribution to $a_{2,4}(\infty)$ for slow quenches is
addressed with the numerical results shown in figure \ref{loga24slow}. In these
plots, the same approximate $1/\a$ scaling was found with an asymptotic
 straight-line fit in $\log\a$ for $\a\gg1$, \ie
\begin{equation}
 \log|a_{2,4}(\infty)| = -\,c - \log\a\,.
 \labell{kapow99}
\end{equation}
The case $\D=7/3$ had the worst fit, with a slope differing from $-1$ by about
$18\%$. We believe that the
slow convergence in our numerical simulations for this case led to this relatively
large error. The other three cases in table \ref{table2} had slopes that differed
from the expected slope by $2\%$ or less. As well as obtaining a fair match in
the slope above, we can compare the intercept $c$ coming from these numerical
results with that calculated from the independently derived results in
eq.~\reef{eq:a24analytic}. We see in table \ref{table2} that the intercepts
derived from the two approaches agree very well. Recall that in figure
\ref{p12hatp0slow}, we also showed that the full time-dependent profile of the
numerical response (after subtracting the adiabatic profile) matched well with
the form derived in section \ref{slowquenches} for $\a\gg1$.

\begin{table}[ht]
\caption{Intercept in eq.~\reef{kapow99} evaluated by two different methods:
$c_{shoot}$ is derived from the results in eq.~\reef{eq:a24analytic}, while
$c_{numer}$ comes from fitting the data in figure \ref{loga24slow}.
\label{table2}} \centering
\begin{tabular}{c c c c}
&&&\\
\hline\hline
$\D$ & $c_{shoot}$ & $c_{numer}$ &
$\frac{c_{shoot}-c_{numer}}{c_{shoot}}$  \\ [0.5ex]
\hline
7/3 & 3.93 & 3.71 & 5.6\% \\
8/3 & 2.96 & 2.98 & -0.68\% \\
10/3 & 2.31 & 2.26 & 2.2\% \\
11/3 & 2.22 & 2.21 & 0.45\% \\ [1ex]
\hline
\end{tabular}
\end{table}

In eq.~\reef{eq:extime}, we found an interesting scaling behaviour for the
excitation time in fast quenches, namely $\t_{ex} \simeq (\DD-2)\,\a \log\a$.
On the one hand, this indicates that the excitation time is longer when the
conformal dimension of the operator is larger but it also shows that $\t_{ex}$
becomes shorter for faster quenches. In particular, $\t_{ex}\to0$ for $\a\to0$,
for which the quench profile \reef{eq:tanh} becomes a step-function at $\t=0$.
In contrast, as shown in eq.~\reef{one}, the relaxation time remains constant
for $\a\ll1$. Hence independent of the precise values of $\D$ and $\a$, the
boundary system relaxes on the thermal timescale $1/T$ for fast quenches.

Perhaps, the most interesting result coming from our analysis was the universal
behaviour found in $|\phi_{{2\D-4}}|$ and $a_{2,4}(\infty)$ for $\a\ll1$. Of
course, in terms of the boundary theory, these results translate into universal
behaviour in the response of $\langle \mathcal{O}_{\Delta} \rangle$ and in the
thermodynamic quantities for fast quenches. First, the results in figure
\ref{p12fast} indicate that the maximum value of $|\phi_{{2\D-4}}|$ scales as
$\a^{-(2\D-4)}$, which with eq.~\reef{newvev}, translates into a scaling for
the expectation value of the quenched operator, \ie
\begin{equation}
{\rm max}\,\langle \mathcal{O}_{\Delta} \rangle \propto \frac{1}{\a^{2\D-4}}
\,. \labell{newvevooo}
\end{equation}
Beyond this scaling, figure \ref{p12p0fast} also indicates that the response
and hence $\langle \mathcal{O}_{\Delta} \rangle$ approach a relatively simple
universal form in the limit $\a\to0$. These results are in agreement with those
found previously in \cite{blm}. However, we might comment that our analysis
here assumed that $\D$ was a fraction, whereas \cite{blm} studied the special
cases $\D=2$ and $3$. In both of these cases, the response also exhibited an
additional contribution which scaled faster than shown in eq.~\reef{newvevooo}
by an extra logarithmic factor. An extra logarithmic factor is also present for
$\D=4$ \cite{blmn}.

The above scaling of $|\phi_{{2\D-4}}|$ also leads to the scaling of
$a_{2,4}(\infty)\propto\a^{-(2\D-4)}$ for fast quenches, as shown with the
numerical data in figure \ref{a24meta}. This contribution then dominates in
eqs.~\reef{eq:sfosi}--\reef{eq:pfpifw} and so the changes of the various
thermodynamic quantities induced by fast quenches exhibit the same scaling, \eg
\begin{equation}
\frac{\Delta\mathcal{E}}{\mathcal{E}_i}\propto \frac{1}{\a^{2\D-4}}
 \labell{newjuice}
\end{equation}
for $\a\ll1$. Again these results are in agreement with those found in
\cite{blm}. However, the results there and in \cite{blmn} indicate that
eq.~\reef{newjuice} is further enhanced by a logarithmic scaling for
$\Delta=2$. While quenches by operators with conformal dimensions in the range
$2\le\D\le4$ are covered by the analysis here and in \cite{blm}, it would be
interesting to understand if this universal behaviour extends to the allowed
regime $1\leq\DD<2$. We will consider this possibility in a later paper
\cite{blmn}.

As noted in \cite{blm}, given the scaling in eqs.~\reef{newvevooo} and
\reef{newjuice}, it appears that `infinitely fast' quenches seem to be
ill-defined because physical quantities are diverging as $\a\to0$. Recall that
in this limit, the quench profile \reef{eq:tanh} becomes a step-function at
$\t=0$. Hence this issue is particularly notable since it is precisely such
`infinitely fast' quenches are studied in the seminal work on this topic
\cite{cc1}. However, we must contrast their description of a quench with the
present approach. In \cite{cc1}, the system is evolved from $t=-\infty$ to
$0^-$ to prepare the system in a far-from-equilibrium state of the `quenched'
Hamiltonian, \ie, the ground state of the initial Hamiltonian. This state is
then used as the initial condition at $t=0^+$ and the subsequent evolution of
the system with the `quenched' Hamiltonian is studied.

Of course, since the present calculations are only perturbative in
$\l_f/T_f^{4-\D}$, one can not take the singularities appearing in
eqs.~\reef{newvevooo} and \reef{newjuice} for $\a\to0$ too seriously. Hence it
would be interesting to study the fast quenches by evolving the full nonlinear
equations of the dual gravity theory. At present, our preliminary analysis
suggests that in fact these singularities are physical \cite{blmn}. In any
event, the present holographic calculations illustrate that the gauge/gravity
correspondence provides a versatile new framework for the study of quantum
quenches. Undoubtedly, interesting new lessons will come from applying
holography to study more general physical quantities and the behaviour of more
complicated systems under a quench. This will help build our intuition for the
behaviour of fast-changing quantum fields that occur when the external
parameters are changed in laboratory experiments.

%%%%%%%%%%%%%%%%%%%%%%%%%%%%%%%%%%%%%%%%
\section*{Acknowledgments}
AvN would like to thank Ross Diener and Alex Yale for helpful discussions.
Research at Perimeter Institute is supported by the Government of Canada
through Industry Canada and by the Province of Ontario through the Ministry of
Research \& Innovation. AB, LL and RCM gratefully acknowledge support from
NSERC Discovery grants. Research by LL and RCM is further supported by funding
from the Canadian Institute for Advanced Research.

\appendix

\section{Coefficients in the metric solution} \labell{coeff}

Here we list the expressions of the coefficients in the metric functions
(\ref{eq:sola}) and (\ref{eq:sols}) in terms of the normalizable and
non-normalizable modes of the scalar, $\phi_{(0)}$ and $\phi_{(2\Delta-4)}$. We
only list the coefficients that are needed (and the first subleading
coefficient) in calculating the boundary stress tensor and the expectation
value of the operator $\mathcal{O}_\D$.

Because $\dot{a}_{2,4}$ depends on it, we will give the expression for $a_{2,5}$,
even though it is subleading and has only vanishing contributions to physical quantities:
\begin{equation}
a_{2,5}=\frac{1}{18} \left(\Delta  (2 \Delta-5) \dot{\phi}_{(0)}\phi_{(2\Delta-2)} -(2\Delta-3 ) (4-\Delta ) \phi_{(0)}\dot{\phi}_{(2\Delta-2)}\right).
\end{equation}

The coefficients of the terms with negative powers of $2\Delta$ in
$A_{\textrm{p}}$ are given by
\begin{eqnarray}
\alpha_{2,4}&=&-\frac{(4-\Delta ) \phi_{(0)}^2}{6 (7-2 \Delta)};\label{alpha4}\\
\alpha_{2,5}&=&\frac{(\Delta -3) \phi_{(0)} \dot{\phi}_{(0)}}{3 (7-2 \Delta)};\\
\alpha_{2,6}&=&\frac{(2\Delta^2-6\Delta +15 ) \dot{\phi}_{(0)}^2
+(2 \Delta^2-13\Delta+24 )  ) \phi_{(0)} \ddot{\phi}_{(0)}}{12 (\Delta-3 ) (9-2 \Delta)};\\
\alpha_{2,7}&=&\frac{(4-\D )\left( -3(\D-2 ) (9-2 \Delta) \dot{\phi}_{(0)} \ddot{\phi}_{(0)}
+(2 \Delta^2-15\Delta+36 )  ) \phi_{(0)} \dddot\phi_{(0)}\right)}{
36 (5-\D ) (\D-3 ) (9-2 \Delta)};\label{alpha7}\\
\alpha_{2,8}&=&-\Big(12 (5-\Delta ) (4-\Delta )^2 (\Delta-3 )^2 (2 \Delta^2-11\Delta+30 ) \phi_{(0)}^2\nonumber\\
&&-3 (7-2 \Delta )^2 (5-\Delta ) (2 \Delta^3-19\Delta^2+56\D-54) \ddot{\phi}_{(0)}^2\nonumber\\
&&-4 (\Delta-3 ) (7-2 \Delta ) (2 \Delta^2-14\Delta+21 )(2 \Delta^2-17\Delta+39 )  \dot{\phi}_{(0)} \phi^{(3)}_{(0)}\nonumber\\
&&+(4-\Delta ) (\Delta-3 ) (9-2 \Delta ) (2 \Delta-7 ) (2 \Delta^2-17\Delta+51 ) \phi_{(0)} \phi^{(4)}_{(0)}\Big)\nonumber\\
&&\Big/\left(288 (5-\Delta ) (4-\Delta ) (\Delta-3 )^2 (4 \Delta^2-36\Delta+77 )\right).
\end{eqnarray}
Of the
coefficients to the terms with positive of powers $2\Delta$ in
$A_{\textrm{p}}$, the coefficient
\begin{equation}
\beta_{2,4}=\frac{\Delta \, \phi_{(2\Delta-2)}^2}{6 (2 \Delta-1 )}
\end{equation}
will be subleading.

The coefficients in $\Sigma_{\textrm{p}}$ corresponding to integer powers of $\rho$ are given by
\begin{eqnarray}
s_{2,5}&=&-\frac{ \Delta \,(4-\Delta )\,  \phi_{(0)} \phi_{(2\D-4)}}{36};\\
s_{2,6}&=&-\frac{1}{60} \left( \Delta\,(5-\Delta )\,  \dot{\phi}_{(0)}\phi_{(2\D-4)} +(4-\Delta ) (\Delta +1)\, \phi_{(0)} \dot{\phi}_{(2\D-4)}\right).
\end{eqnarray}
The coefficients of the terms with negative powers of $2\Delta$ in
$\Sigma_{\textrm{p}}$ are given by
\begin{eqnarray}
\sigma_{2,5}&=&-\frac{(4-\Delta )  \phi_{(0)}^2}{12(7-2\D) };\\
\sigma_{2,6}&=&-\frac{(5-\Delta )  \phi_{(0)} \dot{\phi}_{(0)}}{6  (9-2 \Delta )};\\
\sigma_{2,7}&=&\frac{2 (5-\Delta )^2 (\Delta-3 ) \dot{\phi}_{(0)}^2+(6-\Delta ) (4-\Delta ) (7-2 \Delta )  \phi_{(0)}\ddot{\phi}_{(0)}}{24 (5-\Delta ) (\Delta-3 ) (9-2 \Delta )};\\
\sigma_{2,8}&=&\frac{3 (6-\Delta ) (5-\Delta ) (7-2 \Delta ) \dot{\phi}_{(0)}\ddot{\phi}_{(0)}}{72 (5-\Delta ) (\Delta-3 ) (11-2 \Delta )}\nonumber\\
&&+\frac{(7-\Delta ) (4-\Delta ) (9-2 \Delta ) \phi_{(0)} \phi^{(3)}_{(0)}}{72 (5-\Delta ) (\Delta-3 ) (11-2 \Delta )};\\
\sigma_{2,9}&=&-\Big(12 (8-\Delta ) (4-\Delta )^2 (\Delta-3 )^2 \phi_{(0)}^2-3 (7-2 \Delta )^2 (6-\Delta )^2 \ddot{\phi}_{(0)}^2\nonumber\\
&&-8 (7-\Delta ) (5-\Delta ) (\Delta-3 ) (9-2 \Delta ) \dot{\phi}_{(0)} \phi^{(3)}_{(0)}\nonumber\\
&&-(8-\Delta ) (\Delta-3 ) (11-2 \Delta ) (9-2 \Delta ) \phi_{(0)} \phi^{(4)}_{(0)}\Big)\nonumber\\
&&\Big/\Big({576 (6-\Delta ) (\Delta-3 )^2 (11-2 \Delta )}\Big).
\end{eqnarray}
Of the coefficients to the terms with
positive of powers $2\Delta$ in $\Sigma_{\textrm{p}}$, the coefficient
\begin{equation}
\theta_{2,5}=\frac{\Delta\,  \phi_{(2\Delta-2)}^2}{12 (2 \Delta-1 )}
\end{equation}
is subleading.

\section{Coefficients in the Fefferman-Graham coordinates} \label{fefferman}

\subsection{The time and radial coordinates}

The expansion of the EF time and radial coordinates in terms of the FG time and
radial coordinates was given in eqs.~(\ref{eq:v}) and (\ref{eq:rho}).  The
expressions of the coefficients can be written in terms of the coefficients
$\phi_{(0)}$, $\phi_{(2\Delta-4)}$ and $a_{2,4}$.  First, the terms containing
$v_{n}$ and $\rho_{n}$ are given by the series
\begin{eqnarray}
\sum_{n}v_{n}r^{n} &=&  -r-\frac{3 r^5 \mu ^4}{40}-\frac{11 r^9 \mu ^8}{1152}-\frac{23 r^{13} \mu ^{12}}{13312} + \dots,\\
\sum_{n}\rho_{n}r^{n} &=&-\frac{r^5 \mu ^5}{8}+\frac{3 r^9 \mu ^9}{128}-\frac{5 r^{13} \mu ^{13}}{1024} + \dots.
\end{eqnarray}
Next, the terms of order $\ell^{2}$ are given by
\begin{eqnarray}
\sum_{n=5}\vartheta_{n}r^{n} &=& \frac{3}{40} r^5 \mu ^4 a_{2,4}+\frac{1}{240} r^6 \mu ^4 \left(16 \mu  a_{2,5}-11 \partial_{t}a_{2,4}\right)+ \dots,\\
\sum_{n=5}\chi_{n}r^{n} &=& \frac{1}{8} r^5 \mu ^5 a_{2,4}+\frac{1}{10} r^6 \mu^5  \left(\mu a_{2,5}- \partial_{t}a_{2,4}\right)+ \dots,
\end{eqnarray}
where the terms of order $r^6$ only make subleading contributions in the
calculated quantities. The coefficients of the terms with factors of
$r^{-2\Delta}$ in $\t/\mu$ are given by
\begin{eqnarray}
\nu_{0}&=&\frac{(7-2 \Delta ) \mu ^{2 \Delta-8 } \alpha_{2,4}}{4 (4-\Delta ) (9-2 \Delta )};\nonumber\\
\nu_{1}&=&\frac{\mu ^{8-2 \Delta } \left(\left(4 \Delta ^2-30 \Delta+55\right) \partial_{t}\alpha_{2,4}+4 (4-\Delta )^2 \mu  \alpha_{2,5}\right)}{8 (5-\Delta ) (11-\Delta ) (9-2 \Delta)};\nonumber\\
\nu_{2}&=&\mu ^{8-2 \Delta }\frac{ \left(4 \Delta ^2-32 \Delta+61 \right) \partial^{2}_{t}\alpha_{2,4}}{8 (5-\Delta ) (11-2 \Delta ) (9-2 \Delta )}\nonumber\\
&&+\mu ^{8-2 \Delta }\frac{-2 \left(4 \Delta ^2-34 \Delta+71\right) \mu  \partial_{t}\alpha_{2,5}+2 (9-2 \Delta )^2 \mu ^2 \alpha_{2,6}}{8 (5-\Delta ) (11-2 \Delta ) (9-2 \Delta )};\nonumber\\
\nu_{3}&=&-\mu ^{8-2 \Delta }\frac{\left(4 \Delta ^2-34 \Delta+67 \right) \partial^{3}_{t}\alpha_{2,4}-6(2\Delta^2-18\Delta+39 ) \mu\partial^{2}_{t}\alpha_{2,5}}{48 (6-\Delta ) (5-\Delta ) (11-2 \Delta )}\nonumber\\
&&-\mu ^{8-2 \Delta }\frac{6\left(4 \Delta ^2-38 \Delta+89 \right) \mu^2  \partial_{t}\alpha_{2,6}-24 (5-\Delta )^2 \mu ^3 \alpha_{2,7}}{48 (6-\Delta ) (5-\Delta ) (11-2 \Delta )},
\end{eqnarray}
where $\alpha_{2,n}$ are the
coefficients defined in eq,~(\ref{eq:sola}) and given explicitly in eqs.~\reef{alpha4}--\reef{alpha7} --- implicitly, functions of the FG
time $t$ here. Similarly, the coefficients in $\rho$ with factors of
$r^{-2\Delta}$ are given by
\begin{eqnarray}
\xi_{0}&=&\frac{\mu ^{9-2 \Delta } \alpha_{2,4}}{4 (4-\Delta) };\\
\xi_{1}&=&-\frac{\mu ^{9-2 \Delta } \left(\partial_{t}\alpha_{2,4}-\mu  \alpha_{2,5}\right)}{2 (9-2\Delta) };\\
\xi_{2}&=&\frac{\mu ^{9-2 \Delta } \left(\partial^{2}_{t}\alpha_{2,4}-2 \mu  \partial_{t}\alpha_{2,5}+2 \mu ^2 \alpha_{2,6}\right)}{8 (5-\Delta)};\\
\xi_{3}&=&-\frac{\mu ^{9-2 \Delta } \left(\partial^{3}_{t}\alpha_{2,4}-3 \mu  \partial^{2}_{t}\alpha_{2,5}+6 \mu ^2 \partial_{t}\alpha_{2,6}-6 \mu ^3 \alpha_{2,7}\right)}{12 (11-2 \Delta)};\\
\xi_{4}&=&-\frac{\mu ^{9-2 \Delta } \left( \left(2 \Delta ^2-13 \Delta+30 \right) \mu ^4 \alpha_{2,4}-(4-\D)\partial^{5}_{t}\alpha_{2,4}\right)}{32 (6-\Delta ) (4-\Delta )}\nonumber\\
&&-\frac{\mu ^{9-2 \Delta }\left( 4 \mu  \partial^{3}_{t}\alpha_{2,5}-12 \mu ^2 \partial^{2}_{t}\alpha_{2,6}+24 \mu ^3 \partial_{t}\alpha_{2,7}-24 \mu ^4 \alpha_{2,8}\right)}{96 (6-\Delta ) }.
\end{eqnarray}
The coefficients of the terms with
factors of $r^{2\Delta}$ in $v$ are given by
\begin{eqnarray}
\omega_{1}&=&\frac{(2 \Delta-1 ) \mu ^{2 \Delta } \beta_{2,4}}{4 \Delta  (2 \Delta+1 )};\\
\omega_{2}&=&-\frac{\mu ^{2 \Delta } \left(\left(4 \Delta ^2-2 \Delta-1\right) \partial_{t}\beta_{2,4}-4 \Delta ^2 \mu  \beta_{2,5}\right)}{8 \Delta  (\Delta+1 ) (2 \Delta+1 )};\\
\omega_{3}&=&\mu ^{2 \Delta } \frac{\left(4 \Delta ^2-3\right) \partial^{2}_{t}\beta_{2,4}}{8 \left(4 \Delta ^3+12 \Delta ^2+11 \Delta+3 \right)}\nonumber\\
&&-\mu ^{2 \Delta }\frac{2 \left(4 \Delta ^2+2 \Delta-1\right) \mu  \partial_{t}\beta_{2,5}-2 (2 \Delta+1    )^2\mu^{2} \beta_{2,6}}{8 \left(4 \Delta ^3+12 \Delta ^2+11 \Delta+3 \right)};\\
\omega_{4}&=&-\mu ^{2 \Delta }\frac{(4 \Delta^2+2 \Delta -5) \partial^{3}_{t}\beta_{2,4}-6(2 \Delta^2+2 \Delta-1)\mu\partial^{2}_{t}\beta_{2,5}}{48 (\Delta+1 ) (\Delta+2 ) (2 \Delta+3 )}\nonumber\\
&&-\mu ^{2 \Delta }\frac{ \left(4 \Delta ^2+6 \Delta+1 \right) \mu^2  \partial_{t}\beta_{2,6}-4 (\Delta+1 )^2 \mu ^3 \beta_{2,7}}{12 (\Delta+1 ) (\Delta+2 ) (2 \Delta+3 )},
\end{eqnarray}
where $\beta_{2,n}$ are the
coefficients defined in eq.~(\ref{eq:sola}) --- implicitly, functions of the FG
time $t$ here. Similarly, the coefficients in $\rho$ with factors of
$r^{2\Delta}$ are given by
\begin{eqnarray}
\zeta_{1}&=&\frac{\mu ^{2 \Delta +1} \beta_{2,4}}{4 \Delta };\\
\zeta_{2}&=&-\frac{\mu ^{2 \Delta +1} \left(\partial_{t}\beta_{2,4}-\mu  \beta_{2,5}\right)}{2 (2 \Delta+1 )};\\
\zeta_{3}&=&\frac{\mu ^{2 \Delta +1} \left(\partial^{2}_{t}\beta_{2,4}-2 \mu  \partial_{t}\beta_{2,5}+2 \mu ^2 \beta_{2,6}\right)}{8 (\Delta+1 )};\\
\zeta_{4}&=&-\frac{\mu ^{2 \Delta +1} \left(\partial^{3}_{t}\beta_{2,4}-3 \mu  \partial^{2}_{t}\beta_{2,5}+6 \mu ^2 \partial_{t}\beta_{2,6}-6 \mu ^3 \beta_{2,7}\right)}{12(2 \Delta+3) };\\
\zeta_{5}&=&-\mu ^{2 \Delta +1}\frac{  3\left(2 \Delta ^2-3 \Delta+10 \right) \mu ^4\beta_{2,4} -\D\partial^{4}_{t}\beta_{2,4}}{96 \Delta  (\Delta+2 )}\nonumber\\
&&-\mu ^{2 \Delta +1}\frac{\Delta  \left(4 \mu  \partial^{3}_{t}\beta_{2,5}-12 \mu ^2 \partial^{2}_{t}\beta_{2,6}+24 \mu ^3 \partial_{t}\beta_{2,7}-24 \mu ^4 \beta_{2,8}\right)}{96 \Delta  (\Delta+2 )}\,.
\end{eqnarray}

\subsection{The metric}

The expansion of the metric in the FG coordinates was given in
eq.~(\ref{eq:efmetric}).  The nonzero components for our purposes are the parts
$G_{00}$, which correspond to the EF metric function $-A$, and the diagonal
components of $G_{ij}$, which correspond to $\Sigma^{2}$.  The order $\ell^{0}$
terms in the metric are given by
\begin{eqnarray}
g^{(0)}_{00} + r^{4}\,g^{(4)}_{00} &=& -1+\frac{3 r^4 \mu ^4}{4} + o\left(r^{8}\right),\\
g^{(0)}_{ii} + r^{4}\,g^{(4)}_{ii} &=& 1+\frac{\mu ^4 r^4}{4} + o\left(r^{8}\right),
 \label{bdymetric}
\end{eqnarray}
where in the second line, we are indicating the three individual diagonal
components, \ie there is no implicit sum over $i$. Next we list the terms of
order $\ell^{2}$:
\begin{eqnarray}
c_{(4)00} &=& -\frac{3}{4} \mu ^4 a_{2,4}\,,\\   %r^{4} + \frac{2}{5} \mu ^4 (2 \mu  a_{2,5}-\partial_{t}a_{2,4}),\\
c_{(4)ii} &=& -\frac{\mu ^4}4 \left(  a_{2,4}+\frac{2}{9}  \Delta(4-\Delta )\,
  \phi_{(0)}  \phi_{(2\Delta-4)}\right)\,.
\end{eqnarray}
The coefficients of the terms with factors of $r^{-2\Delta}$ in $G_{00}$ are given by
\begin{eqnarray}
d_{(0)00}&=&-\frac{(7-2 \Delta ) \mu ^{8-2 \Delta }  \alpha_{2,4}}{2 (4-\Delta )};\\
d_{(1)00}&=&-2 \mu ^{8-2 \Delta }\frac{(\Delta-3 ) \partial_{t}\alpha_{2,4}+ (4-\Delta ) \mu  \alpha_{2,5}}{9-2 \Delta };\\
d_{(2)00}&=&\mu ^{8-2 \Delta }\frac{(\Delta-3 ) (7-2 \Delta ) \partial^{2}_{t}\alpha_{2,4}}{4 (5-\Delta ) (4-\Delta )}\nonumber\\
&&+\mu ^{8-2 \Delta }\frac{ 2(7-2 \Delta ) \mu\partial_{t}\alpha_{2,5}-2(9-2 \Delta ) \mu^{2}  \alpha_{2,6}}{4 (5-\Delta ) };\\
d_{(3)00}&=&-\mu ^{8-2 \Delta }\frac{(\Delta-3 ) (7-2 \Delta ) \partial^{3}_{t}\alpha_{2,4}+3(4-\Delta ) (7-2 \Delta ) \mu\partial^{2}_{t}\alpha_{2,5}}{3 ( 4 \Delta^2-40\D+99 )}\nonumber\\
&&+\mu ^{8-2 \Delta }\frac{ (4-\Delta ) \partial_{t}\alpha_{2,6}-2 (9-2 \Delta ) \mu^{2}  \left((5-\Delta ) \mu  \alpha_{2,7}\right)}{ 4 \Delta^2-40\D+99  };\\
d_{(4)00}&=&-\mu ^{8-2 \Delta }\frac{\left(2 \Delta ^3-25 \Delta ^2+107 \Delta-162 \right) \mu ^{4} \alpha_{2,4}}{8 (6-\Delta ) (5-\Delta )}\nonumber\\
&&+\mu ^{8-2 \Delta }\frac{4(4-\Delta ) (7-2 \Delta )\mu \partial^{3}_{t}\alpha_{2,5}-12(9-2 \Delta )(4-\Delta ) \mu^{2}\partial^{2}_{t}\alpha_{2,6}}{48 (6-\Delta ) (5-\Delta )}\nonumber\\
&&+\mu ^{8-2 \Delta }\frac{24 (9-2 \Delta )\mu^{3}  \partial_{t}\alpha_{2,7}-24 (11-2 \Delta ) \mu ^4 \alpha_{2,8}}{48 (6-\Delta )}\nonumber\\
&&+\mu ^{8-2 \Delta }\frac{(\Delta-3 ) (7-2 \Delta ) \partial^{4}_{t}\alpha_{2,4}}{48 (6-\Delta ) (5-\Delta )},
\end{eqnarray}
where $\alpha_{2,n}$ are the
coefficients defined in eq.~(\ref{eq:sola}) --- implicitly, functions of the FG
time $t$ here. Similarly, the coefficients in $G_{ii}$ with factors of
$r^{-2\Delta}$ are given by
\begin{eqnarray}
d_{(0)ii}&=&-\mu ^{8-2 \Delta } \frac{\alpha_{2,4}-4 (4-\Delta ) \sigma_{2,5} }{2 (4-\Delta )};\\
d_{(1)ii}&=&\mu ^{8-2 \Delta }\frac{\partial_{t}\alpha_{2,4}-2 (9-2 \Delta ) \partial_{t}\sigma_{2,5}-\mu  \alpha_{2,5}+2 (9-2 \Delta ) \mu  \sigma_{2,6}}{9-2 \Delta };\\
d_{(2)ii}&=&-\mu ^{8-2 \Delta }\frac{\partial^{2}_{t}\alpha_{2,4}-4(5- \Delta)  \partial^{2}_{t}\sigma_{2,5}}{4 (5-\Delta )}\nonumber\\
&&+\mu ^{8-2 \Delta }\frac{\mu\partial_{t}\alpha_{2,5}-4 (5-\Delta )\mu \partial_{t}\sigma_{2,6}  -\mu^{2}  \alpha_{2,6}+4 (5-\Delta ) \mu^{2}  \sigma_{2,7}}{2 (5-\Delta )};\\
d_{(3)ii}&=&\frac{\mu ^{8-2 \Delta }}{6(11-2\D) }
\Big(\partial^{3}_{t}\alpha_{2,4}-2(11-2 \Delta)  \partial^{3}_{t}\sigma_{2,5}-
3 \mu\partial^{2}_{t}\alpha_{2,5}+6(11-2\Delta)  \mu\partial^{2}_{t}\sigma_{2,6}\nonumber\\
&&\ \ \ +\partial_{t}\alpha_{2,6}-2 (11-2 \Delta ) \partial_{t}\sigma_{2,7}-6 \mu^{2}
\left(\mu  \alpha_{2,7}+2 (11-2 \Delta ) \mu  \sigma_{2,8}\right)\Big),
\end{eqnarray}
where $\sigma_{2,n}$ are the
coefficients defined in (\ref{eq:sols})  --- implicitly, functions of the FG
time $t$ here.

The coefficients of the terms with factors of $r^{2\Delta}$ in $G_{00}$ are given by
\begin{eqnarray}
e_{(0)00}&=&-\frac{(2 \Delta-1 )\, \mu ^{2 \Delta } \beta_{2,4}}{2 \Delta };\\
e_{(1)00}&=&2 \mu ^{2 \Delta }\frac{ (\Delta-1 ) \partial_{t}\beta_{2,4}-\Delta  \mu  \beta_{2,5}}{2 \Delta+1 },
\end{eqnarray}
where $\beta_{2,n}$ is defined in eq.~(\ref{eq:sola}).  Similarly, in $G_{ii}$
the coefficients of the $r^{2\Delta}$ are given by
\begin{eqnarray}
e_{(0)ii}&=&-\mu ^{2 \Delta }\frac{ \beta_{2,4}-4 \Delta  \theta_{2,5}}{2 \Delta  };\\
e_{(1)ii}&=&\mu ^{2 \Delta } \frac{ \partial_{t}\beta_{2,4}-2 (2 \Delta+1 )  \partial_{t}\theta_{2,5}-\mu  \beta_{2,5}+2 (2 \Delta+1) \mu\tau_{2,6}}{2 \Delta+1},
\end{eqnarray}
where $\tau_{2,n}$ was defined in eq.~(\ref{eq:sols}).

\subsection{The scalar field} \label{fgscalar}

The coefficients of the scalar field $f_{(n)}$ and $g_{(n)}$ in  the FG
coordinates can be written in terms of the coefficients $\phi_{(0)}$ and
$\phi_{(2\Delta-4)}$ in eq.~(\ref{eq:phip}).  The coefficients of
$r^{4-\Delta}$ in $\Phi_{\textrm{p}}$ are given by
\begin{eqnarray}
\sum_{n=0}f_{(n)}(t)r^{n} &=& \mu^{4-\D}\Big(\phi_{(0)}+r \partial_{t}\phi_{(0)}-\frac{r^2 (7-2 \Delta ) \partial^{2}_{t}\phi_{(0)}}{4 (\Delta-3 ) }-\frac{r^3 (9-2 \Delta ) \partial^{3}_{t}\phi_{(0)}}{12 (\Delta-3 ) }\nonumber\\
&&+r^4 \left(\frac{1}{8} (4-\Delta )\mu ^4 \phi_{(0)}-\frac{(4\Delta^2-40\Delta+99 ) \partial^{4}_{t}\phi_{(0)}}{96 (4-\Delta) (\Delta-3 ) }\right) + o\left(r^{5}\right)\Big), \nonumber \\
&& \label{eq:source}
\end{eqnarray}
while the only coefficient of
$r^{\Delta}$ that may play a role is
\begin{equation}
g_{(0)} = \mu ^{\Delta }\, \phi_{(2\Delta-4)}.
\end{equation}

\section{Boundary stress-energy tensor and $\langle\mathcal{O}_{\Delta}\rangle$}
\labell{stress}

In section \ref{renorm}, we constructed the holographic action
$S_{reg}=S_{bulk} + S_{GHBY} + S_{count}$ which is now finite, \ie no
divergences appear in the limit $\epsilon\to0$. Hence, so are all quantities
that can be calculated from this action. Of course, the latter includes the
one-point functions of the stress tensor and operator $\mathcal{O}_{\Delta}$.
In order to calculate these expectation values, we need to vary $S_{reg}$ with
respect to the boundary metric and the coupling to boundary operator,
respectively. Recall that the boundary metric $g^{(0)}_{ab}$ appears as the
leading coefficient in the expansion \reef{eq:fgmetric} of the bulk metric.
Similarly, the coupling $\lambda$ is proportional to the leading coefficient
$f_{(0)}$ in the expansion \reef{eq:fgscalar} of the bulk scalar. We will
establish our conventions for the precise normalization of the coupling in
section \ref{finalt} and so at this point, we simply introduce a dimensionless
proportionality constant\footnote{The constant $\pp$ is fixed in
eq.~\reef{ppx}.} with $\ell f_{(0)}=\pp\,\lambda$. Then the desired one-point
functions are given by \cite{cascade}
\begin{eqnarray}
8\pi G^{(5)}_{N}\,\langle T^{ab} \rangle &=& \lim_{\epsilon \to 0}
\frac{16\pi G^{(5)}_{N}}{\sqrt{-g^{(0)}}}
\frac{\delta S_{reg}}{\delta g^{(0)}_{ab}} \nonumber \\
&=&  \lim_{\epsilon \to 0}\frac{16\pi G^{(5)}_{N}}{\sqrt{-\gamma}\epsilon^{4}}
\left(\frac{\delta S_{reg}}{\delta \gamma_{cd}}
\frac{\delta \gamma_{cd}}{\delta g^{(0)}_{ab}} + \frac{\delta S_{reg}}{\delta \Phi}
\frac{\delta \Phi}{\delta g^{(0)}_{ab}}\right) \labell{eq:stress}
\end{eqnarray}
and
\begin{eqnarray}
16\pi G^{(5)}_{N}\,\langle \mathcal{O}_{\Delta} \rangle &=&
\lim_{\epsilon \to 0}\frac{16\pi G^{(5)}_{N}}{\sqrt{-g^{(0)}}}
\frac{\delta S_{reg}}{\delta \lambda}\ =\
\lim_{\epsilon \to 0}\frac{16\pi G^{(5)}_{N}\,\pp}{\sqrt{-g^{(0)}}\,\ell}
\frac{\delta S_{reg}}{\delta f_{(0)}} \nonumber \\
&=&  \lim_{\epsilon \to 0}\frac{16\pi G^{(5)}_{N}\,\pp}{\sqrt{-\gamma}\,\ell\,\epsilon^{4}}
\left(\frac{\delta S_{reg}}{\delta \gamma_{ab}}\frac{\delta \gamma_{ab}}{\delta f_{(0)}}
+ \frac{\delta S_{reg}}{\delta \Phi}\frac{\delta \Phi}{\delta f_{(0)}}\right)
 \labell{eq:vev1}\,.
\end{eqnarray}

We proceed by first evaluating the variations of the action with respect to
$\gamma$ and $\Phi$, then the variations of $\gamma$ and $\Phi$ with respect to
$g^{(0)}_{ab}$ and $f_{(0)}$.  The variation of the action on the cut-off surface
$r=\epsilon$ is:
\begin{align}
\frac{16\pi G^{(5)}_{N}}{\sqrt{-\gamma}}\,\frac{\delta\left( S_{\textrm{bulk}} + S_{\textrm{GHBY}}  \right)}{\delta
\gamma_{ab}} &= \frac{r}{2}\left( \gamma^{ac}\gamma^{bd}\partial_{r}\gamma_{cd}
- \gamma^{ab}\gamma^{cd}\partial_{r}\gamma_{cd} \right)\Big{|}_{r=\epsilon} \,,
 \nonumber \\
\frac{16\pi G^{(5)}_{N}}{\sqrt{-\gamma}}\,\frac{\delta\left( S_{\textrm{bulk}} + S_{\textrm{GHBY}}  \right)}{\delta
\Phi} &= r \partial_{r}\Phi\Big{|}_{r=\epsilon}\,, \nonumber \\
\frac{16\pi G^{(5)}_{N}}{\sqrt{-\gamma}}\,\frac{\delta S_{\textrm{count}} }{\delta \gamma_{ab}}
&= -\frac{1}{2}\gamma^{ab}\left( 6 + \frac{4-\Delta}{2}\Phi^2 -
\frac{1}{4\left(\Delta-3\right)}\left(\partial\Phi\right)^{2} \right)\Big{|}_{r=\epsilon}\, \nonumber\\
 +&\frac{1}{24\left(\Delta-3\right)}\Big(\left(\nabla^{a}\nabla^{b}-\gamma^{ab}\,\Box\right)\Phi^{2}
-6\gamma^{ac}\gamma^{bd}
\partial_{c}\Phi\,\partial_{d}\Phi\Big)\Big{|}_{r=\epsilon}\,, \nonumber\\
\frac{16\pi G^{(5)}_{N}}{\sqrt{-\gamma}}\,\frac{\delta S_{\textrm{count}} }{\delta \Phi}
&= -\left(4-\Delta\right) \Phi
- \frac{1}{2\left(\Delta-3\right)} \Box \Phi \Big{|}_{r=\epsilon}\,.
 \labell{eq:actionvar}
\end{align}
In the above we only showed the terms that would contribute at the orders of $\ell$ we are considering.
Using the results of appendix \ref{fefferman}, we find the variation of the
bulk fields with respect to their boundary values (at leading order) are
\begin{eqnarray}
\frac{\delta \gamma_{cd}}{\delta g^{(0)}_{ab}} &=&  \delta^{a}_{(c}\delta^{b}_{d)}
\,r^{-2}\Big{|}_{r=\epsilon}\,, \qquad\qquad\quad\
\frac{\delta \Phi}{\delta g^{(0)}_{ab}} = \textrm{o}\left(r^{5-\Delta}\right)\,,
\nonumber \\
\frac{\delta \gamma_{ab}}{\delta f_{(0)}} &=& - \frac{1}{6}\eta_{ab}\,\ell^{2}\, f_{(0)} r^{6-2\Delta}
\Big{|}_{r=\epsilon}\,,\qquad\quad
\frac{\delta \Phi}{\delta f_{(0)}} = \ell \,r^{4-\Delta}\Big{|}_{r=\epsilon}\,.
\labell{eq:fieldsvar}
\end{eqnarray}
The second equation in eq.~(\ref{eq:fieldsvar}) leads to a vanishing
contribution to the stress tensor in eq.~(\ref{eq:stress}).  The third equation
above only contributes to eq.~(\ref{eq:vev1})when $\Delta=4$ and hence can be
ignored for our purposes. Inserting the variations from
eqs.~(\ref{eq:actionvar}) and (\ref{eq:fieldsvar}) into eqs.~(\ref{eq:stress})
and (\ref{eq:vev1}), we get the results presented in the main text in
eqs.~\reef{eq:endens}--\reef{eq:vev}. Note that for these final results, we
have replaced $f_{(0)}$ and $g_{(0)}$ by their expressions in terms of
$\phi_{(0)}$ and $\phi_{(2\Delta-4)}$ given in appendix \ref{fgscalar}.

\end{document}